\documentclass[aps, prl, reprint, superscriptaddress]{revtex4-2}
\usepackage{amsmath}
\usepackage{dcolumn}
\usepackage{bm}
\usepackage{epsfig}
\usepackage{epstopdf}
\usepackage{graphicx}
\usepackage{amsfonts}
\usepackage{amssymb}
\usepackage{mathrsfs}
\usepackage{color}
\usepackage{multirow}
\usepackage{natbib}
\usepackage{hyperref}
\usepackage{braket}
\usepackage{xcolor}

\usepackage{appendix}
\usepackage{float}
\usepackage{soul}
\sethlcolor{yellow}

\usepackage[caption=false]{subfig} 

\newcommand{\APM}{Wuhan Institute of Physics and Mathematics, Innovation Academy of Precision Measurement Science and Technology, Chinese Academy of Sciences, Wuhan 430071, China}

\newcommand{\UCAS}{University of Chinese Academy of Sciences, Beijing 100049, China}
\newcommand{\GZIIT}{Research Center for Quantum Precision Measurement, Guangzhou Institute of Industry Co., LTD, Guangzhou, 511458, China}

\newcommand{\NUDT}{Institute for Quantum Science and Technology, College of Science, National University of Defense Technology, Changsha 410073, China}
\newcommand{\HNU}{Key Laboratory of Low-Dimensional Quantum Structures and Quantum Control of Ministry of Education, Department of Physics and Synergetic Innovation Center for Quantum Effects and Applications, Hunan Normal University, Changsha 410081, China}
\newcommand{\ZNU}{Department of Physics, Zhejiang Normal University, Jinhua 321004, China}

\begin{document}

	\title{Finite-Time Electrometry with a Quantum-Regime Single-Ion Phonon Laser}

	\author{Pei-Dong Li}
	  \email{lipeidong21@mails.ucas.ac.cn}
	\affiliation{\APM}
	\affiliation{\UCAS}

    \author{Yuan-Zhang Dong}
	\affiliation{\APM}
	\affiliation{\UCAS}

   \author{Zhuo-Zhu Wu}
	\affiliation{\APM}
	\affiliation{\UCAS}

	\author{Jia-Wei Wang}
	\affiliation{\APM}
	\affiliation{\UCAS}

  	\author{Ji Li}
	\affiliation{\GZIIT}

	\author{Jian-Qi Zhang}
	\affiliation{\APM}

    \author{Zhi-Jiao Deng}
    \affiliation{\NUDT}
   
	\author{Liang Chen}
	\email{liangchen@wipm.ac.cn}
	\affiliation{\APM}

	\author{Mang Feng}
	\email{mangfeng@wipm.ac.cn}
	\affiliation{\APM}
    \affiliation{\GZIIT}
    \affiliation{\HNU}
	\affiliation{\ZNU}

\begin{abstract}
The phonon laser realized in a trapped ion, i.e., a self-sustained mechanical oscillator, has demonstrated the unique characteristics in practically detecting externally applied electric signals without the prerequisite of sideband cooling. Entering the quantum regime via sideband cooling is expected to further improve its sensing performance. Here we report the first experimental realization of a quantum-regime single-ion phonon laser ($\bar{n}<10$) using a trapped $^{40}\mathrm{Ca}^+$ ion and demonstrate electrometry based on its phase-space symmetry-breaking response to weak resonant electric fields. By tuning the phonon-laser parameters, we reveal that the sensing performance is fundamentally governed by the finite-time relaxation dynamics of the underlying open quantum system. We find that a slow Liouvillian relaxation, correlated with the finite experimental interaction window, effectively enhances the dynamic susceptibility while maintaining the structural robustness of the limit cycle. This regime, when applied to the detection of electric fields, produces a shot-noise-limited peak sensitivity of $14.15 \pm 0.77~\mu\mathrm{V/m}/\sqrt{\mathrm{Hz}}$ and a minimum detectable field variation of $\delta E_{\mathrm{min}} \approx 1.83~\mu\mathrm{V/m}$. Our results establish quantum phonon lasers as a practical platform for advanced sensing and highlight the central role of Liouvillian dynamics in non-equilibrium electrometry.
\end{abstract}

\maketitle

\begin{figure*}[htbp]
    \centering
    \includegraphics[width=1.0\textwidth]{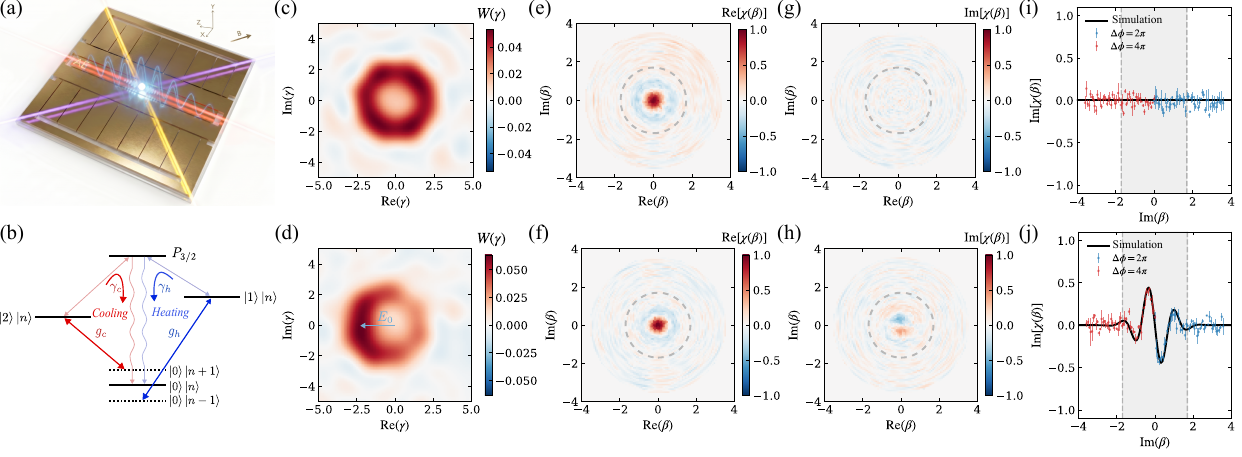}
\caption{\textbf{Experimental platform and phase-space dynamics of the single-ion phonon laser sensor.}
    (a) Schematic of a $^{40}\text{Ca}^{+}$ ion in a surface-electrode trap (SET), driven by an oscillating electric field via the axial electrode (AE).
    (b) Energy-level and laser-coupling scheme for the heating and cooling transitions.
    (c,d) Reconstructed Wigner functions of the undriven phase-isotropic (c) and driven phase-converged (d) states.
    (e-h) Measured real (e,f) and imaginary (g,h) parts of the motional characteristic function $\chi(\beta)$ without (e,g) and with (f,h) the external drive.
    (i,j) Slices of $\text{Im}[\chi(\beta)]$ along $\text{Re}(\beta)=0$ without (i) and with (j) the drive. Circles (s.e.m. from 300 measurements) and solid lines denote experimental data and numerical simulations, respectively. Couplings are fixed at $g_c/2\pi = g_h/2\pi = 4.56$~kHz.}
    \label{fig:1}
\end{figure*}

Phonon lasers, analogous to their optical counterparts, are fundamental tools for coherent mechanical amplification and non-equilibrium physics~\cite{Wallentowitz1996,Jing2014PT}. Although macroscopic and semiclassical phonon lasing has been widely explored on various platforms~\cite{vahala2009phononlaser,2010cavityPL,Pettit2019tweezerPL}, realizing self-sustained mechanical oscillations deep in the quantum regime~\cite{Behrle2023quantumPL}---characterized by vanishingly small mean phonon numbers ($\bar{n} < 10$)---at the single-particle level remains a formidable challenge~\cite{DONG2025singleionPL,baur2026quantumtheoryphononlasing}. Such quantum non-equilibrium states, governed by profound quantum fluctuations, open new frontiers for both fundamental physics and advanced quantum sensing~\cite{degen2017quantumsensing,Burd2019sensing,Gilmore2021sensing}.

Specifically, driven-dissipative systems offer a powerful resource for electrometry through dynamic phase-space symmetry breaking~\cite{Ivanov2015SBsensing,Lorenzo2017DPTsensing}. Unlike conventional protocols detecting the linear coherent displacement of a ground-state wavepacket, the phonon laser's non-equilibrium limit cycle leverages quantum synchronization dynamics~\cite{Lee2013Sync, Walter2014Sync,Lorch2016Sync,Li2025Sync,liu2025Sync}. A weak resonant electric field acts as a symmetry-breaking seed, overcoming quantum phase diffusion to drive this phase-isotropic phonon-laser state into a phase-converged distribution. However, exploiting this non-equilibrium response imposes a fundamental temporal constraint. The dynamic timescale of an open quantum system is dictated by its Liouvillian relaxation rate~\cite{Vicentini2018CriticalSlow,minganti2018spectral,garbe2020critical}. Enhancing the susceptibility physically necessitates a reduced relaxation rate, which inevitably lengthens the time required for the system to respond. Since practical sensing protocols are strictly constrained by finite experimental interrogation windows, the ultimate sensing performance is fundamentally governed by this strict interplay between the amplified susceptibility and the finite-time relaxation.

In this Letter, we report the first experimental realization of a quantum-regime single-ion phonon laser and demonstrate an electrometry protocol governed by the finite-time Liouvillian scaling~\cite{Vicentini2018CriticalSlow,minganti2018spectral}. By engineering tailored heating and cooling channels on a trapped $^{40}\mathrm{Ca}^+$ ion, we deterministically drive the motional mode into a quantum limit cycle. By systematically tuning the phonon-laser parameters, we reveal that the finite-time sensitivity monotonically improves as the Liouvillian gap—the spectral feature dictating the slowest relaxation rate of the open system~\cite{minganti2018spectral}—narrows. We push this dynamic enhancement to a robust practical boundary, where the relaxation timescale approaches the finite experimental interaction window. Beyond this regime, indefinitely weakening the coupling surrenders the dissipative stabilization, rendering the limit cycle inherently vulnerable to trap heating and readout instabilities. By operating near this practical boundary, we effectively maximize the dynamic susceptibility while maintaining a deterministically stable non-equilibrium attractor. By translating microscopic external perturbations into high-contrast phase-space signatures~\cite{fluhmann2020direct}, we apply this protocol to the detection of electric fields, yielding a shot-noise-limited peak sensitivity of $14.15 \pm 0.77~\mu\mathrm{V/m}/\sqrt{\mathrm{Hz}}$ and a minimum detectable field variation of $\delta E_{\mathrm{min}} \approx 1.83~\mu\mathrm{V/m}$.

Our experiment uses a single $^{40}\mathrm{Ca}^+$ ion in a surface-electrode trap (SET)~\cite{Li2025SET,Wu2026LEP} with an axial frequency $\omega_{z}/2\pi = 676.35$~kHz [Fig.~\ref{fig:1}(a)]. A static field of $0.52$~mT defines the quantization axis. The relevant energy levels [Fig.~\ref{fig:1}(b)] comprise $\ket{0} \equiv \ket{4^2S_{1/2}, m_J = +1/2}$, $\ket{1} \equiv \ket{3^2D_{5/2}, m_J = +5/2}$, and $\ket{2} \equiv \ket{3^2D_{5/2}, m_J = +3/2}$.

The system gain and dissipation are engineered using tailored laser configurations~\cite{SM}. Two $729$-nm laser beams independently drive the motional sideband transitions of the ion. Specifically, the $\ket{0} \leftrightarrow \ket{2}$ transition driven on the first red sideband ($-\omega_z$) realizes the Jaynes-Cummings Hamiltonian $\hat{H}_c = \frac{i\hbar}{2} g_c (\hat{a}^\dagger \hat{\sigma}_-^c - \hat{a}\hat{\sigma}_+^c)$~\cite{Leibfried2003RMP}. Concurrently, the $\ket{0} \leftrightarrow \ket{1}$ transition on the first blue sideband ($+\omega_z$) realizes the anti-Jaynes-Cummings Hamiltonian $\hat{H}_h = \frac{i\hbar}{2} g_h (\hat{a}^\dagger \hat{\sigma}_+^h - \hat{a}\hat{\sigma}_-^h)$. The respective coupling strengths are $g_c = \eta \Omega_c$ and $g_h = \eta \Omega_h$, where $\eta = 0.114$ is the Lamb-Dicke parameter.

Dissipation is implemented via $854$-nm quenching beams that primarily couple $\ket{1}$ and $\ket{2}$ to the short-lived $\ket{4^2P_{3/2}, m_J = +3/2}$ state, which rapidly decays to $\ket{0}$. The resulting dissipative dynamics are described by the Lindblad operators $\hat{L}_h = \sqrt{\gamma_h}\,\hat{\sigma}_-^h$ and $\hat{L}_c = \sqrt{\gamma_c}\,\hat{\sigma}_-^c$. Population leakage to other Zeeman levels is repumped by $397$-nm $\sigma_+$-polarized and $866$-nm laser beams.

To investigate the sensing capability, a weak oscillating electric field resonant with $\omega_z$ is applied through the axial electrode (AE). The interaction is governed by $\hat{H}_{\mathrm{drive}} = i\hbar\epsilon(\hat{a}e^{i\phi_d} - \hat{a}^{\dagger}e^{-i\phi_d})$, where $\epsilon = E_0 q z_0 /(2\hbar)$ denotes the drive strength proportional to the target field amplitude $E_0$ with $z_0 = \sqrt{\hbar/(2m\omega_z)}$ the harmonic-oscillator length and $m$ the ion mass. For phase-stable detection, the external driving field is phase-synchronized with the phonon-lasing dynamics, fixing the drive phase at $\phi_d = 0$.

To fully capture the quantum feature of the single-ion phonon-laser dynamics, we model the system density matrix $\rho$ using the Lindblad master equation
\begin{equation}
\label{Eq:master}
\frac{\mathrm{d}\rho}{\mathrm{d}t}
=
\mathcal{L}\rho
=
-\frac{i}{\hbar}[\hat{H},\rho]
+ \mathcal{D}[\hat{L}_h](\rho) + \mathcal{D}[\hat{L}_c](\rho),
\end{equation}
where $\mathcal{L}$ denotes the Liouvillian superoperator, $\hat{H} = \hat{H}_c + \hat{H}_h + \hat{H}_{\text{drive}}$, and the dissipator follows the standard form $\mathcal{D}[\hat{L}_k](\rho) = \hat{L}_k \rho \hat{L}_k^\dagger - \frac{1}{2} \{ \hat{L}_k^\dagger \hat{L}_k, \rho \}$, with fixed decay rates $\gamma_h=15~\mathrm{kHz}$ and $\gamma_c=150~\mathrm{kHz}$. While the long-time asymptotic steady state $\rho_{\mathrm{ss}}$ satisfies $\mathcal{L}\rho_{\mathrm{ss}}=0$, the transient dynamics are governed by the nonzero eigenvalues of $\mathcal{L}$. Crucially, lacking true spectral closing in this finite system, the Liouvillian gap $\Delta_{\mathcal L} = -\mathrm{Re}(\lambda_1)$---with $\lambda_1$ being the nonzero eigenvalue of smallest real part---robustly dictates the open-system relaxation timescale $\tau \sim 1/\Delta_{\mathcal{L}}$. 

\begin{figure*}[tbp]
    \centering
    \includegraphics[width=1.0\textwidth]{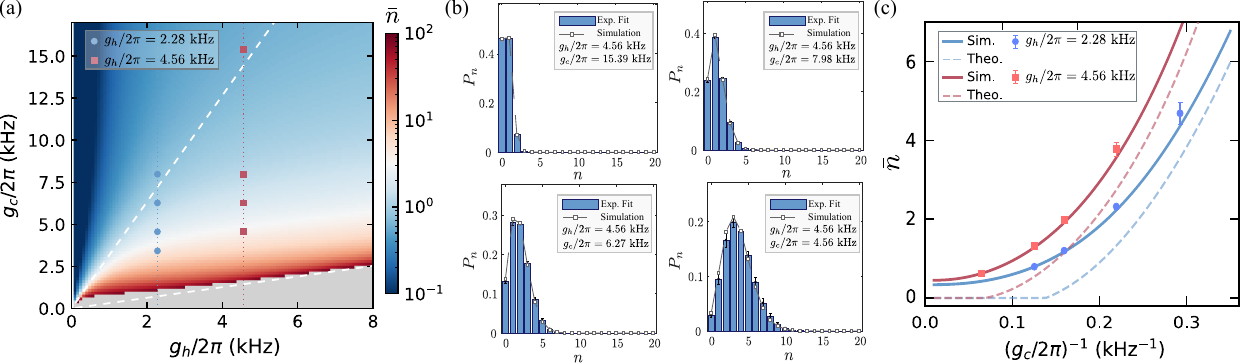}
\caption{\textbf{Observation of phonon lasing and steady-state distributions.}
(a) Phase diagram of steady-state $\bar{n}$ versus $g_h$ and $g_c$ (Hilbert space truncated at $n=100$). Dashed lines show mean-field boundaries separating dark, lasing, and instability regimes; gray areas denote $\bar{n} > 80$. Markers indicate constant-$g_h$ slices.
(b) Measured (bars) and simulated (curves) phonon-number distributions at $g_h/2\pi=4.56$~kHz.
(c) Measured $\bar{n}$ vs $g_c$ along the slices in (a), compared with quantum simulations (solid lines) and mean-field theory (dashed lines). Error bars denote $\pm 1\sigma$ uncertainties via Monte Carlo analysis \cite{SM}.}
    \label{fig:2}
\end{figure*}

To experimentally establish the phonon laser, we initialize the ion near its motional ground state ($\bar{n}_{\mathrm{init}} \approx 0.3$) via resolved sideband cooling. Subsequently, the engineered couplings and dissipations are applied for a fixed interaction time $t_{\mathrm{pl}} = 5.7$~ms. In the semiclassical mean-field limit \cite{SM}, dynamic saturation restricts the mechanical growth and stabilizes the system into a robust limit cycle with a deterministic steady-state phonon occupancy,
\begin{equation}
\label{Eq:n_ss}
n_{\mathrm{ss}} = \frac{ (\gamma_c + \gamma_h) \left(\dfrac{g_h^2}{\gamma_h} - \dfrac{g_c^2}{\gamma_c}\right) }{ \left(\dfrac{g_c^2}{\gamma_h}-\dfrac{g_h^2}{\gamma_c}\right) \left(\dfrac{g_h^2}{\gamma_h}+\dfrac{g_c^2}{\gamma_c}\right) }.
\end{equation}

The sensing mechanism leverages the dynamic symmetry breaking of this state. Without the external drive, the experimentally reconstructed Wigner function exhibits a phase-isotropic limit cycle [Fig.~\ref{fig:1}(c)]. In contrast, under a preliminary calibrated weak signal field ($E_0 \approx 100~\mu\mathrm{V/m}$), the resonant drive shears the isotropic ring into a phase-converged distribution localized along the drive axis [Fig.~\ref{fig:1}(d)].

This field-induced symmetry breaking manifests as high-contrast oscillatory features in the motional characteristic function $\chi(\beta)$. Figures~\ref{fig:1}(e)--\ref{fig:1}(h) present $\chi(\beta)$, mapped via a bichromatic light field implementing a spin-dependent motional displacement~\cite{Leibfried1996Wigner,fluhmann2020direct}. This synthesizes $\beta = -i \eta \Omega t e^{-i\Delta\phi/2}$, with the Rabi frequency fixed at $\Omega/2\pi = 50$~kHz. The profiles are reconstructed across 32 distinct phases, and we adopt a structural cutoff of $|\beta| = 1.7$ during reconstruction to filter out long-term technical fluctuations while preserving low-order quantum dynamics. Compared to the vanishing signal in the unperturbed scenario [Figs.~\ref{fig:1}(g) and \ref{fig:1}(i)], the driven phonon laser exhibits robust oscillations in $\text{Im}[\chi(\beta)]$ against a flat zero baseline [Figs.~\ref{fig:1}(h) and \ref{fig:1}(j)], demonstrating this symmetry-breaking mechanism as an exceptionally sensitive signature for electrometry.

Using Eq.~(\ref{Eq:master}), we numerically compute the steady-state mean phonon number $\bar{n}$ across the parameter space spanned by the coupling strengths $g_h$ and $g_c$, yielding the phase diagram shown in Fig.~\ref{fig:2}(a). Equation~(\ref{Eq:n_ss}) outlines the stable lasing window, bounded by the lasing threshold $g_h/g_c = \sqrt{\gamma_h/\gamma_c}$ and the instability line $g_h/g_c = \sqrt{\gamma_c/\gamma_h}$.

These mean-field boundaries physically correspond to dissipative phase transitions. Strictly speaking, however, a non-analytic transition with a rigorously closing Liouvillian gap emerges only in the thermodynamic limit ($\bar{n} \rightarrow \infty$). In our genuinely quantum single-ion implementation, this boundary is regularized into a smooth crossover by quantum fluctuations. This is clearly evidenced by the excellent quantitative agreement of the reconstructed distributions in Fig.~\ref{fig:2}(b) and the pronounced deviation of the experimental mean phonon numbers from classical mean-field predictions in Fig.~\ref{fig:2}(c).

To translate this highly tunable symmetry-breaking response into a robust electrometry protocol, we systematically investigate how the phonon-laser parameters govern the ultimate sensitivity. The imaginary part of the motional characteristic function is directly mapped onto the experimentally measurable spin population~\cite{SM}:
\begin{equation}
\label{Eq:P_up}
P_\uparrow(t) = \frac{1}{2}\left\{1 - \text{Im}\left[\chi(\beta)\right]\right\}.
\end{equation}

\begin{figure}[tbp]
\centering
\includegraphics[width=1.0\columnwidth]{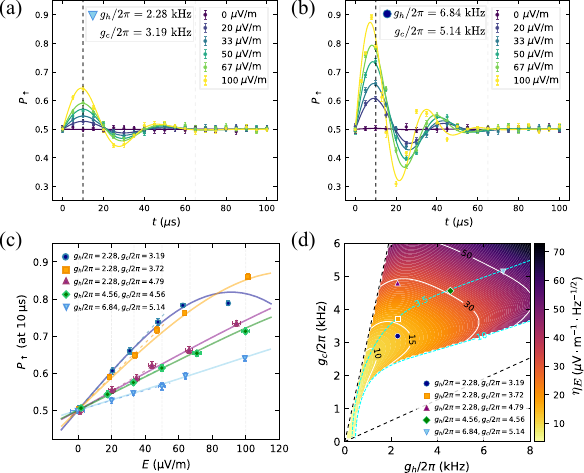}
\caption{\textbf{Phonon-laser electrometry and sensitivity.}
(a,b) Time-resolved spin population $P_\uparrow$ under calibrated electric fields $E$. Markers (with s.e.m. from 3000 cycles) and solid curves denote experimental data and numerical fits, respectively. Vertical dashed lines mark the optimal sensing time $t_s$. 
(c) Measured (symbols) and simulated (solid curves) $P_\uparrow(t_s)$ versus field $E$. Dashed lines are linear fits for sensitivity extraction. 
(d) Simulated sensitivity $\eta_E$ mapped across the phase diagram at $t_{\mathrm{pl}}$. White solid and cyan dashed contours indicate $\eta_E$ and the final mean phonon number, respectively. Parameters in (c,d) are in kHz.}
\label{fig:3}
\end{figure}

Figures~\ref{fig:3}(a) and \ref{fig:3}(b) show the time-resolved spin population $P_\uparrow$ for representative operating parameters, where the relative phase of the bichromatic probe is fixed at $\Delta\phi = 2\pi$ throughout the measurements. To suppress long-term drifts, we extract the effective electric-field strength $E$ by fitting numerical open-system trajectories to these early-time population dynamics ($t < 65~\mu\mathrm{s}$). Remarkably, across a wide range of preliminary calibrated field strengths, the maximal population response consistently occurs at a well-defined optimal sensing time $t_s = 10~\mu\mathrm{s}$. This robustness arises from a competition between signal accumulation via the growing sensing baseline and the geometric reduction of wavepacket overlap at larger displacements \cite{SM}. Using this feature, we implement an efficient single-shot sensing protocol based on a fixed-time readout at $t_s$.

As shown in Fig.~\ref{fig:3}(c), $P_\uparrow(t_s)$ scales linearly with the weak electric fields, matching numerical simulations excellently across five parameter sets. Sensitivity is defined as $\eta_E = (\Delta P_\uparrow / |\partial P_\uparrow / \partial E|) \sqrt{T_{\mathrm{cyc}}}$, where $T_{\mathrm{cyc}} = 20~\mathrm{ms}$ is the total experimental cycle time. The remaining time within each cycle ($\sim 14.3~\mathrm{ms}$) is functionally allocated to Doppler cooling, resolved sideband cooling, as well as state preparation and measurement (SPAM) protocols~\cite{SM}. Near the working point ($P_\uparrow \approx 1/2$), projection noise fundamentally limits the statistics to $\Delta P_\uparrow = \sqrt{P_\uparrow(1-P_\uparrow)} = 1/2$~\cite{Itano1993QPN}. The response slope $|\partial P_\uparrow / \partial E|$ and its uncertainty are extracted from a linear fit to four independently evaluated field strengths near zero field.

For the highest-sensitivity configuration, we achieve a shot-noise-limited peak sensitivity of $14.15 \pm 0.77~\mu\mathrm{V/m}/\sqrt{\mathrm{Hz}}$ (integration time $T_{\mathrm{int}} = 60~\mathrm{s}$, 3000 cycles). This yields a minimum detectable field variation of $\delta E_{\mathrm{min}} \approx 1.83~\mu\mathrm{V/m}$, setting the fundamental sensitivity bound.

Figure~\ref{fig:3}(d) maps the simulated finite-time sensitivity $\eta_E$ across the non-equilibrium phase diagram after the evolution time $t_{\mathrm{pl}}$. To elucidate the underlying mechanism governing this sensitivity landscape, we analyze the dynamic phase-space response of the phonon laser.

\begin{figure}[tbp]
    \centering
    \includegraphics[width=1.0\columnwidth]{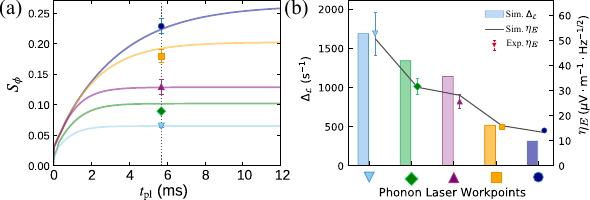}
    \caption{\textbf{Finite-time phase-space dynamics and Liouvillian scaling.}
    (a) Time-resolved phase-space response metric $S_{\phi}$ under a constant external field $E_S$ for various experimental configurations. Curves illustrate numerical simulations, while distinct markers represent experimental results at the interaction time $t_{\mathrm{pl}}$.
    (b) Extracted experimental sensitivities aligned with the corresponding Liouvillian gap $\Delta_{\mathcal{L}}$. Error bars are derived via differential error propagation~\cite{SM}.}
    \label{fig:4}
\end{figure}

We numerically simulate the time-dependent evolution of the phase-space response metric $S_{\phi} = |\langle a \rangle| / \sqrt{\langle a^\dagger a \rangle}$ under a constant weak drive $E_S = 20.0~\mu\mathrm{V/m}$ for the parameter sets highlighted in Fig.~\ref{fig:3}(d). As shown in Fig.~\ref{fig:4}(a), the experimentally extracted values~\cite{SM} at the fixed interaction time $t_{\mathrm{pl}}$ show excellent quantitative agreement with open-system simulations.

This dynamic behavior reflects the fundamental role of the Liouvillian gap $\Delta_{\mathcal{L}}$ in dictating the open-system relaxation. Within the experimentally accessible regime, a reduced gap leads to slower relaxation ($\tau \sim 1/\Delta_{\mathcal{L}}$), which grants the limit cycle a "softer" phase-space profile. This softness accommodates an enhanced dynamic symmetry-breaking convergence within the fixed interaction window. Importantly, because the system operates in a finite-time regime rather than the asymptotic steady state, this enhancement stems directly from the slowed relaxation dynamics.

To explicitly map this mechanism, Fig.~\ref{fig:4}(b) evaluates the finite-time sensitivity against the corresponding Liouvillian gap $\Delta_{\mathcal{L}}$, extracted via full open-system diagonalization~\cite{SM}. The sensitivity exhibits a strict monotonic scaling, improving significantly as the gap narrows. However, exploiting this scaling with arbitrarily small couplings is restricted by three practical constraints. First, the required exceedingly low Rabi frequencies introduce severe experimental control instabilities. Second, weakened dissipative stabilization fails to suppress residual environmental heating, thermalizing the delicate non-equilibrium state and washing out the symmetry-breaking signatures. Third, a highly softened limit cycle becomes overly susceptible to amplitude perturbations; this causes the optimal sensing time $t_s$ [Fig.~\ref{fig:3}(a)] to drift, breaking the stability of the fixed-time readout protocol. Consequently, our highest-sensitivity configuration ($\tau \approx 3.2~\mathrm{ms}$) represents a robust practical boundary, successfully maximizing the dynamic susceptibility while preserving the sensor's structural and readout stability.

In summary, we have experimentally realized a quantum-regime single-ion phonon laser and demonstrated high-sensitivity electrometry. Our results establish a practical platform for exploiting finite-time open-system relaxation dynamics as an advanced resource for quantum sensing. The resonance condition can be dynamically scanned across a wide frequency range by adjusting the trap potentials, enabling frequency-tunable electrometry and quantum noise spectroscopy. Extending this finite-time dynamic framework to multi-ion crystals~\cite{Prototype2023PL} or coupled ion-trap arrays opens an exciting outlook, where collective dissipation engineering and many-body symmetry breaking may enable sensing scalings beyond the standard quantum limit via mechanical squeezing~\cite{baur2026quantumtheoryphononlasing,lee2026bathfreesqueezedphononlasing, Pezze2018RMP, Marciniak2022}. Furthermore, it paves the way for novel non-Hermitian dissipative applications, such as exceptional-point (EP) phonon lasing~\cite{Zhang2018EP} and chiral mechanics~\cite{Jiang2018Nonreciprocal}. More broadly, the interplay between finite-time dynamic stabilization and relaxation-engineered open-system dynamics provides a versatile route toward dissipative quantum sensing in driven open quantum systems.

\textit{Acknowledgements\textemdash}This work was supported by National Natural Science Foundation of China under Grant Nos. 12534020, U24A2015, 12074390, 12304315, 12074346, 12322410, 12374466, by Guangdong Provincial Quantum Science Strategic Initiative under Grant No. GDZX2305004, by Natural Science Foundation of Wuhan under Grant No. 2024040701010063, Nansha Senior Leading Talent Team Project under Grant No. 2021CXTD02, and by the Special Project for Research and Development in Key Areas of Guangdong Province under Grant No. 2020B0303300001. \\


\let\oldaddcontentsline\addcontentsline
\renewcommand{\addcontentsline}[3]{} 

%

\let\addcontentsline\oldaddcontentsline

\clearpage
\onecolumngrid 

\makeatletter
\c@secnumdepth=4 
\renewcommand{\@seccntformat}[1]{\csname the#1\endcsname.\quad} 
\makeatother

\setcounter{section}{0}
\setcounter{equation}{0}
\setcounter{figure}{0}
\setcounter{table}{0}
\setcounter{page}{1}

\renewcommand{\thesection}{S\arabic{section}}
\renewcommand{\theequation}{S\arabic{equation}}
\renewcommand{\thefigure}{S\arabic{figure}}
\renewcommand{\thetable}{S\arabic{table}}

\begin{center}
    \vspace*{0.5cm}
    {\large\bf Supplementary Materials: Finite-Time Electrometry with a Quantum-Regime Single-Ion Phonon Laser}
    
    \vspace{4mm}
    {\small Pei-Dong Li, Yuan-Zhang Dong, Zhuo-Zhu Wu, Jia-Wei Wang, Ji Li, Jian-Qi Zhang, Zhi-Jiao Deng, Liang Chen and Mang Feng}
    \vspace{4mm}
\end{center}


\begin{center}
    \begin{minipage}{1\textwidth} 
        \small 
        
        \makeatletter
        \def\tableofcontents{\@starttoc{toc}}
        \makeatother
        
        \tableofcontents 
    \end{minipage}
\end{center}

\newpage
\section{Derivation of the Hamiltonian and Theoretical Model}

\subsection{System Hamiltonian}

We consider a single trapped $^{40}\mathrm{Ca}^{+}$ ion with a three-level internal structure coupled to a quantized harmonic motion along the axial ($z$) direction. The three internal states are defined as
\begin{equation}
|0\rangle \equiv |100\rangle,\qquad
|1\rangle \equiv |010\rangle,\qquad
|2\rangle \equiv |001\rangle,
\end{equation}
corresponding to three distinct electronic configurations of the ion. The system can be viewed as a $V$-type configuration with two independently driven transitions sharing a common ground state $|0\rangle$.

The total Hamiltonian of the system is
\begin{equation}
\hat{H} =
\hat{H}_0^h + \hat{H}_0^c + \hat{H}_{\mathrm{h.o.}}
+ \hat{H}_{\mathrm{int}}^h + \hat{H}_{\mathrm{int}}^c
+ \hat{H}_{\mathrm{ext}},
\end{equation}
where the free Hamiltonians are given by
\begin{align}
\hat{H}_0^h &= \frac{\hbar \omega_0^h}{2}\hat{\sigma}_z^h, \\
\hat{H}_0^c &= \frac{\hbar \omega_0^c}{2}\hat{\sigma}_z^c, \\
\hat{H}_{\mathrm{h.o.}} &= \hbar \omega_z \hat{a}^\dagger \hat{a}.
\end{align}

Here $\omega_0^{h(c)}$ denote the transition frequencies of the two internal transitions, and $\omega_z$ is the axial trap frequency. The operators $\hat{a}$ and $\hat{a}^\dagger$ are the annihilation and creation operators of the harmonic motion.

The relevant internal-state operators are defined as
\begin{align}
\hat{\sigma}_-^h &= |0\rangle\langle 1|, \qquad
\hat{\sigma}_-^c = |0\rangle\langle 2|, \qquad
\hat{\sigma}_{12} = |1\rangle\langle 2|, \\
\hat{\sigma}_z^h &= |1\rangle\langle 1| - |0\rangle\langle 0|, \qquad
\hat{\sigma}_z^c = |2\rangle\langle 2| - |0\rangle\langle 0|.
\end{align}

These operators describe two effective two-level subsystems embedded in a common three-level manifold.

\subsection{Laser-Ion Interactions}

The two optical transitions are driven by laser fields propagating along the $z$ direction in a counter-propagating configuration. The corresponding interaction Hamiltonians are given by \cite{singleiondynamics}:
\begin{align}
\hat{H}_{\mathrm{int}}^h &=
\frac{\hbar \Omega_h}{2}\hat{\sigma}_x^h
\left[
e^{i(+k_h \hat{z} - \omega_l^h t + \phi_l^h)}
+ e^{-i(+k_h \hat{z} - \omega_l^h t + \phi_l^h)}
\right], \\
\hat{H}_{\mathrm{int}}^c &=
\frac{\hbar \Omega_c}{2}\hat{\sigma}_x^c
\left[
e^{i(-k_c \hat{z} - \omega_l^c t + \phi_l^c)}
+ e^{-i(-k_c \hat{z} - \omega_l^c t + \phi_l^c)}
\right].
\end{align}

The ion position operator along the trap axis is
\begin{equation}
\hat{z} = z_0 (\hat{a} + \hat{a}^\dagger),
\end{equation}
where $z_0 = \sqrt{\hbar/(2m\omega_z)}$ is the zero-point fluctuation, with $m$ the ion mass.

\subsection{External Classical Driving Field}

An oscillating classical electric field applied along the axial ($z$) direction is written as
\begin{equation}
E(t) = E_0 \sin(\omega_d t + \phi_d)
= \frac{E_0}{2i}
\left[
e^{i(\omega_d t + \phi_d)} - e^{-i(\omega_d t + \phi_d)}
\right],
\end{equation}
where $E_0$, $\omega_d$, and $\phi_d$ denote the amplitude, driving frequency, and phase of the field, respectively.

The interaction between the ion and the external field is described by the dipole coupling \cite{wineland1998experimental, vdpcritical}
\begin{equation}
\hat{H}_{\mathrm{ext}} = -q E(t)\,\hat{z},
\end{equation}
where $q$ is the elementary charge of the ion.

\subsection{Interaction-Picture Transformation and Rotating-Wave Approximation}

The total Hamiltonian in the Schr\"odinger picture is given by
\begin{equation}
\hat{H}
=
\hat{H}_0
+
\hat{H}_{\mathrm{int}}^h
+
\hat{H}_{\mathrm{int}}^c
+
\hat{H}_{\mathrm{ext}},
\end{equation}
where the free Hamiltonian reads
\begin{equation}
\hat{H}_0
=
\frac{\hbar \omega_0^h}{2}\hat{\sigma}_z^h
+
\frac{\hbar \omega_0^c}{2}\hat{\sigma}_z^c
+
\hbar \omega_z \hat{a}^\dagger \hat{a}.
\end{equation}

After applying the optical rotating-wave approximation (RWA), the laser-ion interaction Hamiltonians take the form
\begin{align}
\hat{H}_{\mathrm{int}}^h
&=
\frac{\hbar \Omega_h}{2}
\left[
\hat{\sigma}_+^h
e^{i k_h \hat{z}}
e^{-i\omega_l^h t + i\phi_l^h}
+
\mathrm{h.c.}
\right],
\\[4pt]
\hat{H}_{\mathrm{int}}^c
&=
\frac{\hbar \Omega_c}{2}
\left[
\hat{\sigma}_+^c
e^{-i k_c \hat{z}}
e^{-i\omega_l^c t + i\phi_l^c}
+
\mathrm{h.c.}
\right].
\end{align}

To remove fast optical oscillations and define a convenient rotating frame, we introduce the transformation generated by
\begin{equation}
\hat{H}_{00}
=
\frac{\hbar (\omega_l^h-\omega_d)}{2}\hat{\sigma}_z^h
+
\frac{\hbar (\omega_l^c+\omega_d)}{2}\hat{\sigma}_z^c
+
\hbar \omega_d \hat{a}^\dagger \hat{a}.
\end{equation}

The corresponding unitary operator is
\begin{equation}
\hat{U}(t)
=
\exp\!\left(
-\frac{i}{\hbar}\hat{H}_{00} t
\right),
\end{equation}
and the interaction-picture Hamiltonian becomes
\begin{equation}
\hat{H}_I(t)
=\hat{U}^\dagger(t)\hat{H}\hat{U}(t)
-
i\hbar \hat{U}^\dagger(t)\dot{\hat{U}}(t)
=
\hat{U}^\dagger(t)\hat{H}\hat{U}(t)
-
\hat{H}_{00}.
\end{equation}

We decompose the transformed Hamiltonian as
\begin{equation}
\hat{H}_I(t)
=
\hat{H}_{\mathrm{det}} + \hat{H}_{\mathrm{int}}(t),
\end{equation}
where the static part describes residual detunings,
\begin{equation}
\hat{H}_{\mathrm{det}} = \hat{H}_0 - \hat{H}_{00}.
\end{equation}

Assuming sideband-resonance conditions
\begin{equation}
\omega_l^h = \omega_0^h + \omega_z,
\qquad
\omega_l^c = \omega_0^c - \omega_z,
\end{equation}
and defining the vibrational detuning
\begin{equation}
\Delta = \omega_d - \omega_z,
\end{equation}
we obtain
\begin{equation}
\hat{H}_{\mathrm{det}}
=
-\hbar \Delta \hat{a}^\dagger \hat{a}
+
\frac{\hbar \Delta}{2}\hat{\sigma}_z^h
-
\frac{\hbar \Delta}{2}\hat{\sigma}_z^c.
\end{equation}

Under this transformation, the operators evolve as
\begin{equation}
\hat{a}(t)=\hat{a}e^{-i\omega_d t},
\qquad
\hat{\sigma}_+^h(t)=\hat{\sigma}_+^h e^{i(\omega_l^h-\omega_d)t},
\qquad
\hat{\sigma}_+^c(t)=\hat{\sigma}_+^c e^{i(\omega_l^c+\omega_d)t}.
\end{equation}

In the Lamb--Dicke regime,
\begin{equation}
\eta_{h(c)}\sqrt{2\langle n\rangle+1}\ll 1,
\qquad
\eta_{h(c)} = k_{h(c)} z_0,
\end{equation}
the position-dependent exponentials can be expanded as
\begin{align}
e^{ik_h\hat{z}(t)}
&\simeq
1+i\eta_h\left(\hat{a}e^{-i\omega_d t}+\hat{a}^\dagger e^{i\omega_d t}\right),
\\
e^{-ik_c\hat{z}(t)}
&\simeq
1-i\eta_c\left(\hat{a}e^{-i\omega_d t}+\hat{a}^\dagger e^{i\omega_d t}\right).
\end{align}

Substituting these expressions into the interaction Hamiltonians and performing a vibrational rotating-wave approximation (i.e., neglecting terms oscillating at $\pm \omega_d$ and higher frequencies), we retain only near-resonant sideband couplings:
\begin{align}
\hat{H}_{h}
&=
i\frac{\hbar\eta_h\Omega_h}{2}
\left(
\hat{\sigma}_+^h \hat{a}^\dagger
-
\hat{\sigma}_-^h \hat{a}
\right),
\\[4pt]
\hat{H}_{c}
&=
i\frac{\hbar\eta_c\Omega_c}{2}
\left(
\hat{\sigma}_-^c \hat{a}^\dagger
-
\hat{\sigma}_+^c \hat{a}
\right).
\end{align}

For simplicity, we set
\begin{equation}
\phi_l^h=\phi_l^c=0,
\qquad
\eta_h=\eta_c\equiv\eta,
\end{equation}
and define the effective coupling strengths
\begin{equation}
g_h=\eta\Omega_h,
\qquad
g_c=\eta\Omega_c.
\end{equation}

The sideband Hamiltonians reduce to
\begin{align}
\hat{H}_{h}
&=
i\frac{\hbar g_h}{2}
\left(
\hat{\sigma}_+^h \hat{a}^\dagger
-
\hat{\sigma}_-^h \hat{a}
\right),
\\[4pt]
\hat{H}_{c}
&=
i\frac{\hbar g_c}{2}
\left(
\hat{\sigma}_-^c \hat{a}^\dagger
-
\hat{\sigma}_+^c \hat{a}
\right).
\end{align}

We next consider the external classical driving field. Substituting
\(
\hat{z}=z_0(\hat{a}+\hat{a}^\dagger)
\)
into $\hat{H}_{\mathrm{ext}}$ and transforming to the interaction picture yields
\begin{align}
\hat{H}_{\mathrm{ext}}^{(I)}
=
-\frac{qE_0z_0}{2i}
\left(
\hat{a}e^{-i\omega_d t}
+
\hat{a}^\dagger e^{i\omega_d t}
\right)
\left[
e^{i(\omega_d t+\phi_d)}
-
e^{-i(\omega_d t+\phi_d)}
\right].
\end{align}

We apply a second rotating-wave approximation, eliminating fast-oscillating terms at $\pm 2\omega_d$, and obtain a coherent phonon-drive Hamiltonian
\begin{equation}
\hat{H}_{\mathrm{drive}}
=
i\hbar\epsilon
\left(
\hat{a}e^{i\phi_d}
-
\hat{a}^\dagger e^{-i\phi_d}
\right),
\qquad
\epsilon=\frac{qE_0z_0}{2\hbar}.
\label{drivehamiltonian}
\end{equation}

Finally, the full effective interaction-picture Hamiltonian is given by
\begin{align}
\hat{H}_{\mathrm{eff}}
&=
-\hbar \Delta \hat{a}^\dagger \hat{a}
+
\frac{\hbar \Delta}{2}\hat{\sigma}_z^h
-
\frac{\hbar \Delta}{2}\hat{\sigma}_z^c
\notag \\
&\quad
+
i\frac{\hbar g_h}{2}
\left(
\hat{\sigma}_+^h \hat{a}^\dagger
-
\hat{\sigma}_-^h \hat{a}
\right)
+
i\frac{\hbar g_c}{2}
\left(
\hat{\sigma}_-^c \hat{a}^\dagger
-
\hat{\sigma}_+^c \hat{a}
\right)
\notag \\
&\quad
+
i\hbar\epsilon
\left(
\hat{a}e^{i\phi_d}
-
\hat{a}^\dagger e^{-i\phi_d}
\right).
\end{align}

Under resonant driving of the external electric field with the axial vibrational mode ($\Delta=0$), the Hamiltonian reduces to
\begin{equation}
\hat{H}_{\mathrm{res}}
=
i\frac{\hbar g_h}{2}
\left(
\hat{\sigma}_+^h \hat{a}^\dagger
-
\hat{\sigma}_-^h \hat{a}
\right)
+
i\frac{\hbar g_c}{2}
\left(
\hat{\sigma}_-^c \hat{a}^\dagger
-
\hat{\sigma}_+^c \hat{a}
\right)
+
i\hbar\epsilon
\left(
\hat{a}e^{i\phi_d}
-
\hat{a}^\dagger e^{-i\phi_d}
\right),
\end{equation}
which is the form used in the main text.

\subsection{Lindblad Master Equation Formalism}

Having derived the effective Hamiltonian in the interaction picture, we can now describe the open-system dynamics of the single-ion system within the Lindblad master equation formalism.

The time evolution of the system density operator $\rho$ is governed by the Lindblad master equation
\begin{equation}
\frac{\mathrm{d}\rho}{\mathrm{d}t}
=
-\frac{i}{\hbar}
\left[
\hat{H}_{\mathrm{eff}}, \rho
\right]
+
\sum_k
\mathcal{D}[\hat{L}_k](\rho),
\end{equation}
where the Lindblad dissipator is defined as
\begin{equation}
\mathcal{D}[\hat{L}_k](\rho)
=
\hat{L}_k \rho \hat{L}_k^\dagger
-
\frac{1}{2}
\left\{
\hat{L}_k^\dagger \hat{L}_k,\rho
\right\}.
\end{equation}

The dissipative processes associated with the heating and cooling transitions are described by the Lindblad operators
\begin{equation}
\hat{L}_{h}
=
\sqrt{\gamma_h}\,
\hat{\sigma}_{-}^{h},
\qquad
\hat{L}_{c}
=
\sqrt{\gamma_c}\,
\hat{\sigma}_{-}^{c},
\end{equation}
where $\gamma_h$ and $\gamma_c$ denote the effective decay rates of the corresponding transitions.

It is worth noting that the ambient axial heating rate of our surface-electrode trap is independently measured to be approximately $11$ phonons/s. Over the fixed $t_{pl} = 5.7$ ms phonon-lasing interaction window, this ambient heating contributes a negligible thermal occupation of $\sim 0.06$ phonons. This rigorously justifies the exclusion of an environmental thermal-bath dissipator in our current master equation model.

By tuning the Rabi frequencies $\Omega_h$ and $\Omega_c$, together with the dissipation rates $\gamma_h$ and $\gamma_c$, the system can be driven into different dynamical regimes, including dark state region, instability region, and phonon lasing region.

\section{Analytical Derivation of Semiclassical Phonon-Lasing States}
\label{sec:analytical_derivation}

To transparently analyze the competition between gain and loss in the single-ion tripartite system and establish the analytical phonon-lasing threshold, we formulate the open-system dynamics within the Heisenberg picture. The expectation value of a generic system operator $\hat{O}$ evolves according to the Heisenberg-Lindblad master equation:
\begin{equation}
\frac{\mathrm{d}\langle \hat{O} \rangle}{\mathrm{d}t} = \frac{i}{\hbar} \left\langle \left[ \hat{H}_{\mathrm{pl}}, \hat{O} \right] \right\rangle + \sum_k \left\langle \tilde{\mathcal{D}}[\hat{L}_k](\hat{O}) \right\rangle,
\end{equation}
where is governed by the resonant laser-ion coupling Hamiltonian:
\begin{equation}
\hat{H}_{\mathrm{pl}} = i\frac{\hbar g_h}{2} \left( \hat{\sigma}_+^h \hat{a}^\dagger - \hat{\sigma}_-^h \hat{a} \right) + i\frac{\hbar g_c}{2} \left( \hat{\sigma}_-^c \hat{a}^\dagger - \hat{\sigma}_+^c \hat{a} \right),
\end{equation}
and $\tilde{\mathcal{D}}[\hat{L}_k](\hat{O}) = \hat{L}_k^\dagger \hat{O} \hat{L}_k - \frac{1}{2} \{ \hat{L}_k^\dagger \hat{L}_k, \hat{O} \}$ represents the dissipative adjoint superoperator associated with the Lindblad relaxation channels $\hat{L}_h$ and $\hat{L}_c$.

\subsection{Complete Set of Semiclassical Equations of Motion}

Utilizing the three-level manifold, we define the mechanical amplitude and electronic cross-coherences as $A \equiv \langle \hat{a} \rangle$, $S_h \equiv \langle \hat{\sigma}_+^h \rangle$, $S_c \equiv \langle \hat{\sigma}_+^c \rangle$, and $S_{12} \equiv \langle \hat{\sigma}_{12} \rangle = \langle |1\rangle\langle 2| \rangle$. To map the internal population configurations completely, we track both the excited-state populations $P_h \equiv \langle |1\rangle\langle 1| \rangle$, $P_c \equiv \langle |2\rangle\langle 2| \rangle$, and the diagonal population inversions defined relative to the shared ground state $|0\rangle$:
\begin{equation}
W_h \equiv \langle\hat{\sigma}_z^h\rangle = P_h - P_0, \quad 
W_c \equiv \langle\hat{\sigma}_z^c\rangle = P_c - P_0. \label{eq:inversion_definition}
\end{equation}
The global closure relation within the three-level Hilbert subspace dictates $P_0 + P_h + P_c = 1$. Substituting this into Eq.~\eqref{eq:inversion_definition} reveals the explicit bilateral coordinate mapping between the inversions and state populations:
\begin{equation}
W_h = 2P_h + P_c - 1, \quad W_c = 2P_c + P_h - 1 \iff P_h = \frac{1 + 2W_h - W_c}{3}, \quad P_c = \frac{1 + 2W_c - W_h}{3}. \label{eq:population_mapping_complete}
\end{equation}

By executing the explicit trace dynamics and invoking the standard mean-field factorization to decouple multi-particle quantum correlations (e.g., $\langle \hat{a}\hat{\sigma}_z^h \rangle \approx A W_h$), we present the mathematically closed, self-consistent set of coupled semiclassical differential equations capturing the entire tripartite network:
\begin{align}
\dot{A} &= \frac{g_h}{2} S_h + \frac{g_c}{2} S_c^*, \label{eq:EOM_A} \\
\dot{S}_h &= -\frac{g_h}{2} A W_h + \frac{g_c}{2} A^* S_{12} - \frac{\gamma_h}{2} S_h, \label{eq:EOM_Sh} \\
\dot{S}_c &= \frac{g_c}{2} A^* W_c - \frac{g_h}{2} A S_{12}^* - \frac{\gamma_c}{2} S_c, \label{eq:EOM_Sc} \\
\dot{S}_{12} &= \frac{g_h}{2} A S_c^* - \frac{g_c}{2} A S_h - \frac{\gamma_h + \gamma_c}{2} S_{12}, \label{eq:EOM_S12} \\
\dot{P}_h &= \frac{g_h}{2} (A^* S_h + A S_h^*) - \gamma_h P_h, \label{eq:EOM_Ph} \\
\dot{P}_c &= -\frac{g_c}{2} (A^* S_c^* + A S_c) - \gamma_c P_c, \label{eq:EOM_Pc} \\
\dot{W}_h &= g_h(A^* S_h + A S_h^*) - \frac{g_c}{2}(A^* S_c^* + A S_c) - \frac{2\gamma_h}{3} (1 + 2W_h - W_c) - \frac{\gamma_c}{3} (1 - W_h + 2W_c), \label{eq:EOM_Wh} \\
\dot{W}_c &= -g_c(A^* S_c^* + A S_c) + \frac{g_h}{2}(A^* S_h + A S_h^*) - \frac{\gamma_h}{3} (1 + 2W_h - W_c) - \frac{2\gamma_c}{3} (1 - W_h + 2W_c). \label{eq:EOM_Wc}
\end{align}
Note that Eqs.~\eqref{eq:EOM_Wh} and \eqref{eq:EOM_Wc} are structurally locked to Eqs.~\eqref{eq:EOM_Ph} and \eqref{eq:EOM_Pc} via the time derivative of Eq.~\eqref{eq:population_mapping_complete}, where the common ground-state dynamics $\dot{P}_0 = -\dot{P}_h - \dot{P}_c$ introduces cross-channel dissipation feedback.

\subsection{Semiclassical Ansatz and Exact Steady-State Solution}

We evaluate the stationary response by enforcing the steady-state criteria where all coherent and population time derivatives vanish ($\dot{A}=\dot{S}_{12}=\dot{S}_h=\dot{S}_c=\dot{P}_h=\dot{P}_c=\dot{W}_h=\dot{W}_c=0$). Setting $\dot{S}_{12} = 0$ in Eq.~\eqref{eq:EOM_S12} allows the inter-subsystem electronic cross-coherence to be decoupled as:
\begin{equation}
S_{12} = \frac{g_h A S_c^* - g_c A S_h}{\gamma_h + \gamma_c}.
\end{equation}
Substituting $S_{12}$ into Eqs.~\eqref{eq:EOM_Sh} and \eqref{eq:EOM_Sc} with $\dot{S}_h = 0$ and $\dot{S}_c = 0$ yields the cross-locked optical driving configurations:
\begin{align}
S_h &= -\frac{g_h A}{\gamma_h} W_h - \frac{g_c^2 |A|^2}{\gamma_h (\gamma_h + \gamma_c)} S_h + \frac{g_h g_c |A|^2}{\gamma_h (\gamma_h + \gamma_c)} S_c^*, \\
S_c &= \frac{g_c A^*}{\gamma_c} W_c - \frac{g_h^2 |A|^2}{\gamma_c (\gamma_h + \gamma_c)} S_c + \frac{g_h g_c |A|^2}{\gamma_c (\gamma_h + \gamma_c)} S_h^*.
\end{align}

To transition into the semiclassical domain, we introduce the field-synchronized coherence ansatz:
\begin{equation}
S_h = A P, \quad S_c = A^* Q, \label{eq:semiclassical_ansatz}
\end{equation}
where $P$ and $Q$ represent the synchronized coherence coefficients, and $n \equiv |A|^2$ denotes the steady-state mean phonon number. Factoring out the macroscopic complex amplitudes $A$ and $A^*$ reduces the macro-micro coupled dipole relations to a closed linear system:
\begin{subequations}
\label{eq:PQ_coupled_system_native}
\begin{align}
P \left[ 1 + \frac{g_c^2 n}{\gamma_h (\gamma_h + \gamma_c)} \right] - Q \left[ \frac{g_h g_c n}{\gamma_h (\gamma_h + \gamma_c)} \right] &= -\frac{g_h}{\gamma_h} W_h, \\
-P \left[ \frac{g_h g_c n}{\gamma_c (\gamma_h + \gamma_c)} \right] + Q \left[ 1 + \frac{g_h^2 n}{\gamma_c (\gamma_h + \gamma_c)} \right] &= \frac{g_c}{\gamma_c} W_c.
\end{align}
\end{subequations}

Applying Cramer's rule to the coupled system \eqref{eq:PQ_coupled_system_native} explicitly solves for the synchronized dipole profiles $P$ and $Q$ as functions of the internal population inversions:
\begin{align}
P &= \frac{\gamma_h + \gamma_c}{D(n)} \left[ -g_h \gamma_c (\gamma_h + \gamma_c) W_h - g_h^3 n W_h + g_h g_c^2 n W_c \right], \label{eq:P_expanded} \\
Q &= \frac{\gamma_h + \gamma_c}{D(n)} \left[ g_c \gamma_h (\gamma_h + \gamma_c) W_c + g_c^3 n W_c - g_h^2 g_c n W_h \right], \label{eq:Q_expanded}
\end{align}
where $D(n) = [\gamma_h(\gamma_h + \gamma_c) + g_c^2 n] [\gamma_c(\gamma_h + \gamma_c) + g_h^2 n] - (g_h g_c n)^2$. 

Concurrently, by evaluating the population stabilization constraints ($\dot{P}_h = \dot{P}_c = \dot{W}_h = \dot{W}_c = 0$), the steady-state inversion dynamics map directly onto the nonlinearly saturated electronic configurations:
\begin{align}
W_h &= \frac{2g_h n}{\gamma_h} P - \frac{g_c n}{\gamma_c} Q - 1, \label{eq:Wh_PQ_relation} \\
W_c &= \frac{g_h n}{\gamma_h} P - \frac{2g_c n}{\gamma_c} Q - 1. \label{eq:Wc_PQ_relation}
\end{align}

Meanwhile, the macroscopic mechanical rate equation $\dot{n} = \dot{A}A^* + A\dot{A}^*$ incorporates Eq.~\eqref{eq:EOM_A} and the ansatz \eqref{eq:semiclassical_ansatz} to yield $\dot{n} = (g_h P + g_c Q) n$. For a non-trivial self-sustained phonon lasing state ($n > 0$), the vanishing growth rate ($\dot{n}=0$) dictates the exact mechanical gain-loss balance condition:
\begin{equation}
g_h P + g_c Q = 0. \label{eq:balance_condition_native}
\end{equation}

Crucially, simultaneously solving the algebraic network comprised of the Cramer-derived dipole profiles \eqref{eq:P_expanded}--\eqref{eq:Q_expanded} and the nonlinear population constraints \eqref{eq:Wh_PQ_relation}--\eqref{eq:Wc_PQ_relation} self-consistently closes the steady-state inversions $W_h$ and $W_c$, as well as the synchronized coherences $P$ and $Q$, as explicit analytical functions of the intensity $n$. Substituting these fully closed expressions for $P$ and $Q$ into the mechanical balance condition \eqref{eq:balance_condition_native} factors out the non-linear saturation manifolds, yielding the exact analytical solution for the steady-state mean phonon number:
\begin{equation}
\label{eq:n_ss_final_elegant}
n_{\mathrm{ss}} = \frac{ (\gamma_c + \gamma_h) \left(\dfrac{g_h^2}{\gamma_h} - \dfrac{g_c^2}{\gamma_c}\right) }{ \left(\dfrac{g_c^2}{\gamma_h}-\dfrac{g_h^2}{\gamma_c}\right) \left(\dfrac{g_h^2}{\gamma_h}+\dfrac{g_c^2}{\gamma_c}\right) }.
\end{equation}

Remarkably, although the mechanical mode couples linearly to the internal transitions within this first-order LD approximation, the system inherently possesses a powerful nonlinear gain-saturation mechanism originating from the multi-level electronic manifold saturation. Continuous cycling within the three-level electronic structure under competing sideband drives nonlinearly clamps the steady-state atomic population inversions as the phonon amplitude builds up. This intrinsic internal-state saturation directly back-acts on the motional mode, successfully stabilizing the mechanical amplification into a finite, self-sustained steady-state limit cycle $n_{\mathrm{ss}}$, and giving rise to the distinct semiclassical dissipative phase transition boundaries without requiring higher-order structural LD corrections.

\subsection{Phonon-Lasing Threshold and Parametric Instability Boundaries}
The formulation of Eq.~\eqref{eq:n_ss_final_elegant} effortlessly isolates the operational regimes of the single-ion phonon laser:
\begin{enumerate}
    \item \textbf{Phonon Lasing Threshold:} Coherent amplification requires a positive numerator, defining the critical lasing threshold:
    \begin{equation}
    \frac{g_h^2}{\gamma_h} > \frac{g_c^2}{\gamma_c} \implies \frac{g_h}{g_c} > \sqrt{\frac{\gamma_h}{\gamma_c}}.
    \end{equation}
    Below this boundary, cold-channel dissipation dominates, restricting the mechanical steady state to $n_{\mathrm{ss}} \to 0$.
    
    \item \textbf{Instability Boundary:} A semi-classical divergence ($n_{\mathrm{ss}} \to \infty$) occurs when the first factor of the denominator collapses, yielding the exact unstable threshold line:
    \begin{equation}
    \frac{g_c^2}{\gamma_h} = \frac{g_h^2}{\gamma_c} \implies \frac{g_h}{g_c} = \sqrt{\frac{\gamma_c}{\gamma_h}}.
    \end{equation}
\end{enumerate}

Within the stable parameter window $\sqrt{\gamma_h/\gamma_c} < g_h/g_c < \sqrt{\gamma_c/\gamma_h}$ (assuming $\gamma_c > \gamma_h$), multi-level quantum saturation successfully stabilizes the mechanical amplitude, establishing a robust phonon laser. Beyond this boundary, the stationary fixed point collapses, manifesting as a parametric instability driven by internal microscopic coherences.

\subsection{Phase Diagram and Phase-Space Signatures}

\begin{figure}[tb]
    \centering
    \includegraphics[width=\linewidth]{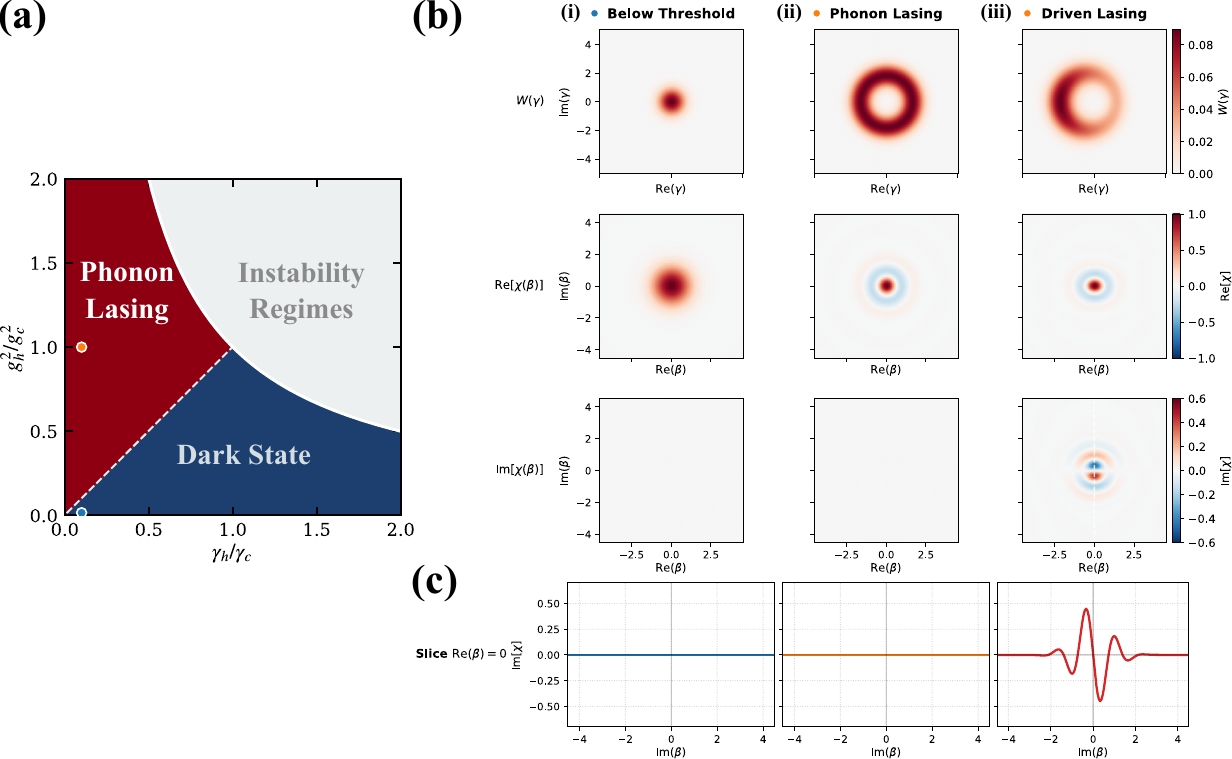} 
    \caption{(a) Phase diagram identifying the dark-state, phonon-lasing, and instability regimes. The blue and orange markers denote the representative operating points selected for the state tomography analysis shown in (b). (b) Numerically reconstructed phase-space quantum-state tomographies in three distinct operational regimes: (i) below threshold [blue marker in (a)], (ii) phonon lasing [orange marker in (a)], and (iii) driven phonon lasing under an external electric field $E_0 = 100~\mu\mathrm{V/m}$. Top row: Wigner functions $W(\gamma)$. Middle and bottom rows: real and imaginary components of the characteristic function, $\mathrm{Re}[\chi(\beta)]$ and $\mathrm{Im}[\chi(\beta)]$, respectively. (c) Numerical sensing signatures extracted from the characteristic-function response. Line slices of $\mathrm{Im}[\chi(\beta)]$ are taken along the optimal detection axis $\mathrm{Re}[\beta]=0$, revealing the field-induced phase-space asymmetry used for electrometry.}
    \label{fig:phase_diagram}
\end{figure}

The steady-state phases of the open-system dynamics are mapped into a comprehensive dynamical phase diagram, as presented in Fig.~\ref{fig:phase_diagram}(a). The analytical boundaries partitioning the parameter space into the dark state, phonon lasing, and the instability regime are defined by the threshold line $g_h^2/g_c^2 = \gamma_h/\gamma_c$ and boundary line $g_h^2/g_c^2 = \gamma_c/\gamma_h$, respectively. This classification is derived via a mean-field semiclassical analysis and aligns with the quantum theory of single-ion phonon lasing established in Ref.~\cite{baur2026quantumtheoryphononlasing}. 

To numerically evaluate the non-equilibrium features and benchmark configurations throughout this work, we select two representative parameter sets. The below-threshold regime, labeled as configuration (i), is located within the dark-state phase boundary, characterized by the coupling and dissipation ratios $g_h^2/g_c^2 = 1/64$ and $\gamma_h/\gamma_c = 0.1$ [blue dot in Fig.~\ref{fig:phase_diagram}(a)]. Conversely, the phonon-lasing phase, encompassing both unperturbed and driven scenarios labeled as configurations (ii) and (iii), is positioned at the balanced coupling point where $g_h^2/g_c^2 = 1$ and $\gamma_h/\gamma_c = 0.1$ [orange dot in Fig.~\ref{fig:phase_diagram}(a)].

A weak resonant oscillating electric field is applied externally. In the rotating frame, this coherent perturbation is described by $\hat{H}_{\mathrm{drive}}$ in Eq.~\ref{drivehamiltonian}, where the driving phase is fixed at $\phi_d = 0$ in the numerical simulations. To theoretically elucidate the underlying metrological mechanism without loss of generality, a weak target electric field with amplitude $E_0 = 100\ \mu\mathrm{V/m}$ is employed throughout the master-equation simulations.

The microscopic manifestation of these driven-dissipative dynamics can be comprehensively mapped via phase-space quantum state tomography. Figure~\ref{fig:phase_diagram}(b) details the steady-state distributions of the motional mode across the three designated regimes. Experimentally, the complex characteristic function $\chi(\beta) = \text{Tr}[\rho \hat{D}(\beta)]$, where $\hat{D}(\beta)$ is the displacement operator, is resolved by mapping the motional states onto the ion's internal electronic population using a state-extracting quantum circuit, following the protocol outlined in Ref.~\cite{fluhmann2020direct} and detailed in Sec.~\ref{subsec:motionaltomo}. The corresponding Wigner functions $W(\gamma)$ are subsequently reconstructed via a full two-dimensional Fourier transform of the measured joint distributions of $\text{Re}[\chi(\beta)]$ and $\text{Im}[\chi(\beta)]$.

This phase-space diagnostic approach highlights the fundamental metrological signature of our protocol [Fig.~\ref{fig:phase_diagram}(c)]. In the absence of an external drive, both the below-threshold dark state (i) and the stable phonon-lasing limit cycle (ii) exhibit strict $U(1)$ rotational symmetry in phase space. Consequently, their respective Wigner functions possess reflection symmetry about any arbitrary axis through the origin, which forces the imaginary part of the characteristic function, $\text{Im}[\chi(\beta)]$, to vanish identically everywhere. This rigorous symmetry constraint provides an exceptionally robust, zero-background baseline for sensing applications. 

The application of the resonant electric field in configuration (iii) explicitly breaks this continuous rotational symmetry by inducing phase convergence and shifting the Wigner distribution along the real axis ($\text{Re}[\gamma]$). To maximize the detection contrast associated with this phase-space symmetry breaking, we extract line slices of the characteristic function along the imaginary axis ($\mathrm{Re}[\beta]=0$). As mathematically demonstrated in Sec.~\ref{subsec:optimal_axis}, this direction constitutes the optimal sensing axis. This optimal projection strategy produces high-contrast, large-amplitude oscillations in $\mathrm{Im}[\chi(\beta)]$. The amplitude of these oscillations directly encodes the external driving strength $E_0$, thereby establishing a highly sensitive and exceptionally clean benchmark for the subsequent experimental demonstrations.

\section{Experimental Implementation and Calibration}

\subsection{Experimental Setup}

As illustrated in main text Fig.~1(a), a single $^{40}\mathrm{Ca}^+$ ion is confined in a surface-electrode trap (SET) at a height of approximately $500~\mu\text{m}$ above the electrode plane, featuring an axial secular frequency of $\omega_{z}/2\pi = 676.35$~kHz. A static magnetic field of $0.52$~mT, generated by a pair of Helmholtz coils, is oriented at $45^\circ$ with respect to the $z$ axis. This field defines the quantization axis and ensures optimal alignment with both a $397$-nm $\sigma_+$-polarized laser and a $729$-nm sideband cooling laser.

The relevant energy level structure and coupling scheme are depicted in Fig.~1(b). We define the state mapping as $\ket{0} \equiv \ket{4^2S_{1/2}, m_J = +1/2}$, $\ket{1} \equiv \ket{3^2D_{5/2}, m_J = +5/2}$, and $\ket{2} \equiv \ket{3^2D_{5/2}, m_J = +3/2}$. The system's gain and dissipation are engineered via tailored laser configurations. Two counter-propagating $729$-nm beams are aligned along the $z$ axis to drive motional sideband transitions. Specifically, the $\ket{0} \leftrightarrow \ket{2}$ transition is driven with a detuning of $-\omega_z$, realizing a red-sideband interaction described by the Jaynes-Cummings (JC) Hamiltonian $\hat{H}_c = i\frac{\hbar}{2} g_c (\hat{a}^\dagger \hat{\sigma}^c_- - \hat{a} \hat{\sigma}^c_+)$ with $g_c = \eta \Omega_c$. Conversely, the $\ket{0} \leftrightarrow \ket{1}$ transition is driven at $+\omega_z$, implementing an anti-Jaynes-Cummings (AJC) interaction $\hat{H}_h = i\frac{\hbar}{2} g_h (\hat{a}^\dagger \hat{\sigma}^h_+ - \hat{a} \hat{\sigma}^h_-)$ with $g_h = \eta \Omega_h$, where $\eta = 0.114$ is the Lamb-Dicke parameter.

Dissipation is introduced by two linearly polarized $854$-nm beams that primarily couple $\ket{1}$ and $\ket{2}$ to the $\ket{4^2P_{3/2}, m_J = +3/2}$ state, followed by rapid spontaneous decay back to $\ket{0}$. The first $854$-nm beam, propagating at $135^\circ$ with respect to the $z$ axis, acts as a pure $\pi$-polarized drive because its polarization vector is orthogonal to the quantization axis. The second $854$-nm beam is directed along the $z$ axis, providing a mixture of $\sigma_\pm$ polarization components. The resulting dissipative dynamics are captured by the Lindblad operators $\hat{L}_{h} = \sqrt{\gamma_{h}} \hat{\sigma}^h_{-}$ and $\hat{L}_{c} = \sqrt{\gamma_{c}} \hat{\sigma}^c_{-}$. By precisely tuning the coupling strengths $\{g_c, g_h\}$ and dissipation rates $\{\gamma_c, \gamma_h\}$, the system can be stabilized to a controllable phonon-lasing steady state.

\subsection{Experimental Sequence for Phonon Lasing and Electric-Field Sensing} \label{subsec:timeline}

\begin{figure}[tb]
    \centering
    \includegraphics[width=\linewidth]{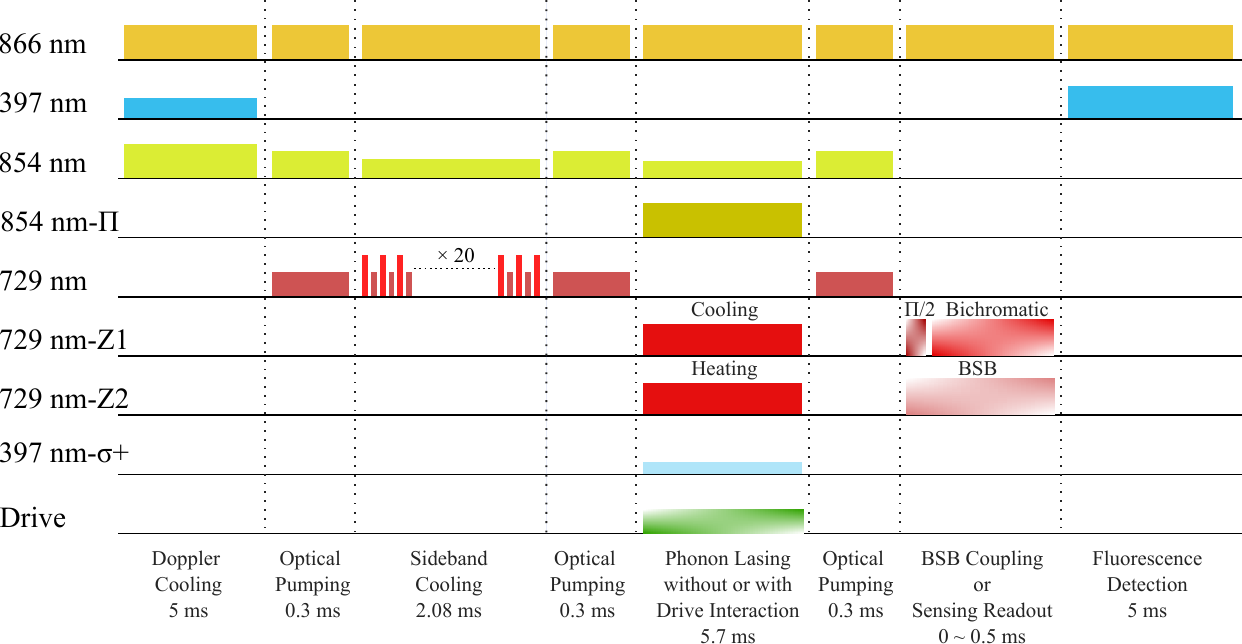} 
    \caption{Experimental pulse sequence. Complete timeline of the experiment, including cooling, state preparation, phonon lasing, and detection stages. The fifth stage, phonon lasing ($5.7\text{ ms}$), is implemented either without (drive off) or with (drive on) the drive interaction. The seventh stage serves as the readout phase, which is configured as either standard blue sideband (BSB) coupling or sensing readout via the characteristic function method, with details described in the Supplementary Text.}
    \label{fig:timeline}
\end{figure}

The experimental protocol for generating a single-ion phonon laser in the quantum regime and its subsequent application in electric-field sensing follows a precise timing sequence, as illustrated in Fig.~\ref{fig:timeline}. 

The sequence initiates with Doppler cooling for $t_{\text{dp}} = 5.0$~ms, followed by an initial optical pumping stage ($t_{\text{op}} = 0.3$~ms) to initialize the ion in the internal ground state $\ket{\downarrow}$. Subsequently, resolved sideband cooling (SBC) is executed for $t_{\text{sb}} = 2.08$~ms to prepare the ion near its motional ground state. Following a second optical pumping pulse ($t_{\text{op}} = 0.3$~ms), the internal state is re-initialized to $\ket{\downarrow}$ to ensure high state purity. 

The phonon lasing interaction is then driven for a fixed duration of $t_{\text{pl}} = 5.7$~ms, either in the presence or absence of a phase-stable external electric field. After this evolution and a third optical pumping pulse ($t_{\text{op}} = 0.3$~ms), the protocol branches into one of two distinct measurement paths:
\begin{itemize}
    \item \textbf{Phonon Number Distribution:} To reconstruct the Fock-state populations, a blue sideband (BSB) coupling is applied for a variable duration up to $t_{\text{bsb}} = 0.5$~ms. The numerical procedure for extracting the phonon distribution from the resulting excitation curves is discussed in Section~\ref{subsec:fock_recon}.
    \item \textbf{Characteristic Function and Sensing:} To extract the characteristic function $\chi(\beta)$ for sensing or Wigner function reconstruction, an optional carrier $\pi/2$-pulse ($t_{\pi/2} = 5~\mu\text{s}$) is employed. The inclusion or exclusion of this pulse allows us to selectively resolve either the imaginary or real component of the characteristic function, respectively. This stage is followed by the application of a bichromatic light field, which implements an effective spin-dependent displacement for a variable duration up to $t_{\text{bic}} = 0.1$~ms. The theoretical framework and detailed derivation of this sensing protocol are provided in Section~\ref{sec:char_func}.
\end{itemize}

In all cases, the sequence concludes with fluorescence detection ($t_{\text{meas}} = 5.0$~ms) to determine the final spin population. While the active experimental operations occupy a total duration of $t_{\text{tot}} \approx 19$~ms, the entire sequence---controlled via an FPGA in coordination with RF switches and AWGs---is synchronized with the 50~Hz AC power line to suppress environmental magnetic and electronic noise. Consequently, the FPGA utilizes a line trigger to ensure a total cycle interval of $T_{\text{cyc}} = 20$~ms. This 20~ms cycle defines the fundamental detection period used to evaluate the sensitivity of our sensing method in subsequent sections.

\subsection{Calibration of the Effective Decay Rates}
\label{decayrate}

In our experimental scheme, the target effective decay rates are set to $\gamma_c = 150$~kHz and $\gamma_h = 15$~kHz. As discussed in the main text, these dissipation rates are induced during the phonon lasing interaction by two 854-nm laser beams. One beam is $\pi$-polarized, while the other is adjusted to couple all polarization components. The two metastable states involved are defined as the cooling branch state $\ket{2} \equiv \ket{3^2D_{5/2}, m_J = +3/2}$ and the heating branch state $\ket{1} \equiv \ket{3^2D_{5/2}, m_J = +5/2}$. 

When these two dissipation channels are active, a small fraction of the population may leak into other Zeeman sublevels. To counteract this, we apply a low-intensity $\sigma_+$-polarized 397-nm laser alongside the 866-nm repumper to maintain a closed transition loop. Given that we operate in the regime $\gamma_c \gg \gamma_h$, the system dynamics can still be accurately described within a simplified three-level scheme. 

\begin{figure}[htbp]
    \centering
    \includegraphics[width=0.48\textwidth]{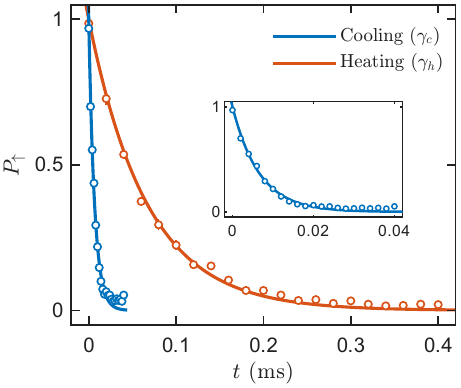}
   \caption{Time evolution of the population $P_{\uparrow}$ during the cooling (blue circles) and heating (red circles) processes. Markers represent experimental data (mean $\pm$ s.e.m.) averaged over 700 measurements per point. Solid lines denote exponential fits used to extract the effective decay rates, $\gamma_c$ and $\gamma_h$, for the two processes.  Inset: A magnified view of the cooling process.}
    \label{fig:gamma}
\end{figure}

By precisely controlling the intensity ratio and total power of these two 854-nm beams, the effective decay rates can be tuned to the desired values. To ensure experimental consistency, we calibrate these rates before each measurement sequence. The calibration procedure is as follows: after initial sideband cooling and optical pumping, a carrier $\pi$-pulse prepares the ion in either state $\ket{1}$ or $\ket{2}$. Subsequently, the 854-nm lasers are applied for a variable duration $t$, while the low-intensity $\sigma_+$-polarized 397-nm laser and the 866-nm repumper remain active to maintain the closed transition loop. The remaining spin population $P_{\uparrow}$ is then measured via fluorescence detection. By fitting the observed population decay to an exponential model $P_{\uparrow}(t) = e^{-\gamma t}$, we extract the effective decay rates. As shown in Fig.~\ref{fig:gamma}, the calibrated values are $\gamma_c = 150.82 \pm 4.41$~kHz and $\gamma_h = 15.05 \pm 0.28$~kHz, showing excellent agreement with our experimental requirements.

\subsection{Reconstruction of Phonon Number Distribution}
\label{subsec:fock_recon}

The phonon number distributions $P_n$ discussed in the main text are extracted by fitting the experimental blue-sideband (BSB) Rabi oscillations. The fitting model for the excited-state population is expressed as:
\begin{equation}
    P_{\uparrow}(t) = \frac{1}{2} \left[ 1 - \sum_{n=0}^{n_{\text{max}}} P_n e^{-\gamma_n t} \cos(\Omega_{n, n+1} t) \right]
\end{equation}
where $\Omega_{n, n+1} = \Omega_{0,1} \sqrt{\frac{n!}{(n+1)!}} L_n^1(\eta^2)$ is the $n$-dependent Rabi frequency and $\eta = 0.114$ is the Lamb-Dicke parameter. The decoherence rate follows the power-law relation $\gamma_n = \gamma(n+1)^{\gamma_k}$ with $\gamma = 0.6501$ and $\gamma_k = 0.6635$ in our experiment setup, as determined in our previous work.

\begin{figure}[tbp]
    \centering
    \includegraphics[width=1.0\columnwidth]{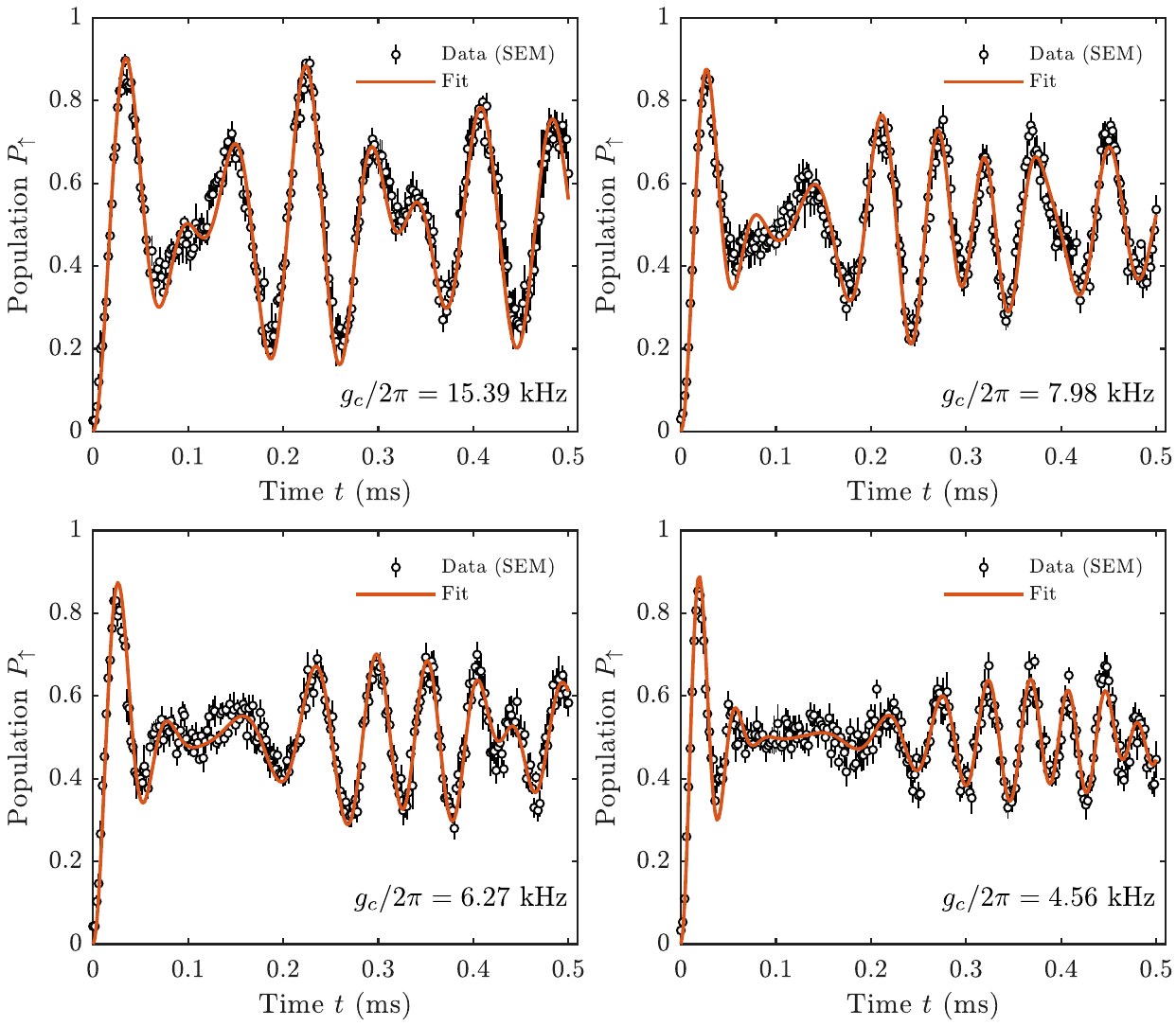}
     \caption{Experimental BSB Rabi oscillations and corresponding numerical fits. The evolution of the internal state population $P_{\uparrow}$ as a function of pulse duration $t$ is shown. Black circles represent experimental data (mean $\pm$ s.e.m.) for 300 measurements per point. Solid orange lines are fits used to extract the phonon statistics. The resulting reconstructed phonon distributions $P_n$ are displayed in Fig.~2 of the main text.}
    \label{fig:SM_Rabi}
\end{figure}

To quantify the uncertainties in the reconstructed $P_n$, a Monte Carlo (MC) procedure is implemented. For each experimental trace, 100 synthetic data sets are generated by adding Gaussian noise (based on the standard deviation of experimental residuals) to the mean population data. Each synthetic set is independently fitted. The standard deviations across these MC iterations define the error bars for $P_n$ and the mean phonon number $\langle n \rangle$ shown in the main text. Representative BSB Rabi fits for conditions in Fig.~2(b) are shown in Fig.~\ref{fig:SM_Rabi}.

The experimental parameter sets corresponding to the circle and square markers in Fig.~2(a) of the main text are summarized in Table~\ref{tab:ExpPara}. For each parameter set, we list the steady-state mean phonon number obtained from the mean-field analytical model, the full numerical calculation, and the reconstruction from the measured BSB Rabi oscillations. The reconstructed values are in good agreement with the numerical predictions over the investigated parameter range.

\begin{table}[htbp]
\caption{
Experimental parameter sets corresponding to Fig.~2(a) of the main text. Mean phonon numbers from the mean-field model, numerical calculation, and experimental reconstruction are listed.
}
\label{tab:ExpPara}
\centering
\begin{tabular}{cccc}
\hline
\hline
$g_h/2\pi \,/\, g_c/2\pi$ (kHz) &
Mean-field $\bar n$ &
Numerical $\bar n$ &
Reconstructed $\bar n$ \\
\hline

2.28 / 7.98
& 0
& 0.792
& $0.793 \pm 0.013$ \\

2.28 / 6.27
& 0.224
& 1.170
& $1.200 \pm 0.027$ \\

2.28 / 4.56
& 1.325
& 2.225
& $2.320 \pm 0.064$ \\

2.28 / 3.42
& 3.549
& 4.362
& $4.686 \pm 0.277$ \\

4.56 / 15.39
& 0
& 0.624
& $0.624 \pm 0.007$ \\

4.56 / 7.98
& 0.541
& 1.283
& $1.314 \pm 0.026$ \\

4.56 / 6.27
& 1.148
& 1.910
& $1.975 \pm 0.053$ \\

4.56 / 4.56
& 2.741
& 3.561
& $3.774 \pm 0.189$ \\

\hline
\hline
\end{tabular}
\end{table}

\subsection{AC Electric-Field Injection and Preliminary Calibration of Field Amplitude}

\begin{figure}[h]
    \centering
    \includegraphics[width=0.6\linewidth]{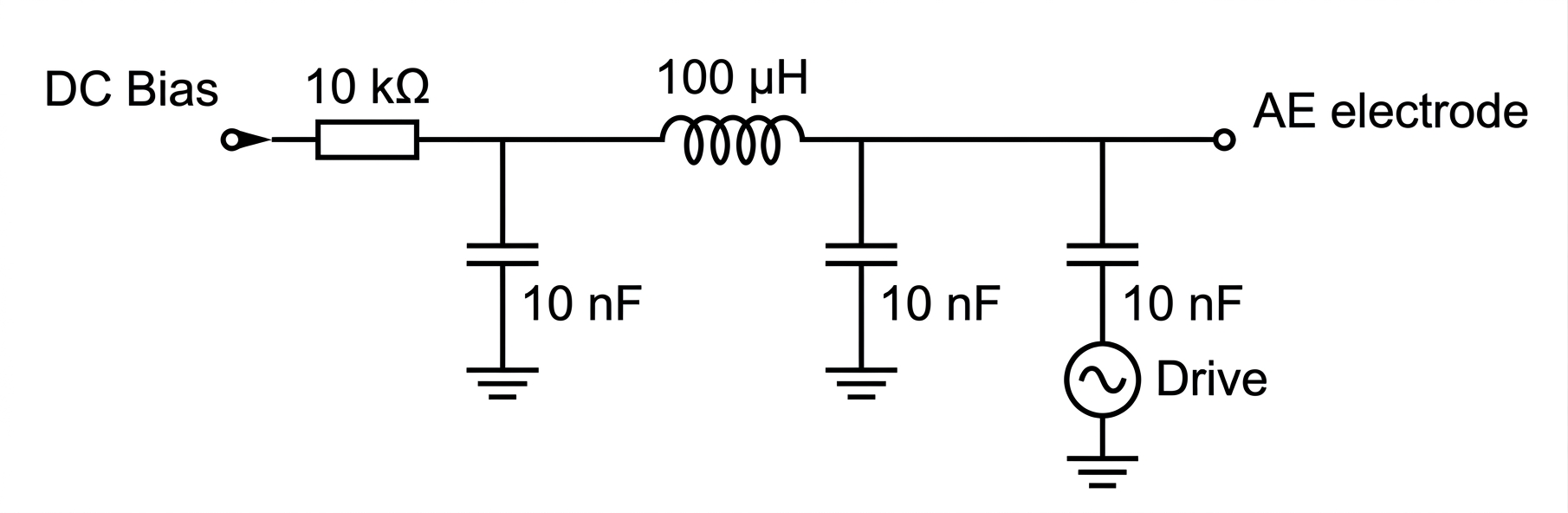}
    \caption{Schematic of the injection and filtering circuit. The DC bias is routed through a $\pi$-filter, while the external AC drive is capacitively coupled to the AE electrode via a 10~nF capacitor for electric field injection.}
    \label{fig:circuit_diagram}
\end{figure}

To couple the external drive to the SET’s AE electrode, we employ a capacitive coupling scheme. Each DC electrode is integrated with a low-pass $\pi$-filter; these are designed to suppress RF pickup and ambient electronic noise near the ion's motional frequencies while preserving the ability to apply DC bias offsets for micromotion compensation. As depicted in the schematic (Fig.~\ref{fig:circuit_diagram}), the drive signal is generated by a waveform generator (RIGOL DG4162), attenuated by $-80$~dB to reach the desired field strength, and subsequently injected onto the AE electrode through a coupling capacitor. This configuration enables the generation of a well-defined external electric field to coherently drive the ion's motional states. To ensure the drive is on resonance with the ion's axial motional frequency $\omega_z$, we implement a ``tickle'' method to scan the excitation frequency and identify the resonance peak.

We perform a preliminary calibration to establish the relationship between the applied voltage and the resulting electric field amplitude $E_0$ by characterizing the displacement of the ion's motional state. The experimental procedure is as follows: 
The ion is first initialized to the motional ground state $\ket{\downarrow, n=0}$ via Doppler cooling, sideband cooling, and optical pumping. Subsequently, the external drive is applied for a fixed duration $t_{\text{drive}} = 1.7$~ms. Following the drive pulse, optical pumping is repeated to ensure the ion remains in the internal state $\ket{\downarrow}$. We then perform blue sideband (BSB) excitation for varying probe durations. By analyzing the resulting BSB Rabi oscillations, we extract the mean phonon number $\bar{n}$ using the phonon reconstruction method described in the previous subsection.

The driven dynamics are governed by the Hamiltonian $\hat{H}_{\mathrm{drive}}$ in Eq.~\ref{drivehamiltonian}. This evolution is equivalent to the displacement operator $\hat{D}(\alpha) = \exp(\alpha\hat{a}^\dagger - \alpha^*\hat{a})$, where the displacement magnitude is given by $\alpha = -\frac{q E_0 z_0}{2\hbar} t e^{-i\phi_d}$. Under this interaction, the ion's motional state evolves into a coherent state $\ket{\alpha}$ with an average phonon occupancy $\bar{n} = |\alpha|^2$. Given the characteristic length scale $z_0 = \sqrt{\hbar / (2m\omega_z)}$, the electric field amplitude $E_0$ is calibrated as:
\begin{equation}
    E_0 = \frac{\sqrt{8 m \omega_z \hbar \bar{n}}}{q t_{\text{drive}}}.
\end{equation}

\begin{figure}[t]
    \centering
    \includegraphics[width=0.48\textwidth]{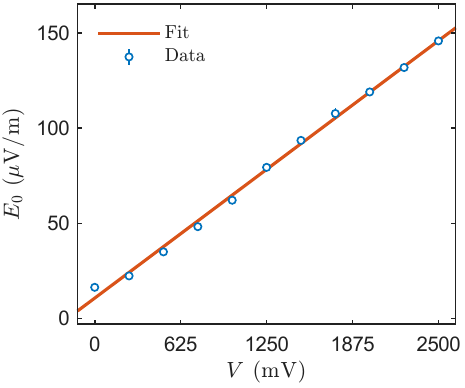}
    \caption{Preliminary calibration of the injected electric field $E_0$. The effective field amplitude is plotted against the signal generator's peak-to-peak voltage $V$. Error bars represent the SEM of three independent fits to the BSB Rabi oscillations, where each fit yields a mean phonon number $\bar{n}$. The red solid line indicates a linear fit.}
    \label{fig:E_field_calibration}
\end{figure}

Figure~\ref{fig:E_field_calibration} demonstrates a clear linear relationship between the waveform generator's output voltage and the effective electric field amplitude $E_0$ sensed by the ion. However, this preliminary calibration serves primarily to verify the system's linearity. In practice, the calibration slope exhibits slight daily fluctuations, likely due to sub-micron shifts in the ion's equilibrium position, which modify the ion-electrode distance and the resulting geometric coupling strength. Furthermore, the sensitivity and accuracy of this phonon-reconstruction-based method are limited. At weak field strengths, the precision becomes highly dependent on the initial ground-state cooling fidelity. Additionally, extending the drive duration to enhance the signal introduces non-negligible heating effects that complicate the dynamics. Finally, the extensive measurement overhead required to fully resolve the blue sideband Rabi oscillations for accurate $\bar{n}$ extraction inherently limits the temporal resolution and sensitivity of this approach.

\section{Characteristic Function and Experimental Reconstruction}
\label{sec:char_func}

To investigate the phase-space signatures of the single ion's motional state, we perform a direct readout of the characteristic function, $\chi(\beta)$. This approach allows us to clearly observe phase-space symmetry breaking, which manifests as an injection-locking-like behavior of the phase. Furthermore, $\chi(\beta)$ serves as a sensitive probe, enabling the quantitative determination of the external electric field strength. 

In addition to these signatures, the corresponding Wigner function is reconstructed by applying a two-dimensional Fourier transform to $\chi(\beta)$, providing a comprehensive visualization of the quantum state in phase space. While our methodology remains consistent with the principles described in Ref.~\cite{fluhmann2020direct}, it departs from the referenced work by employing a distinct initial internal state. The specific experimental configuration and procedures are detailed as follows:

\subsection{Bichromatic Light Field and Parameter Calibration}

To construct the bichromatic light field, we employ a 729-nm laser that simultaneously drives the red and blue motional sidebands of the $\ket{\downarrow}\equiv\ket{4^2S_{1/2}, m_J = +1/2} \leftrightarrow \ket{\uparrow}\equiv\ket{3^2D_{5/2}, m_J = +1/2}$ quadrupole transition. Under the rotating-wave approximation, the interaction Hamiltonians for the resonant red sideband (RSB) and blue sideband (BSB) are given by:
\begin{subequations}
\begin{equation}
    \hat{H}_{\text{rsb}} = i\hbar\eta\frac{\Omega_{r}}{2} \left( e^{i\phi_r}\hat{a}\hat{\sigma}_+ - e^{-i\phi_r}\hat{a}^{\dagger}\hat{\sigma}_- \right),
\end{equation}
\begin{equation}
    \hat{H}_{\text{bsb}} = i\hbar\eta\frac{\Omega_{b}}{2} \left( e^{i\phi_b}\hat{a}^\dagger\hat{\sigma}_+ - e^{-i\phi_b}\hat{a}\hat{\sigma}_- \right).
\end{equation}
\end{subequations}

The total Hamiltonian is obtained by the superposition of these two fields. Assuming equal Rabi frequencies, $\Omega_{r} = \Omega_{b} = \Omega$, and defining the phase difference $\Delta\phi = \phi_r - \phi_b$ and the phase sum $\Phi = \phi_r + \phi_b$, the combined Hamiltonian is expressed as:
\begin{equation}
    \hat{H}_{\text{bic}} = \hbar\eta\frac{\Omega}{2} \left[ i e^{i\frac{\Phi}{2}} \left( e^{i\frac{\Delta\phi}{2}}\hat{a} + e^{-i\frac{\Delta\phi}{2}}\hat{a}^\dagger \right) \hat{\sigma}_+ - i e^{-i\frac{\Phi}{2}} \left( e^{-i\frac{\Delta\phi}{2}}\hat{a} + e^{i\frac{\Delta\phi}{2}}\hat{a}^\dagger \right) \hat{\sigma}_- \right].
\end{equation}
By setting the phase sum to $\Phi = 3\pi$ per experimental convention, the Hamiltonian simplifies to a spin-dependent force:
\begin{equation}
    \hat{H}_{\text{bic}} = \hbar\eta\frac{\Omega}{2} \hat{\sigma}_x \left( \hat{a} e^{i\frac{\Delta\phi}{2}} + \hat{a}^\dagger e^{-i\frac{\Delta\phi}{2}} \right).
\end{equation}

The associated unitary evolution operator, $\hat{U}_{\text{bic}}(t) = \exp(-i \hat{H}_{\text{bic}} t / \hbar)$, acts as a spin-dependent displacement operator:
\begin{equation}
    \hat{U}_{\text{bic}}(t) = \exp \left[ \hat{\sigma}_x \left( \frac{\beta(t)}{2} \hat{a}^\dagger - \frac{\beta^*(t)}{2} \hat{a} \right) \right] = \hat{D}\left(\hat{\sigma}_x \frac{\beta(t)}{2}\right),
\end{equation}
where the complex displacement amplitude is $\beta(t)/2 = -i \eta \Omega t e^{-i\Delta\phi/2} / 2$. 

\begin{figure}[htbp]
    \centering
    \includegraphics[width=0.48\textwidth]{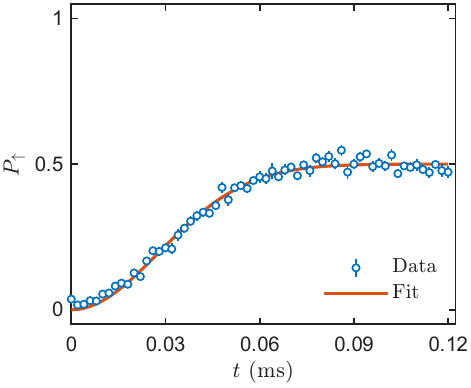}
    \caption{Time evolution of the internal excited-state population $P_\uparrow$ driven by a bichromatic field. Markers represent experimental data (mean $\pm$ s.e.m.) averaged over 800 measurements per point. The solid line denotes a fit to the theoretical model used to calibrate the coupling strength $\Omega$.}
    \label{fig:S1}
\end{figure}

To calibrate the effective coupling strength, we apply this interaction to an initial state $\ket{\Psi_0} = \ket{\downarrow}\ket{n=0}$ with $\Delta\phi=0$. Expressing the internal ground state in the $\hat{\sigma}_x$ basis as $\ket{\downarrow} = (\ket{+_x} - \ket{-_x})/\sqrt{2}$, the system evolves into:
\begin{equation}
    \ket{\Psi(t)} = \frac{1}{\sqrt{2}} \left[ \ket{+_x} \ket{\beta(t)/2} - \ket{-_x} \ket{-\beta(t)/2} \right],
\end{equation}
where $\ket{\pm \beta/2}$ are coherent states. The internal state population is determined by the expectation value $\langle \hat{\sigma}_z(t) \rangle = \bra{\Psi(t)} \hat{\sigma}_z \ket{\Psi(t)}$. Given that $\hat{\sigma}_z \ket{\pm_x} = \ket{\mp_x}$ and $\langle \alpha | \gamma \rangle = e^{-\frac{1}{2}(|\alpha|^2 + |\gamma|^2 - 2\alpha^* \gamma)}$, we find:
\begin{equation}
    \langle \hat{\sigma}_z(t) \rangle = -\frac{1}{2} \left[ \langle \beta(t)/2 | -\beta(t)/2 \rangle + \langle -\beta(t)/2 | \beta(t)/2 \rangle \right] = -e^{-\frac{1}{2}|\beta(t)|^2}.
\end{equation}
Using the relation $P_\uparrow(t) = (1 + \langle \hat{\sigma}_z \rangle) / 2$, the resulting time evolution of the excited-state population follows:
\begin{equation}
    P_\uparrow(t) = \frac{1}{2} \left( 1 - e^{-\frac{1}{2}(ct)^2} \right),
\end{equation}
where $c = \eta \Omega$ is the effective displacement rate. By fitting the experimental data to this model (see Fig.~\ref{fig:S1}), we extract the Rabi frequency $\Omega = 2\pi \times (49.98 \pm 1.28)$~kHz. This calibration ensures the robustness of the quantum state reconstruction, maintaining a stable Rabi frequency of $\Omega \approx 2\pi \times 50$~kHz throughout the measurements.

\subsection{Motional State Tomography via the Characteristic Function}
\label{subsec:motionaltomo}
\begin{figure}[htbp]
    \centering
    \includegraphics[width=0.48\textwidth]{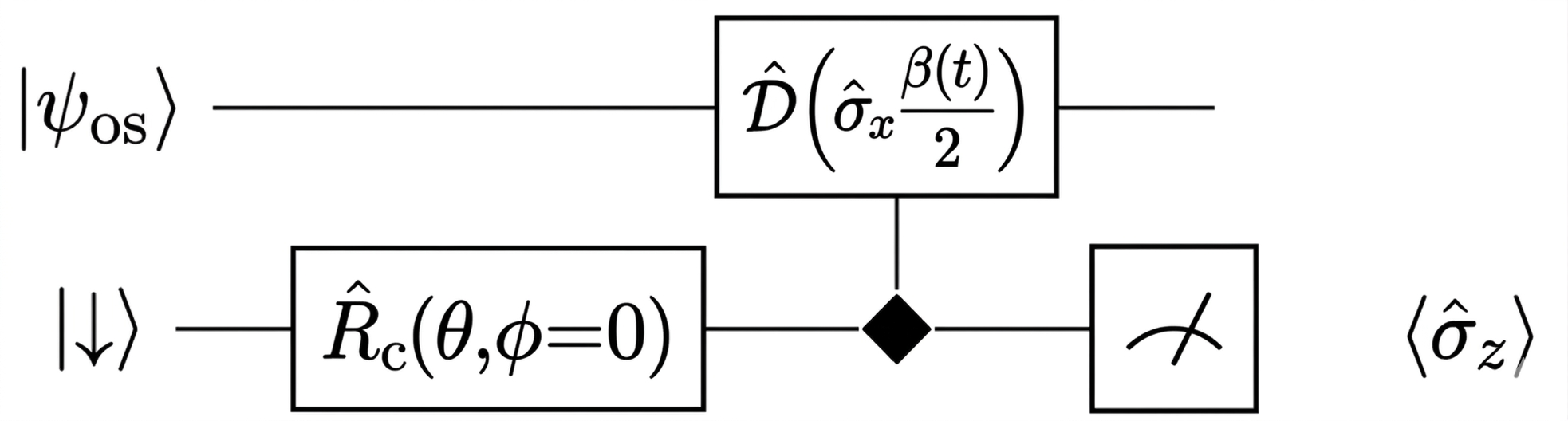} 
    \caption{Quantum circuit for the direct measurement of the motional characteristic function. Starting from $\ket{\downarrow}\ket{\psi_{\text{os}}}$, a carrier rotation $\hat{R}_c(\theta, \phi)$ creates an internal superposition. A subsequent bichromatic interaction applies the spin-dependent displacement $\hat{D}(\hat{\sigma}_x \beta(t)/2)$, followed by a measurement of the population $\langle \hat{\sigma}_z \rangle$ to extract the characteristic function $\chi(\beta)$.}
    \label{fig:S2}
\end{figure}

To reconstruct the motional quantum state, we implement a direct tomography protocol inspired by Ref.~\cite{fluhmann2020direct}. While the original method typically assumes an initial state $\ket{\uparrow}$, our experimental sequence begins with the ion in the internal ground state $\ket{\downarrow}$, as illustrated in the quantum circuit in Fig.~\ref{fig:S2}.

The process initiates with a carrier rotation operator defined as:
\begin{equation}
    \hat{R}_c(\theta, \phi) = \cos\left(\frac{\theta}{2}\right) \hat{I} - i\sin\left(\frac{\theta}{2}\right) \left[ \cos(\phi)\hat{\sigma}_x + \sin(\phi)\hat{\sigma}_y \right],
\end{equation}
where $\theta$ and $\phi$ denote the pulse area and phase, respectively. Setting $\phi = 0$ for the initial rotation, the state after the carrier pulse is $\ket{\Psi_1} = \hat{R}_c(\theta, 0)\ket{\downarrow}\ket{\psi_{\text{os}}}$, where the internal part becomes:
\begin{equation}
    \ket{\psi_{\text{int},1}} = \cos\left(\frac{\theta}{2}\right)\ket{\downarrow} - i\sin\left(\frac{\theta}{2}\right)\ket{\uparrow}.
\end{equation}
In the $\hat{\sigma}_x$ basis, where $\ket{\uparrow, \downarrow} = (\ket{+_x} \pm \ket{-_x})/\sqrt{2}$, this state can be rewritten as:
\begin{equation}
    \ket{\psi_{\text{int},1}} = \frac{1}{\sqrt{2}} \left[ \left(\cos\frac{\theta}{2} - i\sin\frac{\theta}{2}\right)\ket{+_x} - \left(\cos\frac{\theta}{2} + i\sin\frac{\theta}{2}\right)\ket{-_x} \right].
\end{equation}

Next, the bichromatic interaction applies a spin-dependent displacement $\hat{U}_{\text{bic}}(t) = \hat{D}(\hat{\sigma}_x \beta/2)$. Utilizing the property $\hat{D}(\hat{\sigma}_x \alpha)\ket{\pm_x} = \hat{D}(\pm \alpha)\ket{\pm_x}$, the total system state evolves to:
\begin{equation}
    \ket{\Psi(t)} = \frac{1}{\sqrt{2}} \left[ e^{-i\theta/2}\ket{+_x}\hat{D}\left(\frac{\beta}{2}\right)\ket{\psi_{\text{os}}} - e^{i\theta/2}\ket{-_x}\hat{D}\left(-\frac{\beta}{2}\right)\ket{\psi_{\text{os}}} \right].
\end{equation}

The expectation value of the Pauli operator $\hat{\sigma}_z$ is then measured. Using the relation $\hat{\sigma}_z \ket{\pm_x} = \ket{\mp_x}$ and the orthogonality $\langle \pm_x | \mp_x \rangle = 0$, the measurement yields:
\begin{align}
    \langle \hat{\sigma}_z(t) \rangle &= \bra{\Psi(t)} \hat{\sigma}_z \ket{\Psi(t)} \nonumber \\
    &= -\frac{1}{2} \left[ e^{i\theta} \bra{\psi_{\text{os}}} \hat{D}^\dagger\left(\frac{\beta}{2}\right) \hat{D}\left(-\frac{\beta}{2}\right) \ket{\psi_{\text{os}}} + e^{-i\theta} \bra{\psi_{\text{os}}} \hat{D}^\dagger\left(-\frac{\beta}{2}\right) \hat{D}\left(\frac{\beta}{2}\right) \ket{\psi_{\text{os}}} \right].
\end{align}
Since $\hat{D}^\dagger(\alpha)\hat{D}(-\alpha) = \hat{D}(-2\alpha)$, and defining the motional characteristic function as $\chi(\beta) = \langle \hat{D}(\beta) \rangle$, we obtain:
\begin{align}
    \langle \hat{\sigma}_z(t) \rangle &= -\frac{1}{2} \left[ e^{i\theta} \chi^*(\beta) + e^{-i\theta} \chi(\beta) \right] \nonumber \\
    &= -\cos(\theta)\text{Re}[\chi(\beta)] - \sin(\theta)\text{Im}[\chi(\beta)].
\end{align}

Experimentally, we record the excited state population $P_\uparrow = (1 + \langle \hat{\sigma}_z \rangle)/2$. By choosing the carrier pulse area $\theta=0$ or $\theta=\pi/2$, both the real and imaginary parts of $\chi(\beta)$ can be extracted, from which the Wigner function $W(\gamma)$ is reconstructed via the Fourier transform of $\chi(\beta)$\cite{gerry2023introductory}.

$$W(\gamma) = \frac{1}{\pi^2} \int \chi(\beta) e^{\gamma \beta^* - \gamma^* \beta} d^2\beta$$

\section{Physical Foundations of the Sensing Protocol}
\subsection{Optimal Sensing Axis Justification}
\label{subsec:optimal_axis}

To rigorously justify why the line slice along the imaginary axis ($\text{Re}(\beta) = 0$) provides the maximum contrast for electric-field sensing and extracts the most abundant phase-space information, we analyze the fundamental Fourier transform relation between $\chi(\beta)$ and the Wigner function $W(\gamma)$. By definition, the two representations are connected via:
\begin{equation}
\chi(\beta) = \int W(\gamma) e^{\beta \gamma^* - \beta^* \gamma} d^2\gamma.
\end{equation}
Letting $\beta = u + iv$ and $\gamma = x + iy$ (where $x,y$ denote the position and momentum quadratures of the phase space), the exponent simplifies to $\beta \gamma^* - \beta^* \gamma = 2i(vx - uy)$. Therefore, the imaginary part of the characteristic function is explicitly expressed as a two-dimensional sine transform:
\begin{equation}
\text{Im}[\chi(u,v)] = \int_{-\infty}^{\infty} \int_{-\infty}^{\infty} W(x,y) \sin[2(vx - uy)] dxdy.
\end{equation}

In our system, the phase of the external electric-field drive is locked at $\phi_d = 0$. In the absence of the drive ($E_0 = 0$), the steady state of the single-ion phonon laser is a phase-symmetric limit cycle. Its corresponding Wigner function $W_0(x,y)$ exhibits perfect reflection symmetry across both quadratures, satisfying $W_0(x,y) = W_0(-x,y) = W_0(x,-y)$. Because $\sin[2(vx-uy)]$ is an odd function, integrating it against the symmetric distribution $W_0(x,y)$ causes the imaginary part to vanish identically across the entire phase space:
\begin{equation}
\text{Im}[\chi_0(u,v)] = 0 \quad (\forall u, v).
\end{equation}
This mathematically guarantees a robust, zero-background baseline for our electrometry, which is immune to isotropic thermal fluctuations or symmetric phase noise.

Upon injecting the external electric field ($E_0 \neq 0$), this phase symmetry is explicitly broken. As detailed in the main text, the limit cycle collapses into an injection-locked, phase-converged distribution localized around $\phi_d = 0$. This physical transformation implies that the displacement and localization of the Wigner function $W(x,y)$ occur predominantly along the real axis ($x$-direction) of the phase space, thereby generating a significant asymmetry relative to $x=0$, while the reflection symmetry along the $y$-direction ($W(x,y) = W(x,-y)$) is substantially preserved.

To maximize the readout contrast induced by this symmetry breaking, we seek the optimal slice in the $\beta$-space that yields the largest deviation from the zero baseline. By setting $\text{Re}(\beta) = u = 0$, the two-dimensional integral simplifies to a one-dimensional Fourier sine transform of the marginal probability distribution $P(x)$:
\begin{equation}
\text{Im}[\chi(0,v)] = \int_{-\infty}^{\infty} \left[ \int_{-\infty}^{\infty} W(x,y) dy \right] \sin(2vx) dx = \int_{-\infty}^{\infty} P(x) \sin(2vx) dx,
\end{equation}
where $P(x) = \int_{-\infty}^{\infty} W(x,y) dy$ represents the quadrature distribution projected onto the axis of the external drive. Because the driven interaction shifts the center of $P(x)$ away from the origin ($x=0$) and reduces its variance via phase convergence, $P(x)$ becomes highly asymmetric relative to $x=0$. This strong asymmetry is directly mapped onto the amplitude and frequency of the sine transform, unleashing high-contrast, macroscopic oscillations in $\text{Im}[\chi(0,v)]$.

Conversely, if one were to choose a slice along the real axis ($\text{Im}(\beta) = v = 0$), the expression would reduce to:
\begin{equation}
\text{Im}[\chi(u,0)] = \int_{-\infty}^{\infty} \left[ \int_{-\infty}^{\infty} W(x,y) dx \right] \sin(-2uy) dy = \int_{-\infty}^{\infty} P(y) \sin(-2uy) dy,
\end{equation}
where $P(y)$ is the projection along the unperturbed $y$-direction. Since the drive at $\phi_d=0$ does not displace the state along $y$, $P(y)$ remains symmetric ($P(y) = P(-y)$), causing $\text{Im}[\chi(u,0)]$ to stay flat at zero regardless of the field strength.

Therefore, selecting the slice along the pure imaginary axis ($\text{Re}(\beta) = 0$) is the mathematically optimal strategy. It aligns the measurement probe with the direction of maximum phase-space asymmetry, yielding the maximum signal contrast and rendering it the most sensitive observable for reconstructing symmetry-breaking features and performing high-precision electrometry.

\subsection{Optimal Sensing Time Justification}
\label{subsec:optimal_time}

We provide a rigorous theoretical derivation for the existence and analytical scaling of the optimal sensing time $t_0$ based entirely on the phase-space geometry of the limit cycle and its overlap dynamics under the probing fields. In our measurement protocol, the internal electronic population $P_{\uparrow}(t)$ of the ion is fundamentally mapped onto the imaginary part of the reduced motional characteristic function along the pure imaginary axis, $\text{Im}[\chi_E(\beta)]$. The bichromatic sideband drive generates a spin-dependent phase-space probing vector $\beta(t) = -i \eta \Omega t e^{-i\Delta\phi/2}$. By locking the experimental phase difference at $\Delta\phi = 2\pi$, the phase factor reduces to $e^{-i\pi} = -1$, constraining the probing trajectory strictly to the imaginary quadrature:
\begin{equation}
\label{eq:probing_trajectory}
\beta(t) = i y(t) = i \eta \Omega t.
\end{equation}

The external weak electric field $E$ under investigation acts as a symmetry-breaking perturbation to the open-system Liouvillian dynamics. To establish the precise mathematical link between this perturbation and the tracked signal, we analyze the system within the phase-space Wigner representation. In the absence of an external field, the steady state of the phonon laser features a complete phase diffusion, manifesting as a prominent amplitude-stabilized ring structure $W_0(x, p)$ with native $U(1)$ rotational symmetry, where $x = \text{Re}[\gamma]$ and $p = \text{Im}[\gamma]$. Under a weak directional electric field, the dominant first-order effect can be captured by an effective displacement of the limit-cycle distribution along the symmetry-breaking direction. The steady-state distribution thus evolves into a symmetry-broken, crescent-like profile $W_E(x, p)$ due to the competition between the directional field drive and the non-linear amplitude-stabilization forces.

Within the strict weak-field linear-response regime ($E \to 0$), the primary metrological signature of this crescent-like probability aggregation is governed by a first-order shift of the distribution's center-of-mass along the real quadrature axis, parameterized as $\langle \hat{a} \rangle_E = \gamma_E = \kappa_{\text{eff}} E \in \mathbb{R}$. Under this weak perturbation limit, the field-modified Wigner distribution can be modeled as a local spatial displacement along the probing axis, $W_E(x, p) = W_0(x - \gamma_E, p)$. Expanding this distribution via a first-order Taylor series around the unperturbed baseline ($\gamma_E \to 0$) yields the linearized perturbation relation:
\begin{equation}
\label{eq:wigner_taylor}
W_E(x, p) \approx W_0(x, p) + \left. \frac{\partial W_0(x - \gamma_E, p)}{\partial \gamma_E} \right|_{\gamma_E=0} \cdot \gamma_E = W_0(x, p) - \gamma_E \frac{\partial W_0(x, p)}{\partial x}.
\end{equation}
Equation~(\ref{eq:wigner_taylor}) demonstrates that the first-order structural perturbation to the Wigner function is directly proportional to its local spatial gradient along the symmetry-breaking axis.

To evaluate how this structural symmetry breaking manifests in characteristic-function space, we take the two-dimensional Fourier transform of the linearized distribution. Substituting Eq.~(\ref{eq:wigner_taylor}) into the formal definition of the characteristic function yields:
\begin{equation}
\label{eq:characteristic_integral_step}
\chi_E(\beta) = \int_{-\infty}^{\infty} \int_{-\infty}^{\infty} W_E(x, p) e^{\beta \gamma^* - \beta^* \gamma} dx dp \approx \chi_0(\beta) - \gamma_E \int_{-\infty}^{\infty} \int_{-\infty}^{\infty} \frac{\partial W_0(x, p)}{\partial x} e^{\beta \gamma^* - \beta^* \gamma} dx dp.
\end{equation}

The exponential kernel can be explicitly written in quadrature coordinates as $\beta \gamma^* - \beta^* \gamma = 2i(\text{Im}[\beta]x - \text{Re}[\beta]p)$. We perform an integration by parts with respect to $x$ for the second term in Eq.~(\ref{eq:characteristic_integral_step}). Since the unperturbed limit-cycle distribution vanishes at the phase-space boundaries ($W_0 \to 0$ as $x \to \pm\infty$), the boundary term vanishes, and the spatial derivative transfers to the exponential phase factor:
\begin{equation}
\label{eq:integration_by_parts}
-\int_{-\infty}^{\infty} \frac{\partial W_0(x, p)}{\partial x} e^{\beta \gamma^* - \beta^* \gamma} dx = \int_{-\infty}^{\infty} W_0(x, p) \frac{\partial}{\partial x} \left( e^{\beta \gamma^* - \beta^* \gamma} \right) dx = \left( \beta - \beta^* \right) \int_{-\infty}^{\infty} W_0(x, p) e^{\beta \gamma^* - \beta^* \gamma} dx.
\end{equation}

Given that $\gamma_E \in \mathbb{R}$ implies $\gamma_E = \gamma_E^*$, the prefactor can be written as $\gamma_E(\beta - \beta^*) = (\beta \gamma_E^* - \beta^* \gamma_E)$. Reassembling the full two-dimensional integral from Eq.~(\ref{eq:integration_by_parts}) back into Eq.~(\ref{eq:characteristic_integral_step}) directly isolates the first-order mapping of the center-of-mass shift into frequency space:
\begin{equation}
\label{eq:taylor_derivation}
\chi_E(\beta) \approx \chi_0(\beta) + \left( \beta \gamma_E^* - \beta^* \gamma_E \right) \chi_0(\beta) = \chi_0(\beta) \left[ 1 + \left( \beta \gamma_E^* - \beta^* \gamma_E \right) \right].
\end{equation}

This first-order structure is consistent with the displacement covariance of the Weyl characteristic function. By re-expressing the linear response in an exponential form that preserves the correct overlap structure and asymptotic decay behavior, we obtain the compact displacement-induced phase-space mapping:
\begin{equation}
\label{eq:displaced_characteristic}
\chi_E(\beta) \approx \chi_0(\beta) e^{\beta \gamma_E^* - \beta^* \gamma_E} = \chi_0(\beta) e^{\beta \gamma_E - \beta^* \gamma_E}.
\end{equation}

Evaluating Eq.~(\ref{eq:displaced_characteristic}) along our pure imaginary probing trajectory defined in Eq.~(\ref{eq:probing_trajectory}) yields the exponent $\beta \gamma_E - \beta^* \gamma_E = 2iy\gamma_E$. The perturbed characteristic function on the imaginary axis thus evolves as:
\begin{equation}
\label{eq:perturbed_imaginary}
\chi_E(iy) \approx \chi_0(iy) \left[ \cos(2y\gamma_E) + i \sin(2y\gamma_E) \right].
\end{equation}

Crucially, because the unperturbed steady state $W_0(x, p)$ possesses rotational symmetry, its unperturbed characteristic function evaluated along any axis is strictly real valued, forcing $\text{Im}[\chi_0(iy)] = 0$ and $\chi_0(iy) = \text{Re}[\chi_0(iy)]$. Consequently, the field-induced signal resulting from the broken symmetry is isolated entirely within the imaginary component:
\begin{equation}
\label{eq:isolated_imaginary}
\text{Im}[\chi_E(iy)] \approx \chi_0(iy) \sin(2y\gamma_E).
\end{equation}

The spin readout population is given by $P_{\uparrow}(t) = \frac{1}{2}\left(1 - \text{Im}[\chi_E(iy)]\right)$. To quantify the metrological capability of our scheme, we evaluate the differential sensitivity under the weak-field linear-response regime ($E \to 0$, implying $\gamma_E \to 0$):
\begin{equation}
\label{eq:differential_sensitivity}
\frac{\partial P_{\uparrow}}{\partial E} = -\frac{1}{2} \chi_0(iy) \left. \frac{\partial}{\partial E}\sin(2y\gamma_E) \right|_{E \to 0} = -\kappa_{\text{eff}} \cdot y \cdot \chi_0(iy).
\end{equation}

Substituting $y = \eta \Omega t$ from Eq.~(\ref{eq:probing_trajectory}) into Eq.~(\ref{eq:differential_sensitivity}), the absolute metrological susceptibility $\mathcal{S}(t) \equiv \left| \partial P_{\uparrow} / \partial E \right|$ scales as:
\begin{equation}
\label{eq:susceptibility_scaling}
\mathcal{S}(t) \propto t \cdot \chi_0(i\eta \Omega t).
\end{equation}

The mathematical structure of Eq.~(\ref{eq:susceptibility_scaling}) reveals that the existence of an optimal sensing time $t_0$ is a geometric consequence of phase-space overlap. Physically, the characteristic function $\chi_0(iy)$ quantifies the phase-space coherence between the steady-state limit-cycle distribution and its displacement under a momentum-direction shift generated by the displacement operator. As a coordinate-space ring, the two-dimensional Fourier transform yields a smooth overlap profile near the origin of the Fourier-conjugate (characteristic function) space ($y \to 0$). In the experimentally relevant weak-displacement regime ($y \ll 1$), the behavior of $\chi_0(iy)$ is governed by the low-order statistical moments of the phonon distribution. Retaining the exact second moment of the zero-field limit cycle, the characteristic function at small displacements can be expanded as $\chi_0(iy) \approx 1 - (\bar{n} + 1/2)y^2$, where $\bar{n} = \langle \hat{a}^\dagger \hat{a} \rangle_{\text{ss}}$ is the mean steady-state phonon occupation. To derive a physically well-behaved analytical scaling for the optimal sensing time, we employ a standard cumulant expansion, which yields a smooth Gaussian envelope:
\begin{equation}
\label{eq:gaussian_recasting}
\chi_0(iy) \approx \exp\left(-(\bar{n}+\frac{1}{2}) y^2\right) = \exp\left(-(\bar{n}+\frac{1}{2}) \eta^2 \Omega^2 t^2\right).
\end{equation}

We emphasize that Eq.~(\ref{eq:gaussian_recasting}) does not imply that the underlying phonon laser is a Gaussian state; rather, it serves as a generic short-range approximation capturing the local phase-space curvature (the statistical variance) of the limit-cycle state near the origin. This simplifies the time-dependent susceptibility to:
\begin{equation}
\label{eq:final_susceptibility}
\mathcal{S}(t) \propto t \cdot \exp\left(-(\bar{n}+\frac{1}{2}) \eta^2 \Omega^2 t^2\right).
\end{equation}

The optimization of the sensing duration, $t_0 = \text{arg max}_t \mathcal{S}(t)$, is governed by maximizing Eq.~(\ref{eq:final_susceptibility}) with respect to laboratory time $t$. Taking the derivative $\partial \mathcal{S} / \partial t = 0$ via the product rule yields:
\begin{equation}
\label{eq:derivative_derivation}
\frac{\partial \mathcal{S}}{\partial t} \propto \exp\left(-(\bar{n}+\frac{1}{2}) \eta^2 \Omega^2 t^2\right) + t \cdot \left(-2(\bar{n}+\frac{1}{2})\eta^2\Omega^2 t\right)\exp\left(-(\bar{n}+\frac{1}{2}) \eta^2 \Omega^2 t^2\right) = 0.
\end{equation}

Factoring out the non-zero exponential envelope directly yields the extremum condition:
\begin{equation}
\label{eq:extremum_condition}
1 - 2(\bar{n}+\frac{1}{2})\eta^2\Omega^2 t^2 = 0.
\end{equation}

Solving Eq.~(\ref{eq:extremum_condition}) directly establishes the analytical scaling law for the optimal electrometry sensing time:
\begin{equation}
\label{eq:optimal_time_scaling}
t_0 = \frac{1}{\sqrt{2\bar{n}+1}\eta\Omega}.
\end{equation}

Equation~(\ref{eq:optimal_time_scaling}) provides an intuitive physical interpretation based on phase-space matching. At short times ($t \to 0$), the susceptibility grows linearly because the readout displacement $\beta$ is insufficient to map the field-induced drift onto a detectable asymmetry. Conversely, at long times ($t \gg t_0$), the probing displacement grows too large, reducing the overlap with the reference state.

To ground this analytical framework in our experimental realization, we evaluate Eq.~(\ref{eq:optimal_time_scaling}) using the physical parameters of our trapped-ion system. In our setup, the carrier Rabi frequency is $\Omega = 2\pi \times 50\,\text{kHz}$ and the Lamb-Dicke parameter is $\eta = 0.114$, yielding an effective coupling strength $\eta\Omega = 2\pi \times 5.7\,\text{kHz}$. Depending on the experimental sensing operating point, the steady-state mean phonon number is tuned within $\bar{n} \in [2.0, 5.2]$. Substituting these values yields $t_0 \in [8.3, 12.5]\,\mu\text{s}$. For a typical value $\bar{n} \approx 3.5$, Eq.~(\ref{eq:optimal_time_scaling}) gives $t_0 \approx 9.87\,\mu\text{s}$, providing a robust quantitative justification for choosing the experimental sensing time at $t_s = 10\,\mu\text{s}$ to maximize the metrological gain.

\subsection{Comparison with Direct Coherent-State Sensing}
\label{subsec:coherent_comparison}

To further highlight the advantages of our active phonon laser sensing paradigm, we compare its performance against a natural alternative protocol: the direct coherent displacement of a pure motional ground state ($\langle n \rangle \approx 0$) under isolation. While such a passive Hermitian scheme theoretically boasts a minimum quantum noise floor limited only by the standard quantum limit (SQL), it suffers from severe bottlenecks in realistic implementations. 

First, a bare motional ground state is highly vulnerable to ambient heating. During the prolonged $t_{\mathrm{pl}} = 5.7\text{ ms}$ interaction window required for weak-field accumulation, environmental thermal leakage rapidly injects chaotic phonons, severely corrupting the pristine noise floor. Second, the resulting microscopic displacement yields a minute quantum state distinguishability. To extract the average phonon number $\langle n \rangle$ via the conventional blue-sideband (BSB) readout method, one must painstakingly sample the population across varying probe durations to map out the BSB Rabi oscillations. This tedious reconstruction process demands thousands of experimental repetitions, severely compromising the measurement efficiency.

For completeness, it is instructive to contrast our protocol with a direct coherent-state displacement sensing scheme based on an initially prepared motional ground state. In such a Hermitian scenario, the external electric field generates a unitary phase-space displacement that can, in principle, be accessed via the same symmetry-breaking readout in the characteristic-function domain. However, this approach relies on transient quantum coherence in an effectively isolated system and does not support a stabilizing dynamical mechanism against environmental noise. In particular, during the finite interaction time required for weak-field accumulation, residual heating and technical fluctuations continuously degrade the purity of the motional state, leading to rapid deterioration of the phase-space signal.

In contrast, the phonon-laser sensor operates as an intrinsically nonequilibrium steady-state system, where coherent driving and engineered dissipation jointly stabilize a macroscopic limit-cycle attractor. By operating in the weak-coupling regime near the phonon-lasing threshold, the Liouvillian-gap suppression provides an additional dissipative amplification channel, enabling continuous accumulation of weak perturbations within a stable dynamical manifold. This combination of dynamical stabilization and gap-enhanced susceptibility fundamentally distinguishes our protocol from passive coherent-state sensing schemes.

\section{Electric-Field Sensitivity Analysis}
\label{sec:sensitivity_dynamics}

\subsection{Sensitivity Definition}
\label{subsec:sensitivity_def}

We evaluate the electric-field sensing performance using the quantum-projection-noise-limited (or shot-noise-limited) sensitivity. To suppress line-triggering frequency noise, our total experimental cycle time is synchronized with the power line grid, yielding a fixed period of $T_{\mathrm{cyc}} = 20\text{ ms}$. The sensitivity $\eta_E$ is conventionally defined as
\begin{equation}
\eta_E = \delta E_{\mathrm{single}} \sqrt{T_{\mathrm{cyc}}},
\end{equation}
where $\delta E_{\mathrm{single}}$ denotes the minimum detectable electric field achieved within a single experimental shot.

This single-shot minimum detectable field $\delta E_{\mathrm{single}}$ is determined by the linear response of the measured atomic population $P$ to the external electric field $E$, written as
\begin{equation}
\delta E_{\mathrm{single}} = \frac{\Delta P}{\left|\partial P / \partial E\right|},
\end{equation}
where $\Delta P$ represents the statistical fluctuation governed by quantum projection noise. Here, the response slope $\partial P / \partial E$ encapsulates the complete chronological sequence of the sensing protocol detailed above, capturing the population variation at the designated probing snapshot $t_s$ following the $t_{\mathrm{pl}}$ evolution.

Assuming projective binomial measurement statistics, the standard deviation of the population is given by $\Delta P = \sqrt{P(1-P)}$. Around the optimal working point where the response curve exhibits its maximum slope (typically at $P \approx 1/2$), the projection noise reduces to $\Delta P \approx 1/2$. Consequently, the total-cycle experimental sensitivity simplifies to
\begin{equation}
\label{eq:sensitivity}
\eta_E = \frac{\sqrt{T_{\mathrm{cyc}}}}{2 \left|\partial P / \partial E\right|}.
\end{equation}

In our numerical simulations, the response slope $\partial P/\partial E$ is evaluated via the central finite-difference method under a weak electric-field perturbation $\delta e$:
\begin{equation}
\frac{\partial P}{\partial E} = \frac{P(E + \delta e; t_{\mathrm{pl}}, t_s) - P(E - \delta e; t_{\mathrm{pl}}, t_s)}{2\delta e}.
\end{equation}

\subsection{Reconstruction of Experimental Electric Fields}
\label{subsec:field_fitting_protocol}

To quantitatively extract the actual electric field strength $E_{\mathrm{fit}}$ acting on the single-ion sensor from raw experimental data, we employ a self-consistent non-linear least-squares optimization that maps the time-resolved electronic population trajectories directly onto the field amplitude domain. Rather than relying on standard empirical regressions, our protocol constructs a physics-driven forward model where the external field amplitude $E$ serves as the sole free optimization parameter, enforcing strict quantum-mechanical constraints on the fitted observables. 

The forward prediction framework maps each iterative trial field $E$ onto the time-resolved excitation trajectories $P_{\uparrow}(t)$ by executing a full-scale open-quantum-system simulation. The dynamics are initialized from a motional thermal state with a mean phonon number of $\bar{n}_{\mathrm{init}} = 0.3$. This state is prepared via a $2\text{ ms}$ resolved-sideband cooling sequence, which represents an optimized timing budget trade-off constrained by the total experimental cycle duration of $T_{\mathrm{cyc}} = 20\text{ ms}$. To maximize the dynamical susceptibility of the sensor, the cooling duration is purposefully bounded to ensure a sufficient temporal window for the subsequent $5.7\text{ ms}$ phonon-laser evolution and its long-term coherent interaction with the external driving field. While this truncated $2\text{ ms}$ cooling window does not drive the motion completely to the quantum ground state, it steers the initial state sufficiently deep into the resolved-sideband regime to enable robust, high-fidelity coherent manipulation. The value of $\bar{n}_{\mathrm{init}}$ is independently calibrated using the motional state reconstruction via the excitation profiles of blue-sideband Rabi oscillations in Sec.~\ref{subsec:fock_recon}. For each optimization step, the combined system-bath interaction is numerically integrated under the full Lindblad master equation for the fixed evolution duration of $t_{\mathrm{pl}} = 5.7\text{ ms}$. This ensures that the density matrix completes its symmetry-breaking and reaches structural stabilization. The resulting steady-state motional density matrix $\hat{\rho}_m(E) = \hat{\rho}(t_{\mathrm{pl}})$ is then utilized to evaluate the prospective probing population profile $P_{\uparrow}(t)$ as a function of the probing duration $t$.

Specifically, the optimization routine applies a strict physical truncation cutoff at $T_{\mathrm{max}} = 65.0\ \mu\text{s}$. Only the experimental data points recorded within this early stage ($t_j < T_{\mathrm{max}}$) contribute to the cost-function minimization, whereas the post-cutoff trajectories ($t_j \ge 65.0\ \mu\text{s}$) are excluded. This temporal truncation is physically motivated by the underlying parameter-dependent sensing dynamics: the characteristic population oscillations explicitly driven by the external electric field are optimally pronounced within the initial $65.0\ \mu\text{s}$ transient window, which defines the high-susceptibility sensing regime. Beyond this cutoff, the time-resolved trajectories across all field strengths monotonically converge and stabilize near an uninformative baseline of $P_{\uparrow} \approx 0.5$, yielding a vanishing parameter gradient ($\partial P_{\uparrow}/\partial E \to 0$). By confining the fitting domain to the pre-cutoff window ($t_j < T_{\mathrm{max}}$), we ensure that the optimization remains exclusively locked onto the pristine, field-sensitive coherent dynamics of the phonon-laser sensor, while the long-term trailing steady-state data points are preserved solely as a visual benchmark to verify the global boundary conditions of the system.

The statistical uncertainties for both the extracted field strengths and the time-resolved observables are derived rigorously from raw experimental fluctuations without any heuristic scaling. For each specific probing duration $t_j$, the quantum electronic state is repetitively interrogated via projective measurements over $N_{\mathrm{shots}} = 3000$ independent experimental cycles to suppress statistical fluctuations. The experimental mean population $\bar{P}_{\mathrm{exp}}(t_j)$ and its corresponding standard error of the mean (SEM), denoted as $\sigma_j$, are evaluated directly from this ensemble via
\begin{equation}
\sigma_j = \frac{\Delta P_j}{\sqrt{N_{\mathrm{shots}}}},
\end{equation}
where $\Delta P_j$ represents the empirical sample standard deviation of the single-shot projection outcomes. Since each individual measurement is fundamentally governed by quantum projection noise (QPN), these empirical SEM values naturally embody the true quantum statistical variances of our sensor. In the non-linear optimization framework, $\sigma_j$ serves as the exact statistical weight, penalizing deviations via a weighted $\chi^2$ cost function evaluated exclusively over the selected fitting window:
\begin{equation}
\chi^2(E) = \sum_{t_j < T_{\mathrm{max}}} \left[ \frac{\bar{P}_{\mathrm{exp}}(t_j) - P_{\uparrow}(t_j, E)}{\sigma_j} \right]^2.
\end{equation}
Consequently, data points characterized by tighter noise distributions and higher statistical confidence automatically dominate the field-extraction constraints.

The reconstructed electric field strength $E_{\mathrm{fit}}$ is uniquely defined as the argument that minimizes this weighted cost function, such that $E_{\mathrm{fit}} = \arg\min_{E} \chi^2(E)$. Following the convergence of this minimization routine, the final symmetric uncertainty $\Delta E$ attached to the best-fit field is rigorously determined from the asymptotic covariance matrix $\Sigma$ evaluated at the global minimum, which directly yields $\Delta E = \sqrt{\Sigma}$. This matrix product naturally incorporates the localized Jacobian (the susceptibility of the population with respect to the field) and the diagonal weight matrix.

While Figs.~4(a) and 4(b) in the main text focus on two representative regimes to illustrate the core sensing principle, here we present the comprehensive experimental dataset encompassing all five distinct phonon-laser operating parameter sets investigated in this work. This extended gallery showcases the systematic reproducibility of the field-dependent quantum dynamics across diverse structural regimes of the sensor.

\begin{figure}[htbp]
\centering
\includegraphics[width=1\linewidth]{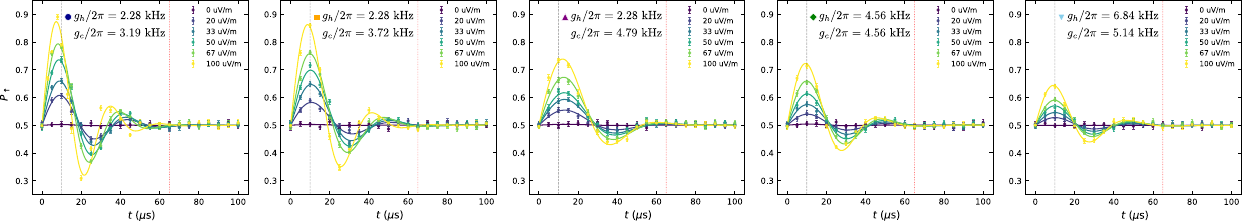}
\caption{Comprehensive time-resolved electrometry dynamics across all five phonon-laser parameter sets.}
\label{fig:sm_all_5_regimes}
\end{figure}

\subsection{Experimental Extraction of Electrometry Sensitivity}
\label{subsec:sensitivity_derivation_multi}

To experimentally evaluate the shot-noise-limited electrometry sensitivity, we investigate the sensor's linear response near the null-field baseline across five distinct phonon laser parameter configurations. For each configuration, a weighted linear regression, $P_{\uparrow}(E) = \alpha E + b$, is applied exclusively to the first four low-field data points to ensure the population response remains bounded within the pristine, linear susceptibility domain before saturation effects manifest. The fitting residuals are weighted by the empirical standard error of the mean ($\sigma_j$) from 3000 independent projection shots per point. 

The experimental response slope $\alpha$ and its associated uncertainty $\sigma_\alpha$ are extracted from the localized covariance matrix of the linear regression. To transform this empirical slope into the shot-noise-limited sensitivity floor, we consider the foundational projection noise. At the optimal linear bias point where the excitation probability hovers near $P_{\uparrow} \approx 0.5$, the single-shot quantum projection noise (QPN) obeys a binomial distribution, yielding a maximum single-shot population standard deviation of $\Delta P_{\uparrow} = \sqrt{P_{\uparrow}(1-P_{\uparrow})} = 1/2$. Incorporating this fundamental noise limit alongside the cycle time $T_{\mathrm{cyc}} = 20\text{ ms}$, the experimental sensitivity $\eta_{E,\mathrm{exp}}$ and its propagated statistical uncertainty $\Delta\eta_{E,\mathrm{exp}}$ are rigorously determined via:
\begin{equation}
\eta_{E,\mathrm{exp}} = \frac{\Delta P_{\uparrow} \sqrt{T_{\mathrm{cyc}}}}{|\alpha|} = \frac{\sqrt{T_{\mathrm{cyc}}}}{2 |\alpha|}, \quad \Delta\eta_{E,\mathrm{exp}} = \eta_{E,\mathrm{exp}} \left( \frac{\sigma_\alpha}{|\alpha|} \right).
\end{equation}

The theoretical sensitivity benchmark $\eta_{E,\mathrm{th}}$ is evaluated by numerically solving the full open-quantum-system Lindblad master equation. The density matrix is dynamically evolved under the combined sideband-lasing and electric-field-driving Hamiltonians for an interaction duration of $5.7\text{ ms}$. To capture the phase-sensitive displacement encoded in the motional state, the sensing sequence concludes with a carrier $\pi/2$ pulse that maps the motional displacement onto the electronic Bloch sphere's equator, followed by a $t_s = 10\ \mu\text{s}$ bichromatic probing pulse. This interferometric detection scheme maps the phase-space displacement onto the electronic excitation probability $P_{\uparrow} = [1 - \mathrm{Im}[\chi(\beta)]]/2$. By scanning the external field, the theoretical susceptibility slope $\alpha_{\mathrm{th}} = \partial P_{\uparrow}/\partial E$ is locally extracted via numerical differentiation at the optimal operational bias point ($P_{\uparrow} \approx 0.5$). The ultimate projection-noise-limited theoretical sensitivity ceiling is then quantified through $\eta_{E,\mathrm{th}} = \sqrt{T_{\mathrm{cyc}}}/(2|\alpha_{\mathrm{th}}|)$.

The quantitative results evaluated with a cycle period $T_{\mathrm{cyc}} = 20\text{ ms}$ are summarized in Table~\ref{tab:sensitivity_comparison}. The excellent agreement between the dynamically simulated $\eta_{E,\mathrm{th}}$ and the fitted $\eta_{E,\mathrm{exp}}$ values across all configurations confirms the high predictive fidelity of our theoretical framework. Crucially, this systematic match demonstrates that adjusting the phonon-laser operational parameters directly modulates the sensor's response profile; by fine-tuning the sideband coupling ratios closer to the threshold and simultaneously lowering the effective coupling strengths, we successfully harness the dynamical susceptibility of the underlying Liouvillian dynamics, triggering a substantial reduction in the electrometry sensitivity floor down to the sub-microvolt regime.

\begin{table}[htbp]
\caption{Electrometry sensitivity under multiple configurations.}
\label{tab:sensitivity_comparison}
\centering
\setlength{\tabcolsep}{4pt} 
\begin{tabular}{lccccc}
\hline
\hline
Configuration & Mean Phonon & Liouvillian Gap & Slope $\alpha$ & \multicolumn{2}{c}{Sensitivity ($\mu\text{V/m}/\sqrt{\text{Hz}}$)} \\ \cline{5-6}
$g_h / g_c$ ($2\pi \times$ kHz) & $\bar{n}$ & $\Delta_{\mathcal{L}}$ ($\text{s}^{-1}$) & ($\text{m/}\mu\text{V}$) & Theory $\eta_{E,\mathrm{th}}$ & Exp. $\eta_{E,\mathrm{exp}}$ \\
\hline
2.28 / 3.19   & 5.16 & 317  & $(5.00 \times 10^{-3}) \pm 2.7 \times 10^{-4}$ & 13.33 & $14.15 \pm 0.77$ \\
2.28 / 3.72   & 3.56 & 519  & $(4.49 \times 10^{-3}) \pm 2.2 \times 10^{-4}$ & 16.05 & $15.74 \pm 0.78$ \\
2.28 / 4.79   & 2.00 & 1142 & $(2.75 \times 10^{-3}) \pm 3.0 \times 10^{-4}$ & 28.18 & $25.71 \pm 2.77$ \\
4.56 / 4.56   & 3.56 & 1341 & $(2.23 \times 10^{-3}) \pm 2.2 \times 10^{-4}$ & 31.41 & $31.73 \pm 3.15$ \\
6.84 / 5.14   & 3.56 & 1688 & $(1.34 \times 10^{-3}) \pm 2.2 \times 10^{-4}$ & 50.83 & $52.80 \pm 8.65$ \\
\hline
\hline
\end{tabular}
\end{table}

\subsection{Quantitative Discussion on Detection Limits}
\label{subsec:detection_limits}

To evaluate the practical sensing capability of our protocol, we present a quantitative comparison between single-shot and multi-shot measurement limits based on the achieved experimental sensitivity of $\eta_E = 14.15 \pm 0.77\ \mu\text{V/m}/\sqrt{\text{Hz}}$.

Within a single experimental cycle bounded by the line-triggering period ($T_{\mathrm{cyc}} = 20\text{ ms}$), the minimum detectable electric field is fundamentally constrained by the single-shot quantum projection noise. Utilizing the relation $\delta E_{\mathrm{single}} = \eta_E / \sqrt{T_{\mathrm{cyc}}}$, the single-shot detection limit is calculated to be
\begin{equation}
\delta E_{\mathrm{single}} = \frac{14.15\ \mu\text{V/m}/\sqrt{\text{Hz}}}{\sqrt{0.02\text{ s}}} = 100.06 \pm 5.44\ \mu\text{V/m}.
\end{equation}

Crucially, this single-shot statistical constraint is rapidly overcome in actual multi-shot scenarios through continuous data acquisition and real-time averaging. Benefiting from the standard $1/\sqrt{T}$ scaling of statistical averaging where $T$ represents the total integration time, the minimum detectable field is significantly suppressed as follows:
\begin{itemize}
    \item \textbf{At $T = 1\text{ s}$:} The integration over 50 consecutive experimental cycles suppresses the noise floor to exactly match the sensitivity value, yielding a minimum detectable field of $\delta E_{1\text{s}} = 14.15 \pm 0.77\ \mu\text{V/m}$. This high sensitivity accounts for the pronounced response observed even under relatively weak external drives (e.g., $E \approx 20\ \mu\text{V/m}$).
    \item \textbf{At $T = 60\text{ s}$~(our experiment):} Extending the data accumulation to 1 minute (3000 cycles) for population detection in actual implementation further suppresses the minimum detectable field into the sub-microvolt regime, achieving $\delta E_{60\text{s}} = \eta_E / \sqrt{60\text{ s}} \approx 1.83\ \mu\text{V/m}$.
\end{itemize}

Furthermore, to reach an absolute target resolution of $\delta E = 1.0\ \mu\text{V/m}$ for ultra-weak stray field characterization, the required total integration time is projected as
\begin{equation}
T = \left( \frac{\eta_E}{\delta E} \right)^2 = \left( \frac{14.15\ \mu\text{V/m}/\sqrt{\text{Hz}}}{1.0\ \mu\text{V/m}} \right)^2 \approx 200.2\text{ s},
\end{equation}
which demands a modest continuous running time of approximately $3.3\text{ minutes}$. This demonstrates that the gap-suppression-induced phase amplification remains highly efficient and practical for long-term high-precision electrometry despite the dead-time overhead introduced by grid synchronization.

\subsection{Numerical Construction of the Sensitivity Map}
\label{sub:sensitivity_map_calc}

The sensitivity map shown in Fig.~4(d) is obtained from full numerical simulations over the two-dimensional parameter space spanned by the effective couplings $g_h$ and $g_c$. The calculations are performed on a uniform $100\times100$ grid using the same master-equation model and sensing protocol described above.

For each parameter point $(g_h,g_c)$, the system is initialized in a thermal motional state with mean phonon number $\bar n=0.3$ and evolved for a fixed phonon-laser preparation time $t_{\mathrm{pl}}=5.7~\mathrm{ms}$. The motional Hilbert space is truncated at $N_{\mathrm{cutoff}}=25$, which provides converged results throughout the parameter region considered here.

The mean phonon number $\bar n$ is first evaluated numerically. To remain within the experimentally relevant operating regime and ensure reliable convergence within the truncated Hilbert space, the sensitivity is calculated only for parameter points satisfying $\bar n \le 10$. For each valid point, a weak resonant electric-field perturbation of amplitude $\delta E = 1~\mu\mathrm{V/m}$ is introduced, and the corresponding change in the readout signal $P_\uparrow$ is used to evaluate the finite-difference response slope $(P_\uparrow^{\delta}-P_\uparrow^{0})/\delta E$. The electric-field sensitivity $\eta_E$ is then obtained using the same sensitivity definition employed in the main text.

To generate the final phase diagram, only the region between the phonon-lasing threshold boundary and the instability boundary,
$
\sqrt{\gamma_h/\gamma_c}\,g_h
\le g_c \le
\sqrt{\gamma_c/\gamma_h}\,g_h,
$
is displayed.

The color map represents the simulated sensitivity $\eta_E$. White contour lines denote iso-sensitivity curves, while cyan dashed contours indicate constant mean-phonon-number lines obtained from the full numerical simulations.

\subsection{Reconstruction of $S_\phi$}\label{subsec:sphi_reconstruction}To evaluate the experimental phase-space response $S_\phi$ presented in Fig.~5(a), we implement a localized numerical reverse-estimation protocol based on the time-dependent master equation. Unlike steady-state approximations, our framework captures the full transient open-quantum-system dynamics, initialized with an average phonon occupancy of $\langle \hat{n}(0) \rangle \approx 0.3$, consistent with the experimental thermal pre-cooling stage. The dimensionless phase-space response metric $S_\phi(t)$ is evaluated at the fixed interaction duration $t_{\mathrm{pl}} = 5.7\text{ ms}$ as\begin{equation}S_\phi(t) = \frac{|\langle\hat{a}(t)\rangle|}{\sqrt{\langle\hat{n}(t)\rangle}},\label{eq:S_phi_def}\end{equation}where $\hat{a}$ ($\hat{n} = \hat{a}^\dagger\hat{a}$) denotes the motional annihilation (number) operator. The underlying dynamics define a non-linear mapping $\mathcal{M}$ from the external field strength $E$ to the phase-space metric, expressed as $S_\phi(t_{\mathrm{pl}}) = \mathcal{M}(E; t_{\mathrm{pl}})$.The central experimental values and their corresponding statistical error bars are reconstructed via a two-step numerical procedure. First, the experimental expectation value $S_{\phi,\mathrm{exp}}$ is extracted by solving the time-dependent master equation trajectory, starting from the measured initial state, at the best-fit electric field amplitude $E_{\mathrm{fit}}$:\begin{equation}S_{\phi,\mathrm{exp}} = \mathcal{M}(E_{\mathrm{fit}}; t_{\mathrm{pl}}).\end{equation}Second, to address the inherent non-linearity of the mapping $\mathcal{M}$ induced by the concurrent parametric sideband coupling and active dissipation, we compute the localized susceptibility coefficient (Jacobian) $\partial S_\phi / \partial E$ via a central-difference numerical derivative:\begin{equation}\frac{\partial S_\phi}{\partial E} \approx \frac{\mathcal{M}(E_{\mathrm{fit}} + dE; t_{\mathrm{pl}}) - \mathcal{M}(E_{\mathrm{fit}} - dE; t_{\mathrm{pl}})}{2dE},\end{equation}where $dE = 0.1\ \mu\text{V/m}$ is a small finite perturbation. Finally, the symmetric experimental uncertainty $\Delta S_\phi$ is rigorously determined via first-order error propagation:\begin{equation}\Delta S_\phi = \left| \frac{\partial S_\phi}{\partial E} \right| \Delta E.\end{equation}This time-dependent differential framework rigorously maps the experimental field uncertainties into the $S_\phi$ domain while explicitly accounting for the initial thermal population and the non-equilibrium evolution of the phonon laser working points, ensuring a statistically self-consistent comparison between theory and experiment.

\section{Spectral Mechanism of Sensitivity Enhancement}
\label{sec:liouvillian_dynamics}

In this section, we present the microscopic theoretical framework underlying the electric-field sensitivity enhancement observed in the trapped-ion phonon laser. By analyzing the spectral structure of the Liouvillian superoperator, we establish a direct connection between the experimentally measured metrological quantities and the Liouvillian-gap suppression, which is achieved by tuning the system near the phonon-lasing threshold and operating in the weak-coupling regime. This framework provides the physical basis for the enhanced response slope, the extended operational timescale, and the sensitivity scaling behavior discussed in the main text.

\subsection{Liouvillian Perturbation Theory}
\label{subsec:liouvillian_origin}

The open-system dynamics of the phonon laser under a weak external electric field are governed by the Lindblad master equation
\begin{equation}
\dot{\rho}
=
\mathcal{L}\rho
=
\mathcal{L}_0 \rho
-
i[\hat H_E,\rho],
\end{equation}
where $\mathcal{L}_0$ describes the unperturbed phonon-lasing dynamics generated by the simultaneous red- and blue-sideband interactions, and $\hat H_E$ denotes the weak electric-field perturbation.

The spectral decomposition of the unperturbed Liouvillian $\mathcal L_0$ yields a set of complex eigenvalues $\lambda_k$, ordered according to their real parts,
\begin{equation}
0=\mathrm{Re}(\lambda_0)
>
\mathrm{Re}(\lambda_1)
\ge
\mathrm{Re}(\lambda_2)
\ge \cdots .
\end{equation}
The zero eigenvalue $\lambda_0=0$ uniquely determines the steady state $\rho_0$, while the Liouvillian gap
\begin{equation}
\Delta_{\mathcal L}
\equiv
-\mathrm{Re}(\lambda_1)
\end{equation}
characterizes the relaxation rate of the slowest dissipative mode.

By operating the phonon laser near the phonon-lasing threshold with weak effective couplings, the Liouvillian gap becomes strongly suppressed, $\Delta_{\mathcal L}\rightarrow0$. This spectral narrowing produces three key physical consequences relevant to the sensing protocol.

\begin{enumerate}

\item
\textbf{Liouvillian relaxation slowing and long-time signal accumulation.}

Near the dissipative crossover, the characteristic relaxation timescale scales as
\begin{equation}
\tau \sim \frac{1}{\Delta_{\mathcal L}}.
\end{equation}
The narrowing of the Liouvillian gap therefore substantially slows the relaxation dynamics and enhances the susceptibility of the limit cycle to weak perturbations. The experimentally chosen phonon-lasing duration ($t_{\mathrm{pl}}=5.7~\mathrm{ms}$) is intentionally longer than this relaxation timescale, ensuring that the weak perturbation $\hat H_E$ can continuously deform the steady-state phase-space distribution through long-time dissipative accumulation. This mechanism enables a sub-microvolt electric field to induce a symmetry breaking of the phonon-laser limit cycle.

\item
\textbf{Susceptibility enhancement from Liouvillian mode amplification.}

The enhanced electric-field response originates from the amplification of the slowest Liouvillian mode under gap suppression. Using first-order perturbation theory for open quantum systems, the perturbed steady state can be expanded as
\begin{equation}
\rho_{\mathrm{ss}}(E)
\approx
\rho_0
-
i
\sum_{k\neq0}
\frac{
\mathrm{Tr}
\left[
\hat v_k
[\hat H_E,\rho_0]
\right]
}{
\lambda_k
}
\hat u_k,
\end{equation}
where $\hat u_k$ and $\hat v_k$ are the right and left eigenmodes of $\mathcal L_0$, respectively. In this gap-suppressed limit, the first excited mode dominates because
\begin{equation}
|\lambda_1|
\ll
|\lambda_{k>1}|.
\end{equation}
Consequently, the steady-state deformation scales approximately as
\begin{equation}
\delta\rho
\sim
\frac{1}{\Delta_{\mathcal L}}
\hat u_1.
\end{equation}
This inverse-gap enhancement strongly amplifies the phase-space displacement generated by the external field and produces the large response slope observed experimentally.

\item
\textbf{Projection noise and dissipative noise filtering.}

Although the gap-suppressed regime amplifies the signal response, it also enhances low-frequency phase fluctuations associated with the soft Liouvillian mode. For a single experimental realization, the dominant uncertainty therefore originates from quantum projection noise,
\begin{equation}
\Delta P_\uparrow
=
\sqrt{
P_\uparrow
(1-P_\uparrow)
},
\end{equation}
which reaches its maximum value near the optimal operating point $P_\uparrow\approx0.5$. However, higher-order Liouvillian modes remain strongly damped ($\mathrm{Re}(\lambda_{k\ge2})\ll0$), continuously suppressing incoherent thermal fluctuations through dissipative stabilization. As a result, repeated measurements preserve coherent signal accumulation over long averaging times, allowing the sensitivity to improve according to the standard statistical scaling
\begin{equation}
\eta_E \propto \frac{1}{\sqrt{T}},
\end{equation}
without encountering a thermal saturation floor within the experimentally accessible timescale.

\end{enumerate}

\subsection{Liouvillian-Gap Suppression and Sensitivity Optimization}
\label{subsec:scaling_laws}

To optimize sensitivity, the system is operated in a regime where the Liouvillian gap $\Delta_{\mathcal L}$ is significantly suppressed. Experimentally, this gap suppression is achieved by tuning the sideband parameters close to the phonon-lasing threshold and concurrently operating with weak effective coupling strengths. In this limit, the response is entirely dominated by the slowest Liouvillian mode.

The steady state satisfies
\begin{equation}
\mathcal L \rho_0 = 0.
\end{equation}
A weak external electric field introduces a perturbation
\begin{equation}
\mathcal L
\rightarrow
\mathcal L + E\mathcal L_E,
\end{equation}
which modifies the steady state according to
\begin{equation}
\rho_{\mathrm{ss}}(E)
=
\rho_0+\delta\rho .
\end{equation}
Keeping only terms linear in $E$ yields
\begin{equation}
\mathcal L \delta\rho
=
-
E\mathcal L_E\rho_0,
\end{equation}
so that
\begin{equation}
\delta\rho
=
-
\mathcal L^{-1}
\mathcal L_E\rho_0 .
\end{equation}

Using the biorthogonal eigenmode decomposition
\begin{equation}
\mathcal L \hat u_n
=
\lambda_n \hat u_n,
\qquad
\mathcal L^\dagger \hat v_n
=
\lambda_n^* \hat v_n,
\end{equation}
with
\begin{equation}
\mathrm{Tr}
[
\hat v_m^\dagger \hat u_n
]
=
\delta_{mn},
\end{equation}
the Liouvillian pseudo-inverse can be expressed as
\begin{equation}
\mathcal L^{-1}
=
\sum_{n\neq0}
\frac{
\hat u_n \otimes \hat v_n^\dagger
}{
\lambda_n
}.
\end{equation}

Near this threshold, the first excited mode dominates the response because $|\lambda_1| \ll |\lambda_{n>1}|$. The steady-state deformation therefore reduces to
\begin{equation}
\delta\rho
\approx
-
\frac{
\mathrm{Tr}
[
\hat v_1^\dagger
\mathcal L_E\rho_0
]
}{
\lambda_1
}
E\,\hat u_1
\propto
\frac{E}{\Delta_{\mathcal L}}
\hat u_1.
\end{equation}

Observables with finite overlap with the slow Liouvillian mode inherit the same inverse-gap enhancement,
\begin{equation}
\delta\langle \hat O\rangle
=
\mathrm{Tr}
[
\hat O\,\delta\rho
]
\propto
\frac{E}{\Delta_{\mathcal L}}.
\end{equation}

Experimentally, the field-induced phase-space deformation is mapped onto an electronic population signal through the bichromatic characteristic-function probing protocol. The measured excitation probability can be written as
\begin{equation}
P_\uparrow
=
\mathrm{Tr}
[
\hat M
\rho_{\mathrm{ss}}
],
\end{equation}
where $\hat M$ denotes the effective measurement operator associated with the probing sequence. Since $P_\uparrow$ depends linearly on the perturbed steady state, its susceptibility inherits the same inverse-gap enhancement,
\begin{equation}
\left|
\frac{\partial P_\uparrow}{\partial E}
\right|
\propto
\frac{1}{\Delta_{\mathcal L}}.
\end{equation}

Substituting this scaling into the sensitivity expression
\begin{equation}
\eta_E
=
\frac{
\Delta P_\uparrow
\sqrt{T_{\mathrm{cyc}}}
}{
|\partial P_\uparrow/\partial E|
},
\end{equation}
gives
\begin{equation}
\eta_E
\propto
\Delta_{\mathcal L}.
\end{equation}

Finally, the experimentally chosen interaction time $t_{\mathrm{pl}}=5.7~\mathrm{ms}$ exceeds the relaxation timescale $\tau\sim1/\Delta_{\mathcal L}$, ensuring that the system remains in a quasi-steady-state regime during the sensing protocol. Operating in this near-threshold region enables strong susceptibility enhancement while avoiding the severe phase diffusion and exceedingly long relaxation times associated with a vanishing Liouvillian gap.

\subsection{Numerical Extraction of the Liouvillian Gap}\label{subsec:numerical_gap_extraction}To quantitatively correlate the sensing performance with the spectral properties of the system, we numerically compute the Liouvillian gap $\Delta_{\mathcal{L}}$ by diagonalizing the Liouvillian superoperator $\mathcal{L}$ defined by the master equation. The Liouvillian superoperator is constructed as a matrix in the vectorized Hilbert space of the combined ion-phonon system, with dimensions $N_{\mathrm{dim}} = d_{\mathrm{ion}} \times N_{\mathrm{trunc}}^2$, where $d_{\mathrm{ion}} = 3$ and $N_{\mathrm{trunc}} = 25$ is the phonon Fock-state truncation limit.The eigenvalue spectrum $\{\lambda_k\}$ is obtained via full diagonalization of the Liouvillian matrix $\mathcal{L}$. To isolate the physical dissipative gap from numerical artifacts, we apply a filtering protocol:\begin{enumerate}\item We identify the zero eigenvalue $\lambda_0 = 0$, which corresponds to the unique steady state $\rho_{\mathrm{ss}}$ of the open system.\item We exclude all eigenvalues $\lambda_k$ that satisfy $|\mathrm{Re}(\lambda_k)| < \epsilon$, where $\epsilon = 10^{-4}~\mathrm{s}^{-1}$ is a threshold determined by the numerical tolerance of the eigensolver.\item The Liouvillian gap is then extracted as the eigenvalue with the smallest non-zero real part:\begin{equation}\Delta_{\mathcal{L}} = \min_{k} { |\mathrm{Re}(\lambda_k)| : |\mathrm{Re}(\lambda_k)| > \epsilon }.\end{equation}\end{enumerate}This spectral gap represents the inverse of the slowest relaxation timescale $\tau$ of the phonon-laser limit cycle. Our numerical results confirm that $\Delta_{\mathcal{L}}$ exhibits a strong dependence on the sideband coupling strengths $\{g_h, g_c\}$, directly scaling with both the proximity to the phonon-lasing threshold and the reduction of the overall effective coupling strengths. The resulting values, ranging from $\sim 317$ to $1688~\mathrm{s}^{-1}$, demonstrate that the system operates in a regime where the interaction time $t_{\mathrm{pl}} = 5.7~\mathrm{ms}$ ensures significant signal accumulation without encountering the exceedingly long relaxation times associated with a vanishing gap.


\begin{thebibliography}{34}%
\makeatletter
\providecommand \@ifxundefined [1]{%
 \@ifx{#1\undefined}
}%
\providecommand \@ifnum [1]{%
 \ifnum #1\expandafter \@firstoftwo
 \else \expandafter \@secondoftwo
 \fi
}%
\providecommand \@ifx [1]{%
 \ifx #1\expandafter \@firstoftwo
 \else \expandafter \@secondoftwo
 \fi
}%
\providecommand \natexlab [1]{#1}%
\providecommand \enquote  [1]{``#1''}%
\providecommand \bibnamefont  [1]{#1}%
\providecommand \bibfnamefont [1]{#1}%
\providecommand \citenamefont [1]{#1}%
\providecommand \href@noop [0]{\@secondoftwo}%
\providecommand \href [0]{\begingroup \@sanitize@url \@href}%
\providecommand \@href[1]{\@@startlink{#1}\@@href}%
\providecommand \@@href[1]{\endgroup#1\@@endlink}%
\providecommand \@sanitize@url [0]{\catcode `\\12\catcode `\$12\catcode `\&12\catcode `\#12\catcode `\^12\catcode `\_12\catcode `\%12\relax}%
\providecommand \@@startlink[1]{}%
\providecommand \@@endlink[0]{}%
\providecommand \url  [0]{\begingroup\@sanitize@url \@url }%
\providecommand \@url [1]{\endgroup\@href {#1}{\urlprefix }}%
\providecommand \urlprefix  [0]{URL }%
\providecommand \Eprint [0]{\href }%
\providecommand \doibase [0]{https://doi.org/}%
\providecommand \selectlanguage [0]{\@gobble}%
\providecommand \bibinfo  [0]{\@secondoftwo}%
\providecommand \bibfield  [0]{\@secondoftwo}%
\providecommand \translation [1]{[#1]}%
\providecommand \BibitemOpen [0]{}%
\providecommand \bibitemStop [0]{}%
\providecommand \bibitemNoStop [0]{.\EOS\space}%
\providecommand \EOS [0]{\spacefactor3000\relax}%
\providecommand \BibitemShut  [1]{\csname bibitem#1\endcsname}%
\let\auto@bib@innerbib\@empty
\bibitem [{\citenamefont {Wallentowitz}\ \emph {et~al.}(1996)\citenamefont {Wallentowitz}, \citenamefont {Vogel}, \citenamefont {Siemers},\ and\ \citenamefont {Toschek}}]{Wallentowitz1996}%
  \BibitemOpen
  \bibfield  {author} {\bibinfo {author} {\bibfnamefont {S.}~\bibnamefont {Wallentowitz}}, \bibinfo {author} {\bibfnamefont {W.}~\bibnamefont {Vogel}}, \bibinfo {author} {\bibfnamefont {I.}~\bibnamefont {Siemers}},\ and\ \bibinfo {author} {\bibfnamefont {P.~E.}\ \bibnamefont {Toschek}},\ }\bibfield  {title} {\bibinfo {title} {Vibrational amplification by stimulated emission of radiation},\ }\href {https://doi.org/10.1103/PhysRevA.54.943} {\bibfield  {journal} {\bibinfo  {journal} {Phys. Rev. A}\ }\textbf {\bibinfo {volume} {54}},\ \bibinfo {pages} {943} (\bibinfo {year} {1996})}\BibitemShut {NoStop}%
\bibitem [{\citenamefont {Jing}\ \emph {et~al.}(2014)\citenamefont {Jing}, \citenamefont {\"Ozdemir}, \citenamefont {L\"u}, \citenamefont {Zhang}, \citenamefont {Yang},\ and\ \citenamefont {Nori}}]{Jing2014PT}%
  \BibitemOpen
  \bibfield  {author} {\bibinfo {author} {\bibfnamefont {H.}~\bibnamefont {Jing}}, \bibinfo {author} {\bibfnamefont {S.~K.}\ \bibnamefont {\"Ozdemir}}, \bibinfo {author} {\bibfnamefont {X.-Y.}\ \bibnamefont {L\"u}}, \bibinfo {author} {\bibfnamefont {J.}~\bibnamefont {Zhang}}, \bibinfo {author} {\bibfnamefont {L.}~\bibnamefont {Yang}},\ and\ \bibinfo {author} {\bibfnamefont {F.}~\bibnamefont {Nori}},\ }\bibfield  {title} {\bibinfo {title} {$\mathcal{PT}$-symmetric phonon laser},\ }\href {https://doi.org/10.1103/PhysRevLett.113.053604} {\bibfield  {journal} {\bibinfo  {journal} {Phys. Rev. Lett.}\ }\textbf {\bibinfo {volume} {113}},\ \bibinfo {pages} {053604} (\bibinfo {year} {2014})}\BibitemShut {NoStop}%
\bibitem [{\citenamefont {Vahala}\ \emph {et~al.}(2009)\citenamefont {Vahala}, \citenamefont {Xu}, \citenamefont {Yang}, \citenamefont {Painter}, \citenamefont {Carmon},\ and\ \citenamefont {Kippenberg}}]{vahala2009phononlaser}%
  \BibitemOpen
  \bibfield  {author} {\bibinfo {author} {\bibfnamefont {K.~J.}\ \bibnamefont {Vahala}}, \bibinfo {author} {\bibfnamefont {Q.}~\bibnamefont {Xu}}, \bibinfo {author} {\bibfnamefont {W.}~\bibnamefont {Yang}}, \bibinfo {author} {\bibfnamefont {O.}~\bibnamefont {Painter}}, \bibinfo {author} {\bibfnamefont {T.}~\bibnamefont {Carmon}},\ and\ \bibinfo {author} {\bibfnamefont {T.~J.}\ \bibnamefont {Kippenberg}},\ }\bibfield  {title} {\bibinfo {title} {A phonon laser},\ }\href {https://doi.org/10.1038/nphys1344} {\bibfield  {journal} {\bibinfo  {journal} {Nature Physics}\ }\textbf {\bibinfo {volume} {5}},\ \bibinfo {pages} {682} (\bibinfo {year} {2009})}\BibitemShut {NoStop}%
\bibitem [{\citenamefont {Grudinin}\ \emph {et~al.}(2010)\citenamefont {Grudinin}, \citenamefont {Lee}, \citenamefont {Painter},\ and\ \citenamefont {Vahala}}]{2010cavityPL}%
  \BibitemOpen
  \bibfield  {author} {\bibinfo {author} {\bibfnamefont {I.~S.}\ \bibnamefont {Grudinin}}, \bibinfo {author} {\bibfnamefont {H.}~\bibnamefont {Lee}}, \bibinfo {author} {\bibfnamefont {O.}~\bibnamefont {Painter}},\ and\ \bibinfo {author} {\bibfnamefont {K.~J.}\ \bibnamefont {Vahala}},\ }\bibfield  {title} {\bibinfo {title} {Phonon laser action in a tunable two-level system},\ }\href {https://doi.org/10.1103/PhysRevLett.104.083901} {\bibfield  {journal} {\bibinfo  {journal} {Phys. Rev. Lett.}\ }\textbf {\bibinfo {volume} {104}},\ \bibinfo {pages} {083901} (\bibinfo {year} {2010})}\BibitemShut {NoStop}%
\bibitem [{\citenamefont {Pettit}\ \emph {et~al.}(2019)\citenamefont {Pettit}, \citenamefont {Ge}, \citenamefont {Kumar}, \citenamefont {Luntz-Martin}, \citenamefont {Schultz}, \citenamefont {Neukirch}, \citenamefont {Bhattacharya},\ and\ \citenamefont {Vamivakas}}]{Pettit2019tweezerPL}%
  \BibitemOpen
  \bibfield  {author} {\bibinfo {author} {\bibfnamefont {R.~M.}\ \bibnamefont {Pettit}}, \bibinfo {author} {\bibfnamefont {W.}~\bibnamefont {Ge}}, \bibinfo {author} {\bibfnamefont {P.}~\bibnamefont {Kumar}}, \bibinfo {author} {\bibfnamefont {D.~R.}\ \bibnamefont {Luntz-Martin}}, \bibinfo {author} {\bibfnamefont {J.~T.}\ \bibnamefont {Schultz}}, \bibinfo {author} {\bibfnamefont {L.~P.}\ \bibnamefont {Neukirch}}, \bibinfo {author} {\bibfnamefont {M.}~\bibnamefont {Bhattacharya}},\ and\ \bibinfo {author} {\bibfnamefont {A.~N.}\ \bibnamefont {Vamivakas}},\ }\bibfield  {title} {\bibinfo {title} {An optical tweezer phonon laser},\ }\href {https://doi.org/10.1038/s41566-019-0395-5} {\bibfield  {journal} {\bibinfo  {journal} {Nature Photonics}\ }\textbf {\bibinfo {volume} {13}},\ \bibinfo {pages} {402} (\bibinfo {year} {2019})}\BibitemShut {NoStop}%
\bibitem [{\citenamefont {Behrle}\ \emph {et~al.}(2023)\citenamefont {Behrle}, \citenamefont {Nguyen}, \citenamefont {Reiter}, \citenamefont {Baur}, \citenamefont {de~Neeve}, \citenamefont {Stadler}, \citenamefont {Marinelli}, \citenamefont {Lancellotti}, \citenamefont {Yelin},\ and\ \citenamefont {Home}}]{Behrle2023quantumPL}%
  \BibitemOpen
  \bibfield  {author} {\bibinfo {author} {\bibfnamefont {T.}~\bibnamefont {Behrle}}, \bibinfo {author} {\bibfnamefont {T.~L.}\ \bibnamefont {Nguyen}}, \bibinfo {author} {\bibfnamefont {F.}~\bibnamefont {Reiter}}, \bibinfo {author} {\bibfnamefont {D.}~\bibnamefont {Baur}}, \bibinfo {author} {\bibfnamefont {B.}~\bibnamefont {de~Neeve}}, \bibinfo {author} {\bibfnamefont {M.}~\bibnamefont {Stadler}}, \bibinfo {author} {\bibfnamefont {M.}~\bibnamefont {Marinelli}}, \bibinfo {author} {\bibfnamefont {F.}~\bibnamefont {Lancellotti}}, \bibinfo {author} {\bibfnamefont {S.~F.}\ \bibnamefont {Yelin}},\ and\ \bibinfo {author} {\bibfnamefont {J.~P.}\ \bibnamefont {Home}},\ }\bibfield  {title} {\bibinfo {title} {Phonon laser in the quantum regime},\ }\href {https://doi.org/10.1103/PhysRevLett.131.043605} {\bibfield  {journal} {\bibinfo  {journal} {Phys. Rev. Lett.}\ }\textbf {\bibinfo {volume} {131}},\ \bibinfo {pages} {043605} (\bibinfo {year} {2023})}\BibitemShut {NoStop}%
\bibitem [{\citenamefont {Dong}\ \emph {et~al.}(2025)\citenamefont {Dong}, \citenamefont {He}, \citenamefont {Deng}, \citenamefont {Li}, \citenamefont {Chen},\ and\ \citenamefont {Feng}}]{DONG2025singleionPL}%
  \BibitemOpen
  \bibfield  {author} {\bibinfo {author} {\bibfnamefont {Y.-Z.}\ \bibnamefont {Dong}}, \bibinfo {author} {\bibfnamefont {S.-W.}\ \bibnamefont {He}}, \bibinfo {author} {\bibfnamefont {Z.-J.}\ \bibnamefont {Deng}}, \bibinfo {author} {\bibfnamefont {P.-D.}\ \bibnamefont {Li}}, \bibinfo {author} {\bibfnamefont {L.}~\bibnamefont {Chen}},\ and\ \bibinfo {author} {\bibfnamefont {M.}~\bibnamefont {Feng}},\ }\bibfield  {title} {\bibinfo {title} {Single-ion phonon laser in quantum region},\ }\href {https://doi.org/10.7498/aps.74.20250603} {\bibfield  {journal} {\bibinfo  {journal} {Acta Physica Sinica}\ }\textbf {\bibinfo {volume} {74}},\ \bibinfo {pages} {193701} (\bibinfo {year} {2025})}\BibitemShut {NoStop}%
\bibitem [{\citenamefont {Baur}\ \emph {et~al.}(2026)\citenamefont {Baur}, \citenamefont {Behrle}, \citenamefont {Rojkov}, \citenamefont {Jeske}, \citenamefont {Yelin}, \citenamefont {Home},\ and\ \citenamefont {Reiter}}]{baur2026quantumtheoryphononlasing}%
  \BibitemOpen
  \bibfield  {author} {\bibinfo {author} {\bibfnamefont {D.}~\bibnamefont {Baur}}, \bibinfo {author} {\bibfnamefont {T.}~\bibnamefont {Behrle}}, \bibinfo {author} {\bibfnamefont {I.}~\bibnamefont {Rojkov}}, \bibinfo {author} {\bibfnamefont {J.}~\bibnamefont {Jeske}}, \bibinfo {author} {\bibfnamefont {S.}~\bibnamefont {Yelin}}, \bibinfo {author} {\bibfnamefont {J.}~\bibnamefont {Home}},\ and\ \bibinfo {author} {\bibfnamefont {F.}~\bibnamefont {Reiter}},\ }\href {https://arxiv.org/abs/2604.18295} {\bibinfo {title} {Quantum theory for phonon lasing and non-classical state generation in mixed-species and single trapped ions}} (\bibinfo {year} {2026}),\ \Eprint {https://arxiv.org/abs/2604.18295} {arXiv:2604.18295 [quant-ph]} \BibitemShut {NoStop}%
\bibitem [{\citenamefont {Degen}\ \emph {et~al.}(2017)\citenamefont {Degen}, \citenamefont {Reinhard},\ and\ \citenamefont {Cappellaro}}]{degen2017quantumsensing}%
  \BibitemOpen
  \bibfield  {author} {\bibinfo {author} {\bibfnamefont {C.~L.}\ \bibnamefont {Degen}}, \bibinfo {author} {\bibfnamefont {F.}~\bibnamefont {Reinhard}},\ and\ \bibinfo {author} {\bibfnamefont {P.}~\bibnamefont {Cappellaro}},\ }\bibfield  {title} {\bibinfo {title} {Quantum sensing},\ }\href {https://doi.org/10.1103/RevModPhys.89.035002} {\bibfield  {journal} {\bibinfo  {journal} {Rev. Mod. Phys.}\ }\textbf {\bibinfo {volume} {89}},\ \bibinfo {pages} {035002} (\bibinfo {year} {2017})}\BibitemShut {NoStop}%
\bibitem [{\citenamefont {Burd}\ \emph {et~al.}(2019)\citenamefont {Burd}, \citenamefont {Srinivas}, \citenamefont {Bollinger}, \citenamefont {Wilson}, \citenamefont {Wineland}, \citenamefont {Leibfried}, \citenamefont {Slichter},\ and\ \citenamefont {Allcock}}]{Burd2019sensing}%
  \BibitemOpen
  \bibfield  {author} {\bibinfo {author} {\bibfnamefont {S.~C.}\ \bibnamefont {Burd}}, \bibinfo {author} {\bibfnamefont {R.}~\bibnamefont {Srinivas}}, \bibinfo {author} {\bibfnamefont {J.~J.}\ \bibnamefont {Bollinger}}, \bibinfo {author} {\bibfnamefont {A.~C.}\ \bibnamefont {Wilson}}, \bibinfo {author} {\bibfnamefont {D.~J.}\ \bibnamefont {Wineland}}, \bibinfo {author} {\bibfnamefont {D.}~\bibnamefont {Leibfried}}, \bibinfo {author} {\bibfnamefont {D.~H.}\ \bibnamefont {Slichter}},\ and\ \bibinfo {author} {\bibfnamefont {D.~T.~C.}\ \bibnamefont {Allcock}},\ }\bibfield  {title} {\bibinfo {title} {Quantum amplification of mechanical oscillator motion},\ }\href {https://doi.org/10.1126/science.aaw2884} {\bibfield  {journal} {\bibinfo  {journal} {Science}\ }\textbf {\bibinfo {volume} {364}},\ \bibinfo {pages} {1163} (\bibinfo {year} {2019})}\BibitemShut {NoStop}%
\bibitem [{\citenamefont {Gilmore}\ \emph {et~al.}(2021)\citenamefont {Gilmore}, \citenamefont {Affolter}, \citenamefont {Lewis-Swan}, \citenamefont {Barberena}, \citenamefont {Jordan}, \citenamefont {Rey},\ and\ \citenamefont {Bollinger}}]{Gilmore2021sensing}%
  \BibitemOpen
  \bibfield  {author} {\bibinfo {author} {\bibfnamefont {K.~A.}\ \bibnamefont {Gilmore}}, \bibinfo {author} {\bibfnamefont {M.}~\bibnamefont {Affolter}}, \bibinfo {author} {\bibfnamefont {R.~J.}\ \bibnamefont {Lewis-Swan}}, \bibinfo {author} {\bibfnamefont {D.}~\bibnamefont {Barberena}}, \bibinfo {author} {\bibfnamefont {E.}~\bibnamefont {Jordan}}, \bibinfo {author} {\bibfnamefont {A.~M.}\ \bibnamefont {Rey}},\ and\ \bibinfo {author} {\bibfnamefont {J.~J.}\ \bibnamefont {Bollinger}},\ }\bibfield  {title} {\bibinfo {title} {Quantum-enhanced sensing of displacements and electric fields with two-dimensional trapped-ion crystals},\ }\href {https://doi.org/10.1126/science.abi5226} {\bibfield  {journal} {\bibinfo  {journal} {Science}\ }\textbf {\bibinfo {volume} {373}},\ \bibinfo {pages} {673} (\bibinfo {year} {2021})}\BibitemShut {NoStop}%
\bibitem [{\citenamefont {Ivanov}\ \emph {et~al.}(2015)\citenamefont {Ivanov}, \citenamefont {Singer}, \citenamefont {Vitanov},\ and\ \citenamefont {Porras}}]{Ivanov2015SBsensing}%
  \BibitemOpen
  \bibfield  {author} {\bibinfo {author} {\bibfnamefont {P.~A.}\ \bibnamefont {Ivanov}}, \bibinfo {author} {\bibfnamefont {K.}~\bibnamefont {Singer}}, \bibinfo {author} {\bibfnamefont {N.~V.}\ \bibnamefont {Vitanov}},\ and\ \bibinfo {author} {\bibfnamefont {D.}~\bibnamefont {Porras}},\ }\bibfield  {title} {\bibinfo {title} {Quantum sensors assisted by spontaneous symmetry breaking for detecting very small forces},\ }\href {https://doi.org/10.1103/PhysRevApplied.4.054007} {\bibfield  {journal} {\bibinfo  {journal} {Phys. Rev. Appl.}\ }\textbf {\bibinfo {volume} {4}},\ \bibinfo {pages} {054007} (\bibinfo {year} {2015})}\BibitemShut {NoStop}%
\bibitem [{\citenamefont {Fern\'andez-Lorenzo}\ and\ \citenamefont {Porras}(2017)}]{Lorenzo2017DPTsensing}%
  \BibitemOpen
  \bibfield  {author} {\bibinfo {author} {\bibfnamefont {S.}~\bibnamefont {Fern\'andez-Lorenzo}}\ and\ \bibinfo {author} {\bibfnamefont {D.}~\bibnamefont {Porras}},\ }\bibfield  {title} {\bibinfo {title} {Quantum sensing close to a dissipative phase transition: Symmetry breaking and criticality as metrological resources},\ }\href {https://doi.org/10.1103/PhysRevA.96.013817} {\bibfield  {journal} {\bibinfo  {journal} {Phys. Rev. A}\ }\textbf {\bibinfo {volume} {96}},\ \bibinfo {pages} {013817} (\bibinfo {year} {2017})}\BibitemShut {NoStop}%
\bibitem [{\citenamefont {Lee}\ and\ \citenamefont {Sadeghpour}(2013)}]{Lee2013Sync}%
  \BibitemOpen
  \bibfield  {author} {\bibinfo {author} {\bibfnamefont {T.~E.}\ \bibnamefont {Lee}}\ and\ \bibinfo {author} {\bibfnamefont {H.~R.}\ \bibnamefont {Sadeghpour}},\ }\bibfield  {title} {\bibinfo {title} {Quantum synchronization of quantum van der pol oscillators with trapped ions},\ }\href {https://doi.org/10.1103/PhysRevLett.111.234101} {\bibfield  {journal} {\bibinfo  {journal} {Phys. Rev. Lett.}\ }\textbf {\bibinfo {volume} {111}},\ \bibinfo {pages} {234101} (\bibinfo {year} {2013})}\BibitemShut {NoStop}%
\bibitem [{\citenamefont {Walter}\ \emph {et~al.}(2014)\citenamefont {Walter}, \citenamefont {Nunnenkamp},\ and\ \citenamefont {Bruder}}]{Walter2014Sync}%
  \BibitemOpen
  \bibfield  {author} {\bibinfo {author} {\bibfnamefont {S.}~\bibnamefont {Walter}}, \bibinfo {author} {\bibfnamefont {A.}~\bibnamefont {Nunnenkamp}},\ and\ \bibinfo {author} {\bibfnamefont {C.}~\bibnamefont {Bruder}},\ }\bibfield  {title} {\bibinfo {title} {Quantum synchronization of a driven self-sustained oscillator},\ }\href {https://doi.org/10.1103/PhysRevLett.112.094102} {\bibfield  {journal} {\bibinfo  {journal} {Phys. Rev. Lett.}\ }\textbf {\bibinfo {volume} {112}},\ \bibinfo {pages} {094102} (\bibinfo {year} {2014})}\BibitemShut {NoStop}%
\bibitem [{\citenamefont {L\"orch}\ \emph {et~al.}(2016)\citenamefont {L\"orch}, \citenamefont {Amitai}, \citenamefont {Nunnenkamp},\ and\ \citenamefont {Bruder}}]{Lorch2016Sync}%
  \BibitemOpen
  \bibfield  {author} {\bibinfo {author} {\bibfnamefont {N.}~\bibnamefont {L\"orch}}, \bibinfo {author} {\bibfnamefont {E.}~\bibnamefont {Amitai}}, \bibinfo {author} {\bibfnamefont {A.}~\bibnamefont {Nunnenkamp}},\ and\ \bibinfo {author} {\bibfnamefont {C.}~\bibnamefont {Bruder}},\ }\bibfield  {title} {\bibinfo {title} {Genuine quantum signatures in synchronization of anharmonic self-oscillators},\ }\href {https://doi.org/10.1103/PhysRevLett.117.073601} {\bibfield  {journal} {\bibinfo  {journal} {Phys. Rev. Lett.}\ }\textbf {\bibinfo {volume} {117}},\ \bibinfo {pages} {073601} (\bibinfo {year} {2016})}\BibitemShut {NoStop}%
\bibitem [{\citenamefont {Li}\ \emph {et~al.}(2025{\natexlab{a}})\citenamefont {Li}, \citenamefont {Xie}, \citenamefont {Yang}, \citenamefont {Li}, \citenamefont {Zhao}, \citenamefont {Cheng}, \citenamefont {Peng}, \citenamefont {Li}, \citenamefont {Lutz}, \citenamefont {Lin},\ and\ \citenamefont {Du}}]{Li2025Sync}%
  \BibitemOpen
  \bibfield  {author} {\bibinfo {author} {\bibfnamefont {Y.}~\bibnamefont {Li}}, \bibinfo {author} {\bibfnamefont {Z.}~\bibnamefont {Xie}}, \bibinfo {author} {\bibfnamefont {X.}~\bibnamefont {Yang}}, \bibinfo {author} {\bibfnamefont {Y.}~\bibnamefont {Li}}, \bibinfo {author} {\bibfnamefont {X.}~\bibnamefont {Zhao}}, \bibinfo {author} {\bibfnamefont {X.}~\bibnamefont {Cheng}}, \bibinfo {author} {\bibfnamefont {X.}~\bibnamefont {Peng}}, \bibinfo {author} {\bibfnamefont {J.}~\bibnamefont {Li}}, \bibinfo {author} {\bibfnamefont {E.}~\bibnamefont {Lutz}}, \bibinfo {author} {\bibfnamefont {Y.}~\bibnamefont {Lin}},\ and\ \bibinfo {author} {\bibfnamefont {J.}~\bibnamefont {Du}},\ }\bibfield  {title} {\bibinfo {title} {Experimental realization and synchronization of a quantum van der pol oscillator},\ }\href {https://doi.org/10.1126/sciadv.ady5649} {\bibfield  {journal} {\bibinfo  {journal} {Science Advances}\ }\textbf {\bibinfo {volume} {11}},\ \bibinfo {pages} {eady5649} (\bibinfo {year}
  {2025}{\natexlab{a}})}\BibitemShut {NoStop}%
\bibitem [{\citenamefont {Liu}\ \emph {et~al.}(2025)\citenamefont {Liu}, \citenamefont {Wu}, \citenamefont {Moore}, \citenamefont {Haeffner},\ and\ \citenamefont {Wächtler}}]{liu2025Sync}%
  \BibitemOpen
  \bibfield  {author} {\bibinfo {author} {\bibfnamefont {J.}~\bibnamefont {Liu}}, \bibinfo {author} {\bibfnamefont {Q.}~\bibnamefont {Wu}}, \bibinfo {author} {\bibfnamefont {J.~E.}\ \bibnamefont {Moore}}, \bibinfo {author} {\bibfnamefont {H.}~\bibnamefont {Haeffner}},\ and\ \bibinfo {author} {\bibfnamefont {C.~W.}\ \bibnamefont {Wächtler}},\ }\href {https://arxiv.org/abs/2509.18423} {\bibinfo {title} {Observation of synchronization between two quantum van der pol oscillators in trapped ions}} (\bibinfo {year} {2025}),\ \Eprint {https://arxiv.org/abs/2509.18423} {arXiv:2509.18423 [quant-ph]} \BibitemShut {NoStop}%
\bibitem [{\citenamefont {Vicentini}\ \emph {et~al.}(2018)\citenamefont {Vicentini}, \citenamefont {Minganti}, \citenamefont {Rota}, \citenamefont {Orso},\ and\ \citenamefont {Ciuti}}]{Vicentini2018CriticalSlow}%
  \BibitemOpen
  \bibfield  {author} {\bibinfo {author} {\bibfnamefont {F.}~\bibnamefont {Vicentini}}, \bibinfo {author} {\bibfnamefont {F.}~\bibnamefont {Minganti}}, \bibinfo {author} {\bibfnamefont {R.}~\bibnamefont {Rota}}, \bibinfo {author} {\bibfnamefont {G.}~\bibnamefont {Orso}},\ and\ \bibinfo {author} {\bibfnamefont {C.}~\bibnamefont {Ciuti}},\ }\bibfield  {title} {\bibinfo {title} {Critical slowing down in driven-dissipative bose-hubbard lattices},\ }\href {https://doi.org/10.1103/PhysRevA.97.013853} {\bibfield  {journal} {\bibinfo  {journal} {Phys. Rev. A}\ }\textbf {\bibinfo {volume} {97}},\ \bibinfo {pages} {013853} (\bibinfo {year} {2018})}\BibitemShut {NoStop}%
\bibitem [{\citenamefont {Minganti}\ \emph {et~al.}(2018)\citenamefont {Minganti}, \citenamefont {Biella}, \citenamefont {Bartolo},\ and\ \citenamefont {Ciuti}}]{minganti2018spectral}%
  \BibitemOpen
  \bibfield  {author} {\bibinfo {author} {\bibfnamefont {F.}~\bibnamefont {Minganti}}, \bibinfo {author} {\bibfnamefont {A.}~\bibnamefont {Biella}}, \bibinfo {author} {\bibfnamefont {N.}~\bibnamefont {Bartolo}},\ and\ \bibinfo {author} {\bibfnamefont {C.}~\bibnamefont {Ciuti}},\ }\bibfield  {title} {\bibinfo {title} {Spectral theory of liouvillians for dissipative phase transitions},\ }\href {https://doi.org/10.1103/PhysRevA.98.042118} {\bibfield  {journal} {\bibinfo  {journal} {Phys. Rev. A}\ }\textbf {\bibinfo {volume} {98}},\ \bibinfo {pages} {042118} (\bibinfo {year} {2018})}\BibitemShut {NoStop}%
\bibitem [{\citenamefont {Garbe}\ \emph {et~al.}(2020)\citenamefont {Garbe}, \citenamefont {Bina}, \citenamefont {Keller}, \citenamefont {Paris},\ and\ \citenamefont {Felicetti}}]{garbe2020critical}%
  \BibitemOpen
  \bibfield  {author} {\bibinfo {author} {\bibfnamefont {L.}~\bibnamefont {Garbe}}, \bibinfo {author} {\bibfnamefont {M.}~\bibnamefont {Bina}}, \bibinfo {author} {\bibfnamefont {A.}~\bibnamefont {Keller}}, \bibinfo {author} {\bibfnamefont {M.~G.~A.}\ \bibnamefont {Paris}},\ and\ \bibinfo {author} {\bibfnamefont {S.}~\bibnamefont {Felicetti}},\ }\bibfield  {title} {\bibinfo {title} {Critical quantum metrology with a finite-component quantum phase transition},\ }\href {https://doi.org/10.1103/PhysRevLett.124.120504} {\bibfield  {journal} {\bibinfo  {journal} {Phys. Rev. Lett.}\ }\textbf {\bibinfo {volume} {124}},\ \bibinfo {pages} {120504} (\bibinfo {year} {2020})}\BibitemShut {NoStop}%
\bibitem [{\citenamefont {Fl\"uhmann}\ and\ \citenamefont {Home}(2020)}]{fluhmann2020direct}%
  \BibitemOpen
  \bibfield  {author} {\bibinfo {author} {\bibfnamefont {C.}~\bibnamefont {Fl\"uhmann}}\ and\ \bibinfo {author} {\bibfnamefont {J.~P.}\ \bibnamefont {Home}},\ }\bibfield  {title} {\bibinfo {title} {Direct characteristic-function tomography of quantum states of the trapped-ion motional oscillator},\ }\href {https://doi.org/10.1103/PhysRevLett.125.043602} {\bibfield  {journal} {\bibinfo  {journal} {Phys. Rev. Lett.}\ }\textbf {\bibinfo {volume} {125}},\ \bibinfo {pages} {043602} (\bibinfo {year} {2020})}\BibitemShut {NoStop}%
\bibitem [{\citenamefont {Li}\ \emph {et~al.}(2025{\natexlab{b}})\citenamefont {Li}, \citenamefont {Ding}, \citenamefont {Zhang}, \citenamefont {Yuan}, \citenamefont {Dai}, \citenamefont {Cui}, \citenamefont {Zhou}, \citenamefont {Chen}, \citenamefont {Zhong}, \citenamefont {Jing}, \citenamefont {{\"O}zdemir},\ and\ \citenamefont {Feng}}]{Li2025SET}%
  \BibitemOpen
  \bibfield  {author} {\bibinfo {author} {\bibfnamefont {P.-D.}\ \bibnamefont {Li}}, \bibinfo {author} {\bibfnamefont {G.-Y.}\ \bibnamefont {Ding}}, \bibinfo {author} {\bibfnamefont {J.-Q.}\ \bibnamefont {Zhang}}, \bibinfo {author} {\bibfnamefont {Q.}~\bibnamefont {Yuan}}, \bibinfo {author} {\bibfnamefont {S.-Q.}\ \bibnamefont {Dai}}, \bibinfo {author} {\bibfnamefont {T.-H.}\ \bibnamefont {Cui}}, \bibinfo {author} {\bibfnamefont {F.}~\bibnamefont {Zhou}}, \bibinfo {author} {\bibfnamefont {L.}~\bibnamefont {Chen}}, \bibinfo {author} {\bibfnamefont {Q.}~\bibnamefont {Zhong}}, \bibinfo {author} {\bibfnamefont {H.}~\bibnamefont {Jing}}, \bibinfo {author} {\bibfnamefont {{\c{S}. K}.}~\bibnamefont {{\"O}zdemir}},\ and\ \bibinfo {author} {\bibfnamefont {M.}~\bibnamefont {Feng}},\ }\bibfield  {title} {\bibinfo {title} {Experimental demonstration of single-spin stirling engine cycles with enhanced efficiency},\ }\href {https://doi.org/10.1103/PhysRevA.111.L010203} {\bibfield  {journal} {\bibinfo  {journal} {Phys. Rev.
  A}\ }\textbf {\bibinfo {volume} {111}},\ \bibinfo {pages} {L010203} (\bibinfo {year} {2025}{\natexlab{b}})}\BibitemShut {NoStop}%
\bibitem [{\citenamefont {Wu}\ \emph {et~al.}(2026)\citenamefont {Wu}, \citenamefont {Li}, \citenamefont {Cui}, \citenamefont {Wang}, \citenamefont {Dong}, \citenamefont {Dai}, \citenamefont {Li}, \citenamefont {Wei}, \citenamefont {Yuan}, \citenamefont {Cai}, \citenamefont {Chen}, \citenamefont {Zhang}, \citenamefont {Jing},\ and\ \citenamefont {Feng}}]{Wu2026LEP}%
  \BibitemOpen
  \bibfield  {author} {\bibinfo {author} {\bibfnamefont {Z.-Z.}\ \bibnamefont {Wu}}, \bibinfo {author} {\bibfnamefont {P.-D.}\ \bibnamefont {Li}}, \bibinfo {author} {\bibfnamefont {T.-H.}\ \bibnamefont {Cui}}, \bibinfo {author} {\bibfnamefont {J.-W.}\ \bibnamefont {Wang}}, \bibinfo {author} {\bibfnamefont {Y.-Z.}\ \bibnamefont {Dong}}, \bibinfo {author} {\bibfnamefont {S.-Q.}\ \bibnamefont {Dai}}, \bibinfo {author} {\bibfnamefont {J.}~\bibnamefont {Li}}, \bibinfo {author} {\bibfnamefont {Y.-Q.}\ \bibnamefont {Wei}}, \bibinfo {author} {\bibfnamefont {Q.}~\bibnamefont {Yuan}}, \bibinfo {author} {\bibfnamefont {X.-M.}\ \bibnamefont {Cai}}, \bibinfo {author} {\bibfnamefont {L.}~\bibnamefont {Chen}}, \bibinfo {author} {\bibfnamefont {J.-Q.}\ \bibnamefont {Zhang}}, \bibinfo {author} {\bibfnamefont {H.}~\bibnamefont {Jing}},\ and\ \bibinfo {author} {\bibfnamefont {M.}~\bibnamefont {Feng}},\ }\bibfield  {title} {\bibinfo {title} {Experimental witness of quantum jump induced high-order liouvillian exceptional
  points},\ }\href {https://doi.org/10.1038/s41467-026-68705-9} {\bibfield  {journal} {\bibinfo  {journal} {Nature Communications}\ }\textbf {\bibinfo {volume} {17}},\ \bibinfo {pages} {1923} (\bibinfo {year} {2026})}\BibitemShut {NoStop}%
\bibitem [{SM()}]{SM}%
  \BibitemOpen
  \href@noop {} {}\bibinfo {note} {See Supplemental Materials for details on the experimental setup and calibrations, theoretical derivations of the semiclassical limits, and numerical simulations of the open-system dynamics.}\BibitemShut {Stop}%
\bibitem [{\citenamefont {Leibfried}\ \emph {et~al.}(2003)\citenamefont {Leibfried}, \citenamefont {Blatt}, \citenamefont {Monroe},\ and\ \citenamefont {Wineland}}]{Leibfried2003RMP}%
  \BibitemOpen
  \bibfield  {author} {\bibinfo {author} {\bibfnamefont {D.}~\bibnamefont {Leibfried}}, \bibinfo {author} {\bibfnamefont {R.}~\bibnamefont {Blatt}}, \bibinfo {author} {\bibfnamefont {C.}~\bibnamefont {Monroe}},\ and\ \bibinfo {author} {\bibfnamefont {D.}~\bibnamefont {Wineland}},\ }\bibfield  {title} {\bibinfo {title} {Quantum dynamics of single trapped ions},\ }\href {https://doi.org/10.1103/RevModPhys.75.281} {\bibfield  {journal} {\bibinfo  {journal} {Rev. Mod. Phys.}\ }\textbf {\bibinfo {volume} {75}},\ \bibinfo {pages} {281} (\bibinfo {year} {2003})}\BibitemShut {NoStop}%
\bibitem [{\citenamefont {Leibfried}\ \emph {et~al.}(1996)\citenamefont {Leibfried}, \citenamefont {Meekhof}, \citenamefont {King}, \citenamefont {Monroe}, \citenamefont {Itano},\ and\ \citenamefont {Wineland}}]{Leibfried1996Wigner}%
  \BibitemOpen
  \bibfield  {author} {\bibinfo {author} {\bibfnamefont {D.}~\bibnamefont {Leibfried}}, \bibinfo {author} {\bibfnamefont {D.~M.}\ \bibnamefont {Meekhof}}, \bibinfo {author} {\bibfnamefont {B.~E.}\ \bibnamefont {King}}, \bibinfo {author} {\bibfnamefont {C.}~\bibnamefont {Monroe}}, \bibinfo {author} {\bibfnamefont {W.~M.}\ \bibnamefont {Itano}},\ and\ \bibinfo {author} {\bibfnamefont {D.~J.}\ \bibnamefont {Wineland}},\ }\bibfield  {title} {\bibinfo {title} {Experimental determination of the motional quantum state of a trapped atom},\ }\href {https://doi.org/10.1103/PhysRevLett.77.4281} {\bibfield  {journal} {\bibinfo  {journal} {Phys. Rev. Lett.}\ }\textbf {\bibinfo {volume} {77}},\ \bibinfo {pages} {4281} (\bibinfo {year} {1996})}\BibitemShut {NoStop}%
\bibitem [{\citenamefont {Itano}\ \emph {et~al.}(1993)\citenamefont {Itano}, \citenamefont {Bergquist}, \citenamefont {Bollinger}, \citenamefont {Gilligan}, \citenamefont {Heinzen}, \citenamefont {Moore}, \citenamefont {Raizen},\ and\ \citenamefont {Wineland}}]{Itano1993QPN}%
  \BibitemOpen
  \bibfield  {author} {\bibinfo {author} {\bibfnamefont {W.~M.}\ \bibnamefont {Itano}}, \bibinfo {author} {\bibfnamefont {J.~C.}\ \bibnamefont {Bergquist}}, \bibinfo {author} {\bibfnamefont {J.~J.}\ \bibnamefont {Bollinger}}, \bibinfo {author} {\bibfnamefont {J.~M.}\ \bibnamefont {Gilligan}}, \bibinfo {author} {\bibfnamefont {D.~J.}\ \bibnamefont {Heinzen}}, \bibinfo {author} {\bibfnamefont {F.~L.}\ \bibnamefont {Moore}}, \bibinfo {author} {\bibfnamefont {M.~G.}\ \bibnamefont {Raizen}},\ and\ \bibinfo {author} {\bibfnamefont {D.~J.}\ \bibnamefont {Wineland}},\ }\bibfield  {title} {\bibinfo {title} {Quantum projection noise: Population fluctuations in two-level systems},\ }\href {https://doi.org/10.1103/PhysRevA.47.3554} {\bibfield  {journal} {\bibinfo  {journal} {Phys. Rev. A}\ }\textbf {\bibinfo {volume} {47}},\ \bibinfo {pages} {3554} (\bibinfo {year} {1993})}\BibitemShut {NoStop}%
\bibitem [{\citenamefont {Lee}\ \emph {et~al.}(2023)\citenamefont {Lee}, \citenamefont {Lin},\ and\ \citenamefont {Lin}}]{Prototype2023PL}%
  \BibitemOpen
  \bibfield  {author} {\bibinfo {author} {\bibfnamefont {C.-Y.}\ \bibnamefont {Lee}}, \bibinfo {author} {\bibfnamefont {K.-T.}\ \bibnamefont {Lin}},\ and\ \bibinfo {author} {\bibfnamefont {G.-D.}\ \bibnamefont {Lin}},\ }\bibfield  {title} {\bibinfo {title} {Prototype of a phonon laser with trapped ions},\ }\href {https://doi.org/10.1103/PhysRevResearch.5.023082} {\bibfield  {journal} {\bibinfo  {journal} {Phys. Rev. Res.}\ }\textbf {\bibinfo {volume} {5}},\ \bibinfo {pages} {023082} (\bibinfo {year} {2023})}\BibitemShut {NoStop}%
\bibitem [{\citenamefont {Lee}\ and\ \citenamefont {Lin}(2026)}]{lee2026bathfreesqueezedphononlasing}%
  \BibitemOpen
  \bibfield  {author} {\bibinfo {author} {\bibfnamefont {C.-Y.}\ \bibnamefont {Lee}}\ and\ \bibinfo {author} {\bibfnamefont {G.-D.}\ \bibnamefont {Lin}},\ }\href {https://arxiv.org/abs/2601.05575} {\bibinfo {title} {Bath-free squeezed phonon lasing via intrinsic ion-phonon coupling}} (\bibinfo {year} {2026}),\ \Eprint {https://arxiv.org/abs/2601.05575} {arXiv:2601.05575 [quant-ph]} \BibitemShut {NoStop}%
\bibitem [{\citenamefont {Pezz\`e}\ \emph {et~al.}(2018)\citenamefont {Pezz\`e}, \citenamefont {Smerzi}, \citenamefont {Oberthaler}, \citenamefont {Schmied},\ and\ \citenamefont {Treutlein}}]{Pezze2018RMP}%
  \BibitemOpen
  \bibfield  {author} {\bibinfo {author} {\bibfnamefont {L.}~\bibnamefont {Pezz\`e}}, \bibinfo {author} {\bibfnamefont {A.}~\bibnamefont {Smerzi}}, \bibinfo {author} {\bibfnamefont {M.~K.}\ \bibnamefont {Oberthaler}}, \bibinfo {author} {\bibfnamefont {R.}~\bibnamefont {Schmied}},\ and\ \bibinfo {author} {\bibfnamefont {P.}~\bibnamefont {Treutlein}},\ }\bibfield  {title} {\bibinfo {title} {Quantum metrology with nonclassical states of atomic ensembles},\ }\href {https://doi.org/10.1103/RevModPhys.90.035005} {\bibfield  {journal} {\bibinfo  {journal} {Rev. Mod. Phys.}\ }\textbf {\bibinfo {volume} {90}},\ \bibinfo {pages} {035005} (\bibinfo {year} {2018})}\BibitemShut {NoStop}%
\bibitem [{\citenamefont {Marciniak}\ \emph {et~al.}(2022)\citenamefont {Marciniak}, \citenamefont {Feldker}, \citenamefont {Pogorelov}, \citenamefont {Kaubruegger}, \citenamefont {Vasilyev}, \citenamefont {van Bijnen}, \citenamefont {Schindler}, \citenamefont {Zoller}, \citenamefont {Blatt},\ and\ \citenamefont {Monz}}]{Marciniak2022}%
  \BibitemOpen
  \bibfield  {author} {\bibinfo {author} {\bibfnamefont {C.~D.}\ \bibnamefont {Marciniak}}, \bibinfo {author} {\bibfnamefont {T.}~\bibnamefont {Feldker}}, \bibinfo {author} {\bibfnamefont {I.}~\bibnamefont {Pogorelov}}, \bibinfo {author} {\bibfnamefont {R.}~\bibnamefont {Kaubruegger}}, \bibinfo {author} {\bibfnamefont {D.~V.}\ \bibnamefont {Vasilyev}}, \bibinfo {author} {\bibfnamefont {R.}~\bibnamefont {van Bijnen}}, \bibinfo {author} {\bibfnamefont {P.}~\bibnamefont {Schindler}}, \bibinfo {author} {\bibfnamefont {P.}~\bibnamefont {Zoller}}, \bibinfo {author} {\bibfnamefont {R.}~\bibnamefont {Blatt}},\ and\ \bibinfo {author} {\bibfnamefont {T.}~\bibnamefont {Monz}},\ }\bibfield  {title} {\bibinfo {title} {Optimal metrology with programmable quantum sensors},\ }\href {https://doi.org/10.1038/s41586-022-04435-4} {\bibfield  {journal} {\bibinfo  {journal} {Nature}\ }\textbf {\bibinfo {volume} {603}},\ \bibinfo {pages} {604} (\bibinfo {year} {2022})}\BibitemShut {NoStop}%
\bibitem [{\citenamefont {Zhang}\ \emph {et~al.}(2018)\citenamefont {Zhang}, \citenamefont {Peng}, \citenamefont {{\"O}zdemir}, \citenamefont {Pichler}, \citenamefont {Krimer}, \citenamefont {Zhao}, \citenamefont {Nori}, \citenamefont {Liu}, \citenamefont {Rotter},\ and\ \citenamefont {Yang}}]{Zhang2018EP}%
  \BibitemOpen
  \bibfield  {author} {\bibinfo {author} {\bibfnamefont {J.}~\bibnamefont {Zhang}}, \bibinfo {author} {\bibfnamefont {B.}~\bibnamefont {Peng}}, \bibinfo {author} {\bibfnamefont {{\c{S}}.~K.}\ \bibnamefont {{\"O}zdemir}}, \bibinfo {author} {\bibfnamefont {K.}~\bibnamefont {Pichler}}, \bibinfo {author} {\bibfnamefont {D.~O.}\ \bibnamefont {Krimer}}, \bibinfo {author} {\bibfnamefont {G.}~\bibnamefont {Zhao}}, \bibinfo {author} {\bibfnamefont {F.}~\bibnamefont {Nori}}, \bibinfo {author} {\bibfnamefont {Y.-x.}\ \bibnamefont {Liu}}, \bibinfo {author} {\bibfnamefont {S.}~\bibnamefont {Rotter}},\ and\ \bibinfo {author} {\bibfnamefont {L.}~\bibnamefont {Yang}},\ }\bibfield  {title} {\bibinfo {title} {A phonon laser operating at an exceptional point},\ }\href {https://doi.org/10.1038/s41566-018-0213-5} {\bibfield  {journal} {\bibinfo  {journal} {Nature Photonics}\ }\textbf {\bibinfo {volume} {12}},\ \bibinfo {pages} {479} (\bibinfo {year} {2018})}\BibitemShut {NoStop}%
\bibitem [{\citenamefont {Jiang}\ \emph {et~al.}(2018)\citenamefont {Jiang}, \citenamefont {Maayani}, \citenamefont {Carmon}, \citenamefont {Nori},\ and\ \citenamefont {Jing}}]{Jiang2018Nonreciprocal}%
  \BibitemOpen
  \bibfield  {author} {\bibinfo {author} {\bibfnamefont {Y.}~\bibnamefont {Jiang}}, \bibinfo {author} {\bibfnamefont {S.}~\bibnamefont {Maayani}}, \bibinfo {author} {\bibfnamefont {T.}~\bibnamefont {Carmon}}, \bibinfo {author} {\bibfnamefont {F.}~\bibnamefont {Nori}},\ and\ \bibinfo {author} {\bibfnamefont {H.}~\bibnamefont {Jing}},\ }\bibfield  {title} {\bibinfo {title} {Nonreciprocal phonon laser},\ }\href {https://doi.org/10.1103/PhysRevApplied.10.064037} {\bibfield  {journal} {\bibinfo  {journal} {Phys. Rev. Appl.}\ }\textbf {\bibinfo {volume} {10}},\ \bibinfo {pages} {064037} (\bibinfo {year} {2018})}\BibitemShut {NoStop}%
\end{thebibliography}

\begin{thebibliography}{6}%
\makeatletter
\providecommand \@ifxundefined [1]{%
 \@ifx{#1\undefined}
}%
\providecommand \@ifnum [1]{%
 \ifnum #1\expandafter \@firstoftwo
 \else \expandafter \@secondoftwo
 \fi
}%
\providecommand \@ifx [1]{%
 \ifx #1\expandafter \@firstoftwo
 \else \expandafter \@secondoftwo
 \fi
}%
\providecommand \natexlab [1]{#1}%
\providecommand \enquote  [1]{``#1''}%
\providecommand \bibnamefont  [1]{#1}%
\providecommand \bibfnamefont [1]{#1}%
\providecommand \citenamefont [1]{#1}%
\providecommand \href@noop [0]{\@secondoftwo}%
\providecommand \href [0]{\begingroup \@sanitize@url \@href}%
\providecommand \@href[1]{\@@startlink{#1}\@@href}%
\providecommand \@@href[1]{\endgroup#1\@@endlink}%
\providecommand \@sanitize@url [0]{\catcode `\\12\catcode `\$12\catcode `\&12\catcode `\#12\catcode `\^12\catcode `\_12\catcode `\%12\relax}%
\providecommand \@@startlink[1]{}%
\providecommand \@@endlink[0]{}%
\providecommand \url  [0]{\begingroup\@sanitize@url \@url }%
\providecommand \@url [1]{\endgroup\@href {#1}{\urlprefix }}%
\providecommand \urlprefix  [0]{URL }%
\providecommand \Eprint [0]{\href }%
\providecommand \doibase [0]{http://dx.doi.org/}%
\providecommand \selectlanguage [0]{\@gobble}%
\providecommand \bibinfo  [0]{\@secondoftwo}%
\providecommand \bibfield  [0]{\@secondoftwo}%
\providecommand \translation [1]{[#1]}%
\providecommand \BibitemOpen [0]{}%
\providecommand \bibitemStop [0]{}%
\providecommand \bibitemNoStop [0]{.\EOS\space}%
\providecommand \EOS [0]{\spacefactor3000\relax}%
\providecommand \BibitemShut  [1]{\csname bibitem#1\endcsname}%
\let\auto@bib@innerbib\@empty
\bibitem [{\citenamefont {Leibfried}\ \emph {et~al.}(2003)\citenamefont {Leibfried}, \citenamefont {Blatt}, \citenamefont {Monroe},\ and\ \citenamefont {Wineland}}]{singleiondynamics}%
  \BibitemOpen
  \bibfield  {author} {\bibinfo {author} {\bibfnamefont {D.}~\bibnamefont {Leibfried}}, \bibinfo {author} {\bibfnamefont {R.}~\bibnamefont {Blatt}}, \bibinfo {author} {\bibfnamefont {C.}~\bibnamefont {Monroe}}, \ and\ \bibinfo {author} {\bibfnamefont {D.}~\bibnamefont {Wineland}},\ }\href {\doibase 10.1103/RevModPhys.75.281} {\bibfield  {journal} {\bibinfo  {journal} {Rev. Mod. Phys.}\ }\textbf {\bibinfo {volume} {75}},\ \bibinfo {pages} {281} (\bibinfo {year} {2003})}\BibitemShut {NoStop}%
\bibitem [{\citenamefont {Wineland}\ \emph {et~al.}(1998)\citenamefont {Wineland}, \citenamefont {Monroe}, \citenamefont {Itano}, \citenamefont {Leibfried}, \citenamefont {King},\ and\ \citenamefont {Meekhof}}]{wineland1998experimental}%
  \BibitemOpen
  \bibfield  {author} {\bibinfo {author} {\bibfnamefont {D.~J.}\ \bibnamefont {Wineland}}, \bibinfo {author} {\bibfnamefont {C.}~\bibnamefont {Monroe}}, \bibinfo {author} {\bibfnamefont {W.~M.}\ \bibnamefont {Itano}}, \bibinfo {author} {\bibfnamefont {D.}~\bibnamefont {Leibfried}}, \bibinfo {author} {\bibfnamefont {B.~E.}\ \bibnamefont {King}}, \ and\ \bibinfo {author} {\bibfnamefont {D.~M.}\ \bibnamefont {Meekhof}},\ }\href@noop {} {\bibfield  {journal} {\bibinfo  {journal} {Journal of research of the National Institute of Standards and Technology}\ }\textbf {\bibinfo {volume} {103}},\ \bibinfo {pages} {259} (\bibinfo {year} {1998})}\BibitemShut {NoStop}%
\bibitem [{\citenamefont {Dutta}\ and\ \citenamefont {Cooper}(2019)}]{vdpcritical}%
  \BibitemOpen
  \bibfield  {author} {\bibinfo {author} {\bibfnamefont {S.}~\bibnamefont {Dutta}}\ and\ \bibinfo {author} {\bibfnamefont {N.~R.}\ \bibnamefont {Cooper}},\ }\href {\doibase 10.1103/PhysRevLett.123.250401} {\bibfield  {journal} {\bibinfo  {journal} {Phys. Rev. Lett.}\ }\textbf {\bibinfo {volume} {123}},\ \bibinfo {pages} {250401} (\bibinfo {year} {2019})}\BibitemShut {NoStop}%
\bibitem [{\citenamefont {Baur}\ \emph {et~al.}(2026)\citenamefont {Baur}, \citenamefont {Behrle}, \citenamefont {Rojkov}, \citenamefont {Jeske}, \citenamefont {Yelin}, \citenamefont {Home},\ and\ \citenamefont {Reiter}}]{baur2026quantumtheoryphononlasing}%
  \BibitemOpen
  \bibfield  {author} {\bibinfo {author} {\bibfnamefont {D.}~\bibnamefont {Baur}}, \bibinfo {author} {\bibfnamefont {T.}~\bibnamefont {Behrle}}, \bibinfo {author} {\bibfnamefont {I.}~\bibnamefont {Rojkov}}, \bibinfo {author} {\bibfnamefont {J.}~\bibnamefont {Jeske}}, \bibinfo {author} {\bibfnamefont {S.}~\bibnamefont {Yelin}}, \bibinfo {author} {\bibfnamefont {J.}~\bibnamefont {Home}}, \ and\ \bibinfo {author} {\bibfnamefont {F.}~\bibnamefont {Reiter}},\ }\href {https://arxiv.org/abs/2604.18295} {\enquote {\bibinfo {title} {Quantum theory for phonon lasing and non-classical state generation in mixed-species and single trapped ions},}\ } (\bibinfo {year} {2026}),\ \Eprint {http://arxiv.org/abs/2604.18295} {arXiv:2604.18295 [quant-ph]} \BibitemShut {NoStop}%
\bibitem [{\citenamefont {Fl\"uhmann}\ and\ \citenamefont {Home}(2020)}]{fluhmann2020direct}%
  \BibitemOpen
  \bibfield  {author} {\bibinfo {author} {\bibfnamefont {C.}~\bibnamefont {Fl\"uhmann}}\ and\ \bibinfo {author} {\bibfnamefont {J.~P.}\ \bibnamefont {Home}},\ }\href {\doibase 10.1103/PhysRevLett.125.043602} {\bibfield  {journal} {\bibinfo  {journal} {Phys. Rev. Lett.}\ }\textbf {\bibinfo {volume} {125}},\ \bibinfo {pages} {043602} (\bibinfo {year} {2020})}\BibitemShut {NoStop}%
\bibitem [{\citenamefont {Gerry}\ and\ \citenamefont {Knight}(2023)}]{gerry2023introductory}%
  \BibitemOpen
  \bibfield  {author} {\bibinfo {author} {\bibfnamefont {C.~C.}\ \bibnamefont {Gerry}}\ and\ \bibinfo {author} {\bibfnamefont {P.~L.}\ \bibnamefont {Knight}},\ }\href@noop {} {\emph {\bibinfo {title} {Introductory quantum optics}}}\ (\bibinfo  {publisher} {Cambridge university press},\ \bibinfo {year} {2023})\BibitemShut {NoStop}%
\end{thebibliography}
\end{document}